\documentstyle[12pt,axodraw,cheqno,titlepage]{article}

\newenvironment{litterature}{\begin{list}{}{\settowidth{\leftmargin}{[99]\ \ }
  \settowidth{\labelsep}{\ \ }\setlength{\listparindent}{0mm}}}{\end{list}}

\newcommand{\papersettings}%
  {\setlength{\textheight}{8.9in} %{24cm}
  \setlength{\textwidth}{6.5in} %{16cm}
  \setlength{\topmargin}{0cm}
  \setlength{\headheight}{0cm}
  \setlength{\headsep}{0cm}
  \setlength{\oddsidemargin}{0.1in}% {0.5cm}}
  \baselineskip=16pt plus 2pt minus 1pt}
\papersettings

\newcommand{\dbar}{\makebox[0mm]{\hspace{3.4mm}\rule[2.35mm]{1.4mm}{0.05mm}}d}

\newcommand{\refno}[1]{\rm \mbox{$[#1]$}}

\newcommand{\identity}{{\rm 1\hspace{-3.6pt}I}}
\newcommand{\extraline}{\rule[0.36pt]{0.73pt}{7.5pt}}

\newcommand{\extraC}{{\rm \,\extraline\hspace{-3.7pt}C}}

\newcommand{\extraE}{{\rm I\hspace{-2.4pt}E}}

\newcommand{\extraN}{{\rm I\hspace{-2.3pt}I\hspace{-3.6pt}N}}

\newcommand{\extraR}{{\rm I\hspace{-2.3pt}R}}

\newcommand{\om}{\omega}

\newcommand{\be}{\begin{equation}}
\newcommand{\ee}{\end{equation}}
\newcommand{\bea}{\begin{eqnarray}}
\newcommand{\eea}{\end{eqnarray}}

\def\be{\begin{equation}}
\def\ee{\end{equation}}
\def\H{{\cal H}}
\def\L{\Lambda}
\def\t{\tau}
\def\b{\beta}
\def\m{\mu}

\newcommand{\figureBA}{
\parbox[c]{300pt}{
\setlength{\unitlength}{1pt}
\begin{picture}(300,205)(-150,-110)
  \CArc(0,0)(70,0,360)
  \CArc(0,0)(100,0,360)
  \DashCArc(0,0)(85,0,360){2}
  \LongArrow(42.5,73.6)(49.3,85.3)
  \LongArrow(42.5,73.6)(35.8,61.9)
  \put(42.5,68.6){\scriptsize\makebox(0,0)[tl]{$\frac{k_F}{\lambda}$}}
  \put(0,0){\makebox(0,0)[cc]{$>$}}
  \put(-92,92){\makebox(0,0)[cc]{$>$}}
  \put(-58.1,66){\makebox(0,0)[br]{$<$}}
\end{picture}}}

\newcommand{\figureBB}{
\setlength{\unitlength}{1pt}
\parbox[c]{420pt}{
\begin{picture}(420,230)(-65,-150)
  \DashCArc(0,-320)(320,68.5,111.5){2}
  \CArc(0,-320)(360,68.5,111.5)
  \CArc(0,-320)(280,68.5,111.5)
  \CArc(0,-320)(330,90,104.3)
  \CArc(0,-320)(310,90,104.3)
  \Line(-69.3,-48.7)(-89.1,28.8)
  \Line(69.3,-48.7)(89.1,28.8)
  \Line(0,-40)(0,40)
  \Line(-19.4,-10.6)(-20.6,9.4)
  \Line(-38.6,-12.4)(-41.1,7.4)
  \Line(-57.8,-15.4)(-61.5,4.2)
  \LongArrow(-31.1,-122.4)(-49.5,-5.9)
  \put(-37.5,-63.2){\scriptsize\makebox(0,0)[bl]{$k_F\vec{\omega}$}}
  \put(-51.3,9.97){\scriptsize\makebox(0,0)[b]{
    $B_{\vec{\omega}}$}}
  \LongArrow(150,0)(190,0)
  \put(300,0){\SetScale{2}
    \put(0,0){\circle*{4}}
    \put(0,-6){\makebox(0,0)[t]{$0$}}
    \Curve{(-46.9,32.5)(-6.2,39.5)(34.6,45.3)}
    \Curve{(-32,-46.1)(6.2,-39.5)(44.5,-34.1)}
    \DashCurve{(-39.5,-6.8)(0,0)(39.6,4.6)}{8}
    \Line(44.5,-34.1)(34.6,45.3)
    \Line(-32,-46.1)(-46.9,32.5)
    \LongArrow(-51,11.4)(-54.8,31)
    \LongArrow(-43.6,-27.9)(-39.9,-47.6)
    \put(-98,-18){$k_F$}
    \LongArrow(25.7,-45.1)(45.5,-42)
    \LongArrow(-10.8,-50.9)(-30.5,-54)
    \put(12,-99){$k_F$}
    \put(-10,40){\makebox(0,0)[bl]{$\tilde{B}_{\vec{\omega}}$}}}
\end{picture}}}

\newcommand{\figureBC}{
\setlength{\unitlength}{1pt}
\parbox[c]{110pt}{
\begin{picture}(110,80)(-55,-40)
  \ArrowLine(-25,0)(-55,30)
  \ArrowLine(-55,-30)(-25,0)
  \ArrowLine(25,0)(55,30)
  \ArrowLine(55,-30)(25,0)
  \Photon(-25,0)(25,0){2}{7}
\end{picture}}}

\newcommand{\figureCA}{
\setlength{\unitlength}{1pt}
\parbox[c]{100pt}{
\begin{picture}(100,120)(0,-20)
  \ArrowLine(0,0)(50,0)
  \ArrowLine(50,0)(100,0)
  \ArrowArcn(50,60)(20,270,-90)
  \Photon(50,0)(50,40){2}{5.5}
  \put(25,-5){\scriptsize\makebox(0,0)[t]{$\vec{\omega},p$}}
  \put(75,-5){\scriptsize\makebox(0,0)[t]{$\vec{\omega},p$}}
  \put(50,85){\scriptsize\makebox(0,0)[b]{$\vec{\omega}',k$}}
\end{picture}}
\raisebox{-10mm}{$\hspace{10mm}\hspace{10mm}$}
\parbox[c]{140pt}{
\begin{picture}(140,90)(0,-10)
  \ArrowLine(0,0)(50,0)
  \ArrowLine(90,0)(140,0)
  \ArrowArcn(70,0)(20,180,0)
  \Photon(50,0)(90,0){-2}{5.5}
  \put(25,-5){\scriptsize\makebox(0,0)[t]{$\vec{\omega},p$}}
  \put(115,-5){\scriptsize\makebox(0,0)[t]{$\vec{\omega},p$}}
  \put(70,25){\scriptsize\makebox(0,0)[b]{$\vec{\omega}',k$}}
\end{picture}}}

\newcommand{\figureCB}{
\setlength{\unitlength}{1pt}
\parbox[c]{100pt}{
\begin{picture}(100,80)(0,-20)
  \ArrowLine(0,0)(50,0)
  \ArrowLine(50,0)(100,0)
  \ArrowArcn(50,20)(20,270,-90)
  \put(25,-5){\scriptsize\makebox(0,0)[t]{$\vec{\omega},p$}}
  \put(75,-5){\scriptsize\makebox(0,0)[t]{$\vec{\omega},p$}}
  \put(50,45){\scriptsize\makebox(0,0)[b]{$\vec{\omega}',k$}}
\end{picture}}}

\newcommand{\figureCC}{
\setlength{\unitlength}{1pt}
\parbox[c]{100pt}{
\begin{picture}(100,180)(0,-10)
  \ArrowLine(0,0)(50,0)
  \ArrowLine(50,0)(100,0)
  \multiput(50,20)(0,40){3}{\ArrowArcn(0,0)(20,270,90)
    \ArrowArcn(0,0)(20,90,-90)}
  \ArrowArcn(50,140)(20,270,-90)
\end{picture}}
\hspace{20mm}
\parbox[c]{100pt}{
\begin{picture}(100,180)(0,-10)
  \ArrowLine(0,0)(50,0)
  \ArrowLine(50,0)(100,0)
  \multiput(50,20)(0,40){2}{\ArrowArcn(0,0)(20,270,90)}
  \multiput(50,20)(0,40){3}{\ArrowArcn(0,0)(20,90,-90)}
  \ArrowArcn(50,140)(20,270,-90)
  \ArrowArcn(50,100)(20,270,180)
  \ArrowArcn(50,100)(20,180,90)
  \ArrowArcn(10,100)(20,0,180)
  \ArrowArcn(10,100)(20,180,0)
  \ArrowArcn(-30,100)(20,360,0)
\end{picture}}}

\newcommand{\figureCD}{
\setlength{\unitlength}{1pt}
\parbox[c]{220pt}{
\begin{picture}(220,60)(0,-30)
  \ArrowLine(0,0)(50,0)
  \ArrowLine(90,0)(130,0)
  \ArrowLine(170,0)(220,0)
  \CArc(70,0)(20,0,360)
  \CArc(150,0)(20,0,360)
  \put(25,-5){\scriptsize\makebox(0,0)[t]{$\vec{\omega},p$}}
  \put(110,-5){\scriptsize\makebox(0,0)[t]{$\vec{\omega}',k$}}
  \put(195,-5){\scriptsize\makebox(0,0)[t]{$\vec{\omega},p$}}
\end{picture}}}

\newcommand{\figureCE}{
\setlength{\unitlength}{1pt}
\parbox[c]{140pt}{
\begin{picture}(140,70)(0,-40)
  \ArrowLine(0,0)(50,0)
  \ArrowLine(50,0)(90,0)
  \ArrowLine(90,0)(140,0)
  \ArrowArcn(70,0)(20,360,180)
  \ArrowArcn(70,0)(20,180,0)
  \put(25,-3){\scriptsize\makebox(0,0)[t]{$\vec{\omega}$}}
  \put(70,-3){\scriptsize\makebox(0,0)[t]{$\vec{\omega}_2$}}
  \put(115,-3){\scriptsize\makebox(0,0)[t]{$\vec{\omega}$}}
  \put(70,-23){\scriptsize\makebox(0,0)[t]{$\vec{\omega}_1$}}
  \put(70,17){\scriptsize\makebox(0,0)[t]{$\vec{\omega}_3$}}
\end{picture}}}

\newcommand{\figureCF}{
\setlength{\unitlength}{1pt}
\parbox[c]{220pt}{
\begin{picture}(220,120)(0,-60)
  \Line(0,0)(50,0)
  \Line(55.9,14.1)(20.5,49.5)
  \Line(55.0,-14.1)(20.5,-49.5)
  \Line(90,0)(130,0)
  \Line(170,0)(220,0)
  \CArc(70,0)(20,0,360)
  \CArc(150,0)(20,0,360)
\end{picture}}}

\newcommand{\figureCG}{
\setlength{\unitlength}{1pt}
\parbox[c]{115pt}{
\begin{picture}(115,80)(-15,-40)
  \ArrowLine(30,0)(0,30)
  \ArrowLine(0,-30)(30,0)
  \ArrowArcn(50,0)(20,0,180)
  \ArrowArcn(50,0)(20,180,0)
  \ArrowLine(70,0)(100,30)
  \ArrowLine(100,-30)(70,0)
  \put(7,15){\scriptsize\makebox(0,0)[r]{$\vec{\omega}_f,p_f$}}
  \put(7,-15){\scriptsize\makebox(0,0)[r]{$\vec{\omega}_i,p_i$}}
  \put(50,-25){\scriptsize\makebox(0,0)[t]{$\vec{\omega},k$}}
\end{picture}}
\mbox{$\hspace{10mm}+\hspace{10mm}$}
\parbox[c]{80pt}{
\begin{picture}(80,130)(-40,-60)
  \ArrowLine(-30,-50)(0,-20)
  \ArrowLine(30,-50)(0,-20)
  \ArrowArc(0,0)(20,270,90)
  \ArrowArcn(0,0)(20,270,90)
  \ArrowLine(0,20)(-30,50)
  \ArrowLine(0,20)(30,50)
  \put(-21,-35){\scriptsize\makebox(0,0)[r]{$\vec{\omega}_i,p_i$}}
  \put(21,-35){\scriptsize\makebox(0,0)[l]{$\vec{\omega}'_i,p'_i$}}
  \put(25,0){\scriptsize\makebox(0,0)[l]{$\vec{\omega},k$}}
\end{picture}}}

\newcommand{\figureCH}{
\setlength{\unitlength}{1pt}
\parbox[c]{120pt}{
\begin{picture}(120,100)(0,-50)
  \ArrowLine(0,-40)(0,0)
  \ArrowLine(0,0)(0,40)
  \ArrowLine(120,-40)(120,0)
  \ArrowLine(120,0)(120,40)
  \ArrowArcn(60,0)(20,180,360)
  \ArrowArcn(60,0)(20,0,180)
  \Photon(0,0)(40,0){2}{5.5}
  \Photon(80,0)(120,0){2}{5.5}
\end{picture}}
\mbox{$\hspace{20pt}+\hspace{20pt}$}
\parbox[c]{60pt}{
\begin{picture}(60,140)(0,-70)
  \ArrowLine(0,-40)(0,0)
  \ArrowLine(0,0)(0,40)
  \ArrowArcn(60,0)(20,270,180)
  \ArrowArcn(60,0)(20,180,90)
  \Photon(0,0)(40,0){2}{5.5}
  \Photon(60,-20)(60,20){-2}{5.5}
  \ArrowLine(60,20)(60,60)
  \ArrowLine(60,-60)(60,-20)
\end{picture}}
\mbox{$\hspace{20pt}+\hspace{20pt}$}
\parbox[c]{40pt}{
\begin{picture}(40,140)(0,-70)
  \ArrowLine(0,-60)(0,-20)
  \ArrowLine(0,-20)(0,20)
  \ArrowLine(0,20)(0,60)
  \ArrowLine(40,-60)(40,-20)
  \ArrowLine(40,-20)(40,20)
  \ArrowLine(40,20)(40,60)
  \Photon(0,20)(40,-20){2}{7}
  \Photon(0,-20)(40,20){2}{7}
\end{picture}}}

\newcommand{\figureCI}{
\setlength{\unitlength}{1pt}
\parbox[c]{120pt}{
\begin{picture}(120,70)(0,-40)
  \ArrowLine(0,20)(40,20)
  \ArrowLine(40,20)(80,20)
  \ArrowLine(80,20)(120,20)
  \ArrowLine(0,-20)(40,-20)
  \ArrowLine(40,-20)(80,-20)
  \ArrowLine(80,-20)(120,-20)
  \put(20,15){\scriptsize\makebox(0,0)[t]{$\vec{\omega}_i,p_i$}}
  \put(20,-25){\scriptsize\makebox(0,0)[t]{$\vec{\omega}'_i,p'_i$}}
  \put(60,-25){\scriptsize\makebox(0,0)[t]{$\vec{\omega},k$}}
  \Photon(40,-20)(40,20){2}{5.5}
  \Photon(80,-20)(80,20){2}{5.5}
\end{picture}}}

\newcommand{\figureCJ}{
\parbox[c]{200pt}{
\setlength{\unitlength}{1pt}
\begin{picture}(200,230)(-100,-110)
  \CArc(0,0)(100,0,360)
  \put(16.1,59.9){\circle*{3}}
  \Oval(16.1,59.9)(26.3,74.9)(-15)
  \LongArrow(0,0)(-25.1,45.8)
  \LongArrow(0,0)(56.9,70.7)
  \LongArrow(0,0)(32.2,119.8)
  \DashLine(-26,47.6)(32.2,119.8){2}
  \DashLine(58.2,72.2)(32.2,119.8){2}
  \DashLine(-26,47.6)(58.2,72.2){2}
  \LongArrow(0,0)(60.6,27.4)
  \LongArrow(0,0)(-29.6,89.7)
  \DashLine(62.4,28.2)(-30.2,91.6){2}
  \DashLine(62.4,28.2)(32.2,119.8){2}
  \DashLine(-30.2,91.6)(32.2,119.8){2}
\end{picture}}}

\newcommand{\figureCK}{
\parbox[c]{140pt}{
\setlength{\unitlength}{1pt}
\begin{picture}(140,60)(-70,-30)
  \ArrowLine(-70,0)(-20,0)
  \ArrowLine(20,0)(70,0)
  \CArc(0,0)(20,0,360)
  \put(0,0){\scriptsize\makebox(0,0)[c]{$\vec{\omega},p$}}
\end{picture}}}

\newcommand{\figureCL}{
\parbox[c]{260pt}{
\setlength{\unitlength}{1pt}
\begin{picture}(260,70)(-130,-40)
  \ArrowLine(-130,20)(-80,20)
  \ArrowLine(-130,-20)(-80,-20)
  \ArrowLine(-80,20)(-40,20)
  \ArrowLine(-80,-20)(-40,-20)
  \Line(-40,20)(-20,20)
  \Line(-40,-20)(-20,-20)
  \Line(20,20)(40,20)
  \Line(20,-20)(40,-20)
  \ArrowLine(40,20)(80,20)
  \ArrowLine(40,-20)(80,-20)
  \ArrowLine(80,20)(130,20)
  \ArrowLine(80,-20)(130,-20)
  \multiput(-6,0)(6,0){3}{\circle*{2}}
  \Photon(-80,-20)(-80,20){2}{5.5}
  \Photon(-40,-20)(-40,20){2}{5.5}
  \Photon(40,-20)(40,20){2}{5.5}
  \Photon(80,-20)(80,20){2}{5.5}
  \put(-102,16){\scriptsize\makebox(0,0)[tr]{$\vec{\omega}'$}}
  \put(-102,-24){\scriptsize\makebox(0,0)[tr]{$-\vec{\omega}'$}}
  \put(-57,16){\scriptsize\makebox(0,0)[tr]{$\vec{\omega}_1$}}
  \put(-57,-24){\scriptsize\makebox(0,0)[tr]{$-\vec{\omega}_1$}}
  \put(63,16){\scriptsize\makebox(0,0)[tr]{$\vec{\omega}_n$}}
  \put(63,-24){\scriptsize\makebox(0,0)[tr]{$-\vec{\omega}_n$}}
  \put(108,16){\scriptsize\makebox(0,0)[tr]{$\vec{\omega}$}}
  \put(108,-24){\scriptsize\makebox(0,0)[tr]{$-\vec{\omega}$}}
\end{picture}}}

\newcommand{\figureCM}{
\parbox[c]{140pt}{
\setlength{\unitlength}{1pt}
\begin{picture}(140,60)(-70,-30)
  \ArrowLine(-70,0)(-20,0)
  \ArrowLine(20,0)(70,0)
  \GCirc(0,0){20}{0.9}
  \Text(0,0)[c]{\scriptsize $\vec{\omega},p$}
\end{picture}}}

\newcommand{\figureCN}{
\parbox[c]{360pt}{
\setlength{\unitlength}{1pt}
\begin{picture}(360,100)(-210,-50)
  \ArrowLine(-210,30)(-160,30)
  \ArrowLine(-210,-30)(-160,-30)
  \Text(-185,25)[t]{\scriptsize $\vec{\omega}'$}
  \Text(-185,-35)[t]{\scriptsize $-\vec{\omega}'$}
  \Photon(-160,-30)(-160,30){2}{7}
  \ArrowLine(-160,30)(-140,30)
  \ArrowLine(-160,-30)(-140,-30)
  \GCirc(-130,30){10}{0.9}
  \GCirc(-130,-30){10}{0.9}
  \Text(-130,30)[c]{\scriptsize $\vec{\omega}_1$}
  \Text(-130,-30)[c]{\scriptsize $-\vec{\omega}_1$}
  \ArrowLine(-120,30)(-100,30)
  \ArrowLine(-120,-30)(-100,-30)
  \Photon(-100,-30)(-100,30){2}{7}
  \ArrowLine(-100,30)(-80,30)
  \ArrowLine(-100,-30)(-80,-30)
  \GCirc(-70,30){10}{0.9}
  \GCirc(-70,-30){10}{0.9}
  \Text(-70,30)[c]{\scriptsize $\vec{\omega}_2$}
  \Text(-70,-30)[c]{\scriptsize $-\vec{\omega}_2$}
  \ArrowLine(-60,30)(-40,30)
  \ArrowLine(-60,-30)(-40,-30)
  \Photon(-40,-30)(-40,30){2}{7}
  \ArrowLine(-40,30)(-20,30)
  \ArrowLine(-40,-30)(-20,-30)
  \multiput(-6,0)(6,0){3}{\circle*{2}}
  \ArrowLine(20,30)(40,30)
  \ArrowLine(20,-30)(40,-30)
  \Photon(40,-30)(40,30){2}{7}
  \ArrowLine(40,30)(60,30)
  \ArrowLine(40,-30)(60,-30)
  \GCirc(70,30){10}{0.9}
  \GCirc(70,-30){10}{0.9}
  \Text(70,30)[c]{\scriptsize $\vec{\omega}_n$}
  \Text(70,-30)[c]{\scriptsize $-\vec{\omega}_n$}
  \ArrowLine(80,30)(100,30)
  \ArrowLine(80,-30)(100,-30)
  \Photon(100,-30)(100,30){2}{7}
  \ArrowLine(100,30)(150,30)
  \ArrowLine(100,-30)(150,-30)
  \Text(125,25)[t]{\scriptsize $\vec{\omega}$}
  \Text(125,-35)[t]{\scriptsize $-\vec{\omega}$}
\end{picture}}}

\newcommand{\figureCOa}{
\parbox[c]{60pt}{
\setlength{\unitlength}{1pt}
\begin{picture}(60,40)(-30,-20)
  \ArrowLine(-30,0)(-10,0)
  \ArrowLine(10,0)(30,0)
  \GCirc(0,0){10}{0.9}
  \Text(0,0)[c]{\scriptsize $\vec{\omega},k$}
\end{picture}}}

\newcommand{\figureCOb}{
\parbox[c]{60pt}{
\setlength{\unitlength}{1pt}
\begin{picture}(60,40)(-30,-20)
  \ArrowLine(-30,0)(-10,0)
  \ArrowLine(10,0)(30,0)
  \GCirc(0,0){10}{0.9}
  \Text(0,0)[c]{\small 1PI}
  \Text(-20,-5)[t]{\scriptsize $\vec{\omega},k$}
  \Text(20,-5)[t]{\scriptsize $\vec{\omega},k$}
\end{picture}}}

\newcommand{\figureCP}{
\parbox[c]{60pt}{
\setlength{\unitlength}{1pt}
\begin{picture}(60,40)(-30,-20)
  \ArrowLine(-30,0)(-10,0)
  \ArrowLine(10,0)(30,0)
  \GCirc(0,0){10}{0.9}
  \Text(0,0)[c]{\scriptsize $\vec{\omega},k$}
\end{picture}}}

\newcommand{\figureCQ}{
\setlength{\unitlength}{1pt}
\parbox[c]{80pt}{
\begin{picture}(80,130)(-40,-60)
  \ArrowLine(-30,-50)(0,-20)
  \ArrowLine(30,-50)(0,-20)
  \ArrowArc(0,0)(20,270,90)
  \ArrowArcn(0,0)(20,270,90)
  \ArrowLine(0,20)(-30,50)
  \ArrowLine(0,20)(30,50)
\end{picture}}
\mbox{$\hspace{10mm}+\hspace{10mm}$}
\parbox[c]{115pt}{
\begin{picture}(115,80)(-15,-40)
  \ArrowLine(30,0)(0,30)
  \ArrowLine(0,-30)(30,0)
  \ArrowArcn(50,0)(20,0,180)
  \ArrowArcn(50,0)(20,180,0)
  \ArrowLine(70,0)(100,30)
  \ArrowLine(100,-30)(70,0)
\end{picture}}}

\newcommand{\figureCR}{
\setlength{\unitlength}{1pt}
\parbox[c]{210pt}{
\begin{picture}(210,225)(-105,370)
  \CArc(0,0)(600,80,100)
  \CArc(0,0)(400,80,100)
  \CArc(0,0)(470,80,100)
  \CArc(0,0)(530,80,100)
  \DashCArc(0,0)(500,80,100){3}
  \DashCArc(0,0)(490,80,100){3}
  \LongArrow(0,500)(0,491.5)
  \LongArrow(34.9,498.8)(34.3,490.3)
  \LongArrow(-34.9,498.8)(-34.3,490.3)
  \LongArrow(69.6,495.1)(68.4,486.7)
  \LongArrow(-69.6,495.1)(-68.4,486.7)
  \Text(94.5,485.9)[l]{\}\ \scriptsize $-\frac{\delta\mu^{(j)}}{v_F^{(j)}}$}
  \LongArrow(-94.5,485.9)(-100.8,518.8)
  \LongArrow(-94.5,485.9)(-90,462.8)
  \Text(-96.5,485.9)[r]{\scriptsize $\frac{1}{\lambda_j}$}
  \LongArrow(-112.5,487.2)(-134.6,583.2)
  \LongArrow(-112.5,487.2)(-90.3,391.2)
  \Text(-114.5,487.2)[r]{\scriptsize $\frac{1}{\lambda_{j-1}}$}
\end{picture}}}  

\newcommand{\figureCS}{
\setlength{\unitlength}{1pt}
\parbox[c]{240pt}{
\begin{picture}(240,130)(-120,-65)
  \CArc(-90,0)(15,90,270)
  \CArc(90,0)(15,270,90)
  \Line(-90,15)(90,15)
  \Line(-90,-15)(90,-15)
  \ArrowLine(-90,15)(-120,45)
  \ArrowLine(-120,-45)(-90,-15)
  \ArrowLine(90,15)(120,45)
  \ArrowLine(120,-45)(90,-15)
  \ArrowLine(-60,-15)(-60,-45)
  \ArrowLine(-30,-45)(-30,-15)
  \multiput(-10,-30)(10,0){3}{\circle*{2}}
  \ArrowLine(60,-15)(60,-45)
  \ArrowLine(30,-45)(30,-15)
  \Text(-125,50)[br]{$\scriptsize 2k$}
  \Text(-120,-50)[tl]{$\scriptsize \vec{\omega}_i$}
  \Text(-125,-45)[br]{$\scriptsize 1$}
  \Text(-60,-50)[t]{$\scriptsize 2$}
  \Text(-30,-50)[t]{$\scriptsize 3$}
  \Text(125,-50)[tl]{$\scriptsize \vec{\omega}_i'$}
\end{picture}}}

\newcommand{\figureCT}{
\setlength{\unitlength}{1pt}
\parbox[c]{240pt}{
\begin{picture}(240,130)(-120,-65)
  \CArc(-90,0)(15,90,270)
  \CArc(90,0)(15,270,90)
  \Line(-90,15)(90,15)
  \Line(-90,-15)(90,-15)
  \ArrowLine(-90,15)(-120,45)
  \ArrowLine(-120,-45)(-90,-15)
  \ArrowLine(90,15)(120,45)
  \ArrowLine(120,-45)(90,-15)
  \ArrowLine(-60,-15)(-60,-45)
  \ArrowLine(-30,-45)(-30,-15)
  \multiput(-10,-30)(10,0){3}{\circle*{2}}
  \ArrowLine(60,-15)(60,-45)
  \ArrowLine(30,-45)(30,-15)
  \ArrowArc(-45,-45)(15,180,360)
  \ArrowArc(45,-45)(15,180,360)
  \Text(-125,50)[br]{$\scriptsize 2k$}
  \Text(-125,-50)[tr]{$\scriptsize \vec{\omega}_i$}
  \Text(125,-50)[tl]{$\scriptsize \vec{\omega}_i'$}
\end{picture}}}

\begin{document}

\baselineskip=16pt plus 2pt minus 1pt

\title{Renormalization Group Methods:\\
  Landau-Fermi Liquid and BCS Superconductor}
\author{Research Group in Mathematical Physics
  \thanks{\normalsize\ T. Chen, J. Fr\"ohlich, M. Seifert}\\
  Theoretical Physics\\
  ETH-H\"onggerberg\\
  CH--8093 Z\"urich}
\date{}
\maketitle

\tableofcontents

\vspace{\baselineskip}
This is the second part of the notes to the course on quantum theory of large
systems of non-relativistic matter taught by J. Fr\"{o}hlich at the 1994 Les
Houches summer school. It is devoted to a sketchy exposition of some of the
beautiful and important, recent results of J.Feldman and E.Trubowitz, and
J. Feldman, H. Kn\"{o}rrer, D. Lehmann, J. Magnen, V. Rivasseau and
E. Trubowitz. Their results are about a mathematical analysis of
non-relativistic many-body theory, in particular of the Landau-Fermi liquid
and BCS superconductivity, using Wilson's renormalization group methods and
the techniques of the $\frac{1}{N}$-expansion. While their work is ultimately
aimed at a complete mathematical control (beyond perturbative expansions) of
systems of weakly coupled electron gases at positive density and small or zero
temperature, we can only illustrate some of their ideas within the
context of perturbative solutions  of Wilson-type renormalization group flow
equations (we calculate leading-order terms in a $\frac{1}{N}$-expansion,
where $N$ is an energy scale) and of one-loop effective potential
calculations of the BCS superconducting ground state. We therefore urge the
reader to consult the following original articles :

\begin{litterature}
\item[\refno{1}] J. Feldman and E. Trubowitz, Helv. Phys. Acta {\bf 63}, 156 
  (1990)
  \\
  G. Benfatto and G. Gallavotti, J. Stat. Phys. {\bf 59}, 541 (1991)
\item[\refno{2}] J. Feldman and E. Trubowitz, Helv. Phys. Acta {\bf 64}, 214 
  (1991) 
\item[\refno{3}] J. Feldman, J. Magnen, V. Rivasseau and E. Trubowitz,
  Helv. Phys. Acta {\bf 65}, 679 (1992)
\item[\refno{4}] J. Feldman, J. Magnen, V. Rivasseau and E. Trubowitz,
  Europhysics Letters {\bf 24}, 437 (1993)
\item[\refno{5}] J. Feldman, J. Magnen, V. Rivasseau and E. Trubowitz,
  Europhysics Letters {\bf 24}, 521 (1993)
\item[\refno{6}] J. Feldman, J. Magnen, V. Rivasseau and E. Trubowitz,
  Fermionic Many-Body Models, in {\sl Mathematical Quantum Theory I: Field
  Theory and Many-Body Theory}, J. Feldman, R. Froese and L. Rosen eds, CRM
  Proceedings and Lecture Notes, 1994.
\item[\refno{7}] J. Feldman, D. Lehmann, H. Kn\"{o}rrer and E. Trubowitz, Fermi
  Liquids in Two Space Dimensions, in ``Constructive Physics'', V. Rivasseau
  (ed.), Lecture Notes in Physics vol. 446, Berlin, Heidelberg, New York:
  Springer-Verlag, 1995.
\end{litterature}

In Chapter 1, we review some of the standard material on thermodynamics,
quantum statistical mechanics and functional integration. Sources for this
material can be found in

\begin{litterature}
\item[\refno{8}] D. Ruelle, Statistical Mechanics (Rigorous results), 
  London and Amsterdam: Benjamin, 1969.
\item[\refno{9}] O. Bratteli and D. W. Robinson, Operator Algebras and Quantum
  Statistical Mechanics, Berlin, Heidelberg and New York: Springer-Verlag, 1987
  (I, second edition), and 1981 (II).
\item[\refno{10}] J. W. Negele and H. Orland, Quantum Many-Particle Systems,
  Frontiers in Physics vol 68, New York: Addison-Wesley, 1987.
\item[] Refs. 6 and 7 above.
\end{litterature}

In Chapter 2, we consider weakly coupled electron gases. We recall the notion
of scaling limit (large distance scales and low frequencies). We show that, in
the scaling limit, systems of free electrons at positive density and small
temperature in $d=1,2,3,...$ space dimensions can be mapped onto systems of
multi-flavour, free chiral Dirac fermions in $1+1$ space-time dimensions, the
flavour index corresponding to a direction on the Fermi sphere. This is the
first manifestation of a ``principle of dimensional reduction'' which says
that, in the scaling limit (infrared domain), electron gases at positive
density and very small temperatures have properties analogous to those of
certain $1+1$ dimensional (relativistic) models of Dirac fermions with
four-fermion interactions.

We then introduce the basic ideas underlying a renormalization group analysis
of weakly coupled electron gases and show that there is a parameter, related
to a (dimensionless) ratio of energy scales, that plays a r\^{o}le analogous to
the number, $N$, of fermion flavours in $1+1$ dimensional models of Dirac
fermions with four-fermion interactions, such as the Gross-Neveu model. This
enables us to study the renormalization flow in weakly coupled electron gases
by using $\frac{1}{N}$-techniques. We follow ideas first presented in refs.\
[1], [4] and [5].

Useful additional references to Chapter 2 are:

\begin{litterature}
\item[\refno{11}] K. G. Wilson and J. B. Kogut, Phys. Reports {\bf 12}, 7 
  (1974)
\item[\refno{12}] J. Fr\"{o}hlich, R. G\"{o}tschmann and P. A. Marchetti, ``The
  Effective Gauge Field Action of a System of Non-Relativistic Electrons'', to
  appear in Commun. Math. Phys. 1995
\item[](See also Chapter 5 of Part I.)
\end{litterature}

In Chapter 3, we derive the renormalization group flow equations, to leading
order in a $\frac{1}{N}$-expansion, for systems of non-relativistic electrons
at positive density and zero temperature interacting through weak two-body
potentials of short range (``screening''), following ideas in refs.\
[1] - [5]. Special attention is given to understanding the striking
similarities  
with $1+1$ dimensional Gross-Neveu type models and to an analysis of the flow
of BCS couplings (``BCS instability''). Besides refs.\ [1] - [7] and [11], the
following references will prove useful:

\begin{litterature}
\item[\refno{13}] A. L. Fetter and J.D. Walecka, Quantum Theory of 
  Many-Particle Systems, New York: McGraw-Hill, 1971.
\item[\refno{14}] J. Solyom, Adv. Phys. {\bf 28}, 201 (1979); (RG methods for
  one-dimensional systems).
\item[\refno{15}] J. Polchinski, TASI Lectures 1992; (effective actions).
\item[\refno{16}] R. Shankar, Rev. Mod. Phys. {\bf 66}, 129 (1994); 
  (RG methods for
  ($d\geq 2$)-dimensional, non-relativistic many-particle systems).
\item[\refno{17}] D. J. Gross and A. Neveu, Phys. Rev. D {\bf 10}, 3235 (1974);
  ($\frac{1}{N}$-expansion in the Mitter-Weisz-Gross-Neveu model).
\item[\refno{18}] K. Gaw\c{e}dzki and A. Kupiainen, Commun. Math. Phys. 
  {\bf 102}, 1
  (1985); (rigorous renormalization group analysis of the Gross-Neveu model).
\item[\refno{19}] G. Gallavotti, Rev. Mod. Phys. {\bf 57}, 471 (1985); 
  (RG methods in scalar field theories).
\item[\refno{20}] J. Polchinski, Nucl. Phys. B {\bf 231}, 269 (1984)
\end{litterature}

In Chapter 4, we study the BCS superconducting ground state, the spontaneous
breaking of the U(1) global gauge symmetry (the particle number symmetry) in
the superconducting ground state and the emergence of a massless Goldstone
boson associated with the broken symmetry. We make use of the Nambu-Gorkov
formalism and large-$N$ techniques, (refs.\ [2] - [5]). We emphasize the
striking 
analogies between weakly coupled BCS s-wave superconductors and the $1+1$
dimensional, large-$N$ chiral Gross-Neveu model. Our discussion of symmetry
breaking is based on the loop expansion of the effective action of a charged,
scalar field describing Cooper pairs. In the mean-field approximation, it
reduces to the calculation of an effective potential that proceeds along the
lines of the calculations in ref.\ [17]. In order to understand the dependence
of symmetry breaking on dimension and temperature and the dynamics of
Goldstone bosons, we calculate leading corrections to mean-field theory. 

Useful additional references for Chapter 4 are:

\begin{litterature}
\item[\refno{21}] P. -G. de Gennes, Superconductivity of Metals and Alloys, New
  York: Benjamin, 1966.
\item[\refno{22}] J. R. Schrieffer, Theory of Superconductivity, Menlo Park:
  Benjamin-Cummings, 1964.
\item[\refno{23}] A. A. Abrikosov, L. P. Gorkov and I. E. Dzyaloshinsky, 
  Methods of
  Quantum Field Theory in Statistical Physics, New York: Dover, 1975.
\item[\refno{24}] S. Coleman and E. Weinberg, Phys. Rev. D {\bf 7}, 1888 
  (1973); and ref.\ [17].
\item[\refno{25}] Y. Nambu and G. Jona-Lasinio, Phys. Rev. {\bf 122}, 345 
  (1961)
\item[\refno{26}] J. Goldstone, Nuovo Cimento {\bf 19}, 154 (1961);

  J. Goldstone, A. Salam and S. Weinberg, Phys. Rev. {\bf 127}, 965 (1961)
\item[\refno{27}] D. Mermin and H. Wagner, Phys. Rev. Letters {\bf 17}, 1133 
  (1966);

  D. Mermin, J. Math. Phys. {\bf 8}, 1061 (1967)
\end{litterature}

For rigorous results on continuous symmetry breaking, see also
\begin{litterature}
\item[\refno{28}] J. Fr\"ohlich, B. Simon and T. Spencer, Commun. Math. Phys.
  {\bf 50}, 79 (1976); 

  J. Fr\"ohlich and T. Spencer, 
  in ``Scaling and Self-Similarity in Physics'',\\
  Progress in Physics, Basel, Boston: Birkh\"auser, 1983.
\end{litterature}

{\bf Acknowledgements :} J. F. thanks J. Magnen and E. Trubowitz for very
helpful discussions on the results in refs.\ 1 - 7. We thank A. Alekseev,
P. Boschung, J. Hoppe and A. Recknagel for their help in the preparation of 
these
notes. Special thanks are due to R. G\"{o}tschmann with whom we had countless,
illuminating discussions on renormalization group methods in many-body theory,
and who participated in the work underlying these notes.

%%%%%%%%%%%%%%%%%%%%%%%%%%%%%%%%%%%%%%%%%%%%%%%%%%%%%%%%%%%%%%%%%%%%%%%%%%
\section{Background material}
%%%%%%%%%%%%%%%%%%%%%%%%%%%%%%%%%%%%%%%%%%%%%%%%%%%%%%%%%%%%%%%%%%%%%%%%%%

In this chapter we recall some basic definitions
of quantum statistical mechanics, introduce 
ensembles of identical particles, fermions and bosons,
and express Euclidean correlation functions in terms of functional integrals. 

%%%%%%%%%%%%%%%%%%%%%%%%%%%%%%%%%%%%%%%%%%%%%%%%%%%%%%%%%%%%%%%%%%%%%%%%%%%
\subsection{Thermodynamics and quantum statistical mechanics}

We first recall the definition of {\em statistical
ensemble}. We assume that a quantum mecha\-ni\-cal system is confined
to a box $\Lambda\subset \extraE^{3}$ in 3-dimensional
Euclidean space. Important characteristics of the 
system are its volume $|\Lambda|$
and the particle number $N= n|\Lambda|$, where $n$ is the particle density. 

The quantum-mechanical state space of an N-particle system in the box
$\Lambda$ is a separable Hilbert space ${\cal H}^{(N)}_{\Lambda}$, its
dynamics is given by a selfadjoint Hamilton operator $H^{(N)}_{\Lambda}$
acting on ${\cal H}^{(N)}_{\Lambda}$, whose energy spectrum is discrete and
bounded from below.

In statistical mechanics, one considers 
two standard ensembles: One either fixes the number of particles in the system
({\em canonical ensemble}), or
one permits the number of particles to vary and keeps fixed
only its  mean value $\langle N\rangle$ ({\em grand canonical
ensemble}). Here, N denotes the particle number operator, and
$\langle (\cdot) \rangle$ is the
expectation in some state of the system.

Quantities of interest in thermodynamics
are {\em thermodynamic potentials}. One of the most
important such potential  is the free energy $F$:
\begin{equation}
  \b F(\b,\L,N)=- \log Z(\b,\L,N)
\end{equation}
where the canonical partition function $Z$ is defined by
\begin{equation}
  Z(\b,\L,N)=\mbox{Tr}_{\H^{(N)}_{\L}}\ \left( e^{-\b H^{(N)}_{\L}}\right) 
\end{equation}
The free energy refers to the canonical ensemble, since
the number of particles is fixed. One also 
introduces a free energy per unit volume (per particle) by setting
\be 
f(\b,\L,N)=\frac{1}{|\L|}F(\b,\L,N)\ \ 
(\ = \frac{1}{N} F(\b,\L,N)\ ,\ \mbox{resp.}\ )
\ee

To an ordered sequence, 
$0<\tau_n <\cdots <\tau_2 <\tau_1 <\b$, of {\em imaginary times}, and an 
n-tuple 
of observables (bounded operators on ${\cal H}_{\Lambda}^{(N)}$), 
$a_1, a_2 \dots a_n$, one associates a {\em
temperature-ordered Green (correlation) function}
\be \label{correlation}
\frac{1}{Z}
\mbox{Tr}\, \left( e^{-\b H} e^{\t_1 H} a_1 e^{-(\t_1-\t_2)H} a_2
\dots a_n e^{-\t_n H}\right)
\ee
(where we have omitted the subscripts $\Lambda$ and superscripts $N$).

The grand canonical partition function is defined by
\be
\Xi(\b,\L,\m)=\mbox{Tr}\,_{\H_{\L}}\ \left( e^{-\b (H-\m N)}\right)
=e^{\b \Omega (\b,\L,\m)}
\ee
The coefficient $\m$ is called {\em chemical potential}. 
The thermodynamic potential $\Omega$ is the so called {\em grand potential}. 
We denote by $\H_{\Lambda}=\oplus_N \H^{(N)}_{\Lambda}$ the direct sum of
Hilbert spaces corresponding to different numbers of identical particles
(bosons or fermions). The Hamiltonian
$H$ is then a direct sum of $N$-particle Hamiltonians.
In each $N$-particle subspace, $H$ coincides with $H^{(N)}$ (we again omit the
subscript $\Lambda$).
In nonrelativistic quantum theory, we assume that the
particle number operator, $N$, and the Hamiltonian commute;
in relativistic theories this is usually not the case because
of particle creation-annihilation processes.

The laws of thermodynamics are described by differential
{\em thermodynamic relations}. In the grand canonical ensemble the most 
important such relation is
\be
d\Omega (\b,\L,\m )=\langle N\rangle d\m +S_{\L }dT
\ee
The differential, $d$, is taken with respect to $\beta=\frac{1}{k_BT}$ (where
$k_{B}$ is Boltzmann's constant)
and $\m$;
the box $\L$ is kept fixed. The coefficient in front of $d\m$
is the mean value (in an equilibrium state at inverse temperature $\beta$ and
chemical potential $\mu$) of the particle number operator already encountered
in the definition of the grand canonical ensemble.
The coefficient of $dT$ is the {\em entropy} of the system.

In the study of macroscopically large systems it is useful to consider the
{\em thermodynamic limit}, $\Lambda\nearrow\extraE^3$, and to pass from 
{\em extensive} quantities ($F,\Omega ,N,\ldots$) to {\em intensive} ones
($f,p,n,\d\ldots$). For example, in the grand canonical ensemble one defines
the {\em pressure}, $p$, by setting
\be
p(\b, \m)=\lim_{\L\nearrow \extraE^3} \ \frac{1}{|\L|} \Omega(\b,\L,\m)
\ee
(where $\Lambda\nearrow\extraE^3$ in the sense of van Hove or Fisher).
The particle density and the specific entropy are then given by
\be
n=\frac{\partial p}{\partial \m} \ , \ s=\frac{\partial p}{\partial T}
\ee

Grand canonical correlation functions are defined for systems in a box
$\Lambda$  in the
same way as for the canonical ensemble.
They satisfy the
{\em KMS (Kubo-Martin-Schwinger) condition} which we now describe: Given a
*algebra ${\cal A}$ of quantum-mechanical observables, we define
a {\em state} on ${\cal A}$
as a normalized $(\rho (\identity )=1)$, positive $\bigl( \rho
(a^* a) \geq 0 ,\; \forall a\in{\cal A}\bigr)$, linear functional
$\bigl(\rho(\alpha a + \beta b) = \alpha \rho (a) + \beta \rho (b), \;
\forall a,b\in{\cal A}, \alpha,\beta\in\extraC\bigr)$. An example of a state
is an equilibrium state at some inverse temperature $\beta$ defined by
\be \label{Gibbs}
\rho_{\b}(a)=\frac{1}{\Xi}\ \mbox{Tr}\,(e^{-\b (H-\mu N)} a)
\ee
It is commonly called {\em Gibbs state}.
Introducing time-dependent observables in the Heisenberg
picture by setting 
\be \label{Heis}
a(t)=e^{it(H-\mu N)} a e^{-it(H-\mu N)} \ \ \ ,\ a\in {\cal A}
\ee
we can formulate the KMS condition as follows:
For arbitrary elements a and b of $\cal{A}$, there exists a function
$F_{ab}^{\b}(z)$ analytic in the strip $-\beta < \mbox{Im}\ z < 0$ and
continuous on its closure such that 
\begin{eqnarray} \label{KMS}
F_{ab}^{\b}(t)=  \rho_{\b}(a(t)\ b)\nonumber\\
F_{ab}^{\b}(t-i\b) =  \rho_{\b}(b\ a(t))
\end{eqnarray}
Using the cyclic property of the trace
\be
\mbox{Tr}\, (ab)=\mbox{Tr}\, (ba)
\ee
one can  check that the KMS condition is satisfied
for the Gibbs state
(\ref{Gibbs}). In the formulation (\ref{KMS}), the KMS condition can be
transferred to the thermodynamic limit.
Actually, one can view the KMS condition as an equation for equilibrium states
of the system.
Any state $\rho_{\b}$ satisfying (\ref{KMS}) is called an {\em equilibrium
state} at inverse temperature $\beta$.

%%%%%%%%%%%%%%%%%%%%%%%%%%%%%%%%%%%%%%%%%%%%%%%%%%%%%%%%%%%%%%%%%%%%%%%%%%
\subsection{Systems of identical particles}
 
In this section we consider a gas of identical quantum-mechanical particles,
bosons or fermions.
The {\em second quantization} of
a system of identical particles is conventionally described by starting
with the {\em Fock space}
\begin{equation} \label{J1}
{\cal H}\;=\;\oplus_{N=0}^\infty\;{\cal H}^{(N)}
\end{equation}
where $\H^{(N)}$ is the $N$-particle Hilbert space,
{\em i.e.}, 
\begin{eqnarray} 
\H^{(N)}=(\H^{(1)})^{\otimes_s N} & \mbox{, for bosons} \nonumber \\
\H^{(N)}=(\H^{(1)})^{\otimes_a N} & \mbox{, for fermions} 
\end{eqnarray}
and $\H^{(1)}$ is the Hilbert space of pure states of a single particle.
The subscripts $s$ and $a$ indicate completely
symmetric and antisymmetric tensor products, respectively, 
according to whether the particles are bosons or fermions. Furthermore, 
$({\cal H}^{(1)})^{\otimes_s 0}\cong ({\cal H}^{(1)})^{\otimes_a 0}\cong
\extraC |0\rangle$, where $|0\rangle$ is the vacuum vector in ${\cal H}$.

Let $\xi$ denote the position and spin of a particle. Creation-  and
annihilation-operators,  $\Psi^*(\xi)$ and
$\Psi (\xi)$, are defined as operator-valued distributions on ${\cal H}$. They
satisfy the commutation relations
\begin{eqnarray}
  \bigl[ \Psi^\#(\xi), \Psi^\#(\eta)\bigr]_\pm & = & 0 \nonumber\\
  \bigl[\Psi(\xi), \Psi^\star(\eta)\bigr]_\pm &  = & \delta (\xi-\eta)
  \label{J2}
\end{eqnarray}
where the commutator $[A,B]_-:=AB-BA$ \ is chosen for bosons
(integer spin particles), and the anti-commutator $[A,B]_+:=AB+BA$ is chosen
for
fermions (particles with half-integer spin). Moreover, $\Psi(\xi)|0\rangle=0$. 

The operators
\begin{eqnarray}\label{J3}
\Psi(f) &:=& \int d\xi\, \Psi (\xi)\bar f (\xi) \nonumber\\
\Psi^*(f) &:=& \int d\xi\, f(\xi) \Psi^* (\xi)
\end{eqnarray}
$f\in {\cal H}^{(1)}$,\ then generate a *algebra 
${\cal A}_{{\cal H}^{(1)}}$, with
$\bigl(\Psi (f)\bigr)^* = \Psi^*(f)$, and
\begin{eqnarray}
  \bigl[ \Psi^{\#}(f),\Psi^{\#}(g)\bigr]_\pm & = & 0 \nonumber\\
  \bigl[ \Psi (f), \Psi^\star(g)\bigr]_\pm & = & \langle f,g\rangle \identity
  \label{J4}
\end{eqnarray}
where $\langle f,g\rangle = \int d \xi\, \bar f (\xi) g (\xi)$ denotes
the scalar product on ${\cal H}^{(1)}$. For fermions, $\Psi^*(f)$ and $\Psi
(f)$ are bounded operators: $\|\Psi(f)\|=\|\Psi^*(f)\|=
\sqrt{\langle f,f\rangle}$.

Rather than starting from the Hilbert space ${\cal H}$ and a concrete
realization of (\ref{J2}) by operators acting on ${\cal H}$, one can
start from the abstract *algebra ${\cal A}_{{\cal H}^{(1)}}$ \ generated by
the identity and creation- and annihilation operators  (\ref{J3}),
with relations (\ref{J4}). We 
obtain representations of ${\cal A}_{{\cal H}^{(1)}}$ from states via 
the so-called GNS
(Gelfand-Naimark-Segal) construction, as follows: Consider a state
$\rho$ on a *algebra ${\cal A}$, i.e., $\rho$ is a normalized 
$(\rho (\identity )=1)$, positive $\bigl( \rho
(a^* a) \geq 0, \; \forall a\in{\cal A}\bigr)$, linear functional
$\bigl(\rho(\alpha a + \beta b) = \alpha \rho (a) + \beta \rho (b), \;
\forall a,b\in{\cal A}, \alpha,\beta\in\extraC\bigr)$ on ${\cal
  A}$. Divide ${\cal A}$ by the left-ideal 
\begin{equation} \label{J5}
{\cal K}\;:=\;\bigl\{ a\in {\cal A} | \rho (ba) = 0, \;
\forall b\in{\cal A}\bigr\}
\end{equation}
Define a Hilbert space ${\cal H}$ as the norm-closure of $V := {\cal
A}/{\cal K}$ (where $\langle [a],[b]\rangle := \rho(a^* b)$,
i.e., $\Vert [a]\Vert^2 := \rho (a^*  a)$) and 
a representation $\Pi_\rho$ of ${\cal A}$ via $\langle
[a], \Pi_\rho (c) [b]\rangle := \rho (a^*  cb)$. For the {\em cyclic vector}
$\Omega_\rho := [\identity ] \in V$, one has that $\rho (c) = \langle
\Omega_\rho, \Pi_\rho (c) \Omega_\rho\rangle,\; \forall c\in{\cal A}$. 

The usual Fock space (cf. (\ref{J1})) is obtained by choosing
$\rho=\rho_0$, where $\rho_0$ is uniquely determined by the equations
$\rho_0 \bigl( a \Psi (f)\bigr) \;=\;0\;=\;\rho_0\bigl(\Psi^*  (f)b\bigr)$
,\ $\forall a,b\in{\cal A}_{{\cal H}^{(1)}}$. \  They imply 
{\em Wick's theorem},
\begin{equation} \label{J6}
\rho_0 \bigl(\prod_{i=1}^m \Psi (f_i) \prod_{j=n}^1 \Psi^*  (g_j)\bigr)=
\left\{\begin{array}{lcl} 
         0 & : &  m \neq n \\
         \displaystyle\sum_{\rm permutations\ P}\epsilon_P\prod_{i=1}^n\ 
         \langle f_i, g_{P(i)}\rangle & : & m = n 
       \end{array}\right.
\end{equation}
where
\be
\epsilon_P=\left\{
\begin{array}{lcl}
  1 & , & \mbox{for bosons} \\
  {\rm sign \ } P  & , & \mbox{\rm for fermions}
\end{array} \right.
\ee
In this example, the Hilbert space $\H$ obtained from the GNS construction is
the Fock space, and the cyclic vector $\Omega_{\rho_{0}}$ is the vacuum vector
$|0\rangle$.

A useful operation is
\begin{equation}
a\;=\;\displaystyle\mathop{\prod}_i\;\Psi^{\#} (f_i)\;\rightarrow\;
\displaystyle\mathop{:\,a\,:}
\end{equation} 
the so-called {\em Wick-ordering}. Up to a sign, $:\,a\,:$ is the same monomial
as $a$, but with all $\Psi^*$'s written to the left of all $\Psi$'s. 
As usual,
the sign is  equal to $1$, for bosons, and to $(-1)^t$, for
fermions, where $t$ denotes the number of transpositions
necessary to move all $\Psi^*$'s to the left of all $\Psi$'s.

{\em Remark}. In order to become familiar with the GNS-construction, one
may consider, as a  trivial example, the *algebra 
${\cal A}:={\rm M}_n (\extraC)$ of 
complex $n\times n$ matrices: Choosing  $
\rho (a)\;:=\;\langle x, ax\rangle\;,$  where $ a \in
{\rm M}_n (\extraC)$, \ 
$\langle\, .\, ,\, .\,\rangle$ denotes the usual scalar product on $\extraC^n$,
and $x$ is an arbitrary unit vector in $\extraC^n$, one easily
recovers the representation of ${\rm M}_n(\extraC)$ on $\extraC^n$.
In this example, ${\cal K}= \{ a\mid ax = 0\}$ \ (consisting of all
matrices having their $n$-th column identically equal to zero if $x$ is the 
$n$-th basis-vector of $\extraC^n$), and ${\cal A}/{\cal K}\cong\extraC^n$ 
has complex dimension $n$. Another example is obtained by considering a
density matrix (e.g.\ a Gibbs state) on ${\rm M}_n(\extraC)$. It leads to a
reducible representation of ${\rm M}_n(\extraC)$.

Returning to ${\cal A}_{{\cal H}^{(1)}}$, it is often
convenient to choose an orthonormal basis \ $\bigl\{
h_n\bigr\}_{n=1}^\infty$ \ in ${\cal H}^{(1)}$, and to expand
annihilation- and creation operators according to
\begin{eqnarray}
\Psi (\xi)  &=&\sum_1^\infty \Psi_n\;h_n (\xi)\nonumber \\
\Psi^*  (\xi)&=&\sum_1^\infty \Psi_n^*
\;\overline{h_n} (\xi)\label{J7}  
\end{eqnarray}
with
\begin{eqnarray}
\bigl[ \Psi_n^\#, \Psi_m^\#\bigr]_\pm &=& 0 \nonumber\\ 
\bigl[ \Psi_n, \Psi_m^* \bigr]_\pm     &=& \delta_{mn}\label{J8} 
\end{eqnarray}
We assume that every basis element $h_{n}$, $n= 1,2,3,...$, is a $C^{\infty}$-
vector for the one-particle Hamiltonian $H_{\Lambda}^{(1)}$. 
A typical second-quantized Hamiltonian has the form
\begin{eqnarray}
H &= &\int d\xi\, \Psi^*  (\xi) \biggl( -
\,\frac{\hbar^2 \triangle_{\Lambda}}{2m}\,\Psi\biggr) (\xi) \nonumber \\
&& \qquad +\;g \int \Psi^*  (\xi_1) \Psi^*  (\xi_2)
V(\xi_1,\xi_2)\, \Psi 
(\xi_2)\,\Psi(\xi_1)\; d\xi_1\, d\xi_2 \label{J9}
\end{eqnarray}
where $\triangle_{\Lambda}$ is the Laplacian with, e.g., Dirichlet
boundary conditions at $\partial \Lambda$, and $V$ is a two-body potential. 
One can rewrite $H$  as
\begin{equation}
H\;=\;T (\Psi^* ,\Psi)\;+\;V (\Psi^* ,\Psi)
\end{equation}
where $T(\Psi^* ,\Psi)\,\equiv\,\sum_{m,n} \Psi^*_m\,A_{mn}\,\Psi_n$
is a positive-definite quadratic form in $\Psi=(\Psi_m)$ and
$\Psi^*=(\Psi^*_m)$ corresponding to the kinetic energy of particles (first
term on the right side of (\ref{J9})),
and $V$ involves terms of higher order in
the operators $\Psi$ and $\Psi^*$ and is even in $\Psi,\Psi^*$ and 
gauge-invariant. 

The time-evolution of $a\in{\cal A}_{{\cal H}^{(1)}}$ in the Heisenberg
picture is given by
\begin{equation}
\dot a \;=\; i\; [H-\mu N, a]
\end{equation}
(the term proportional to $\mu$ drops out for gauge-invariant observables)
implying that
\begin{equation}
i\;\dot\Psi_n\;=\;A_{nm}\;\Psi_m\;+\;\frac{\partial^L V}{\partial
  \Psi_n^*}\;-\;\mu \Psi_n
\end{equation}
and
\begin{equation}
 -i\;\dot \Psi_n^*\;=\;\Psi_m^*\;A_{mn}\;+\;\frac{\partial^R
  V}{\partial \Psi_n}\;-\;\mu \Psi_n^*
\end{equation}
where repeated indices are to be summed over, and $\frac{\partial^LV}{\partial
\Psi^*_n}$ is obtained from $V$ by (anti-\nolinebreak[4]) 
commuting any factor of $\Psi^*_n$
in $V$ to the very left and then omitting it and $\frac{\partial^RV}{\partial
\Psi_n}$ is obtained by (anti-) commuting any factor of $\Psi_n$ in V to the
very right and then omitting it.

%%%%%%%%%%%%%%%%%%%%%%%%%%%%%%%%%%%%%%%%%%%%%%%%%%%%%%%%%%%%%%%%%%%%%%%%%%%%%%
\subsection{Functional integrals: Bosons}

The goal of this section is to rewrite the correlation functions (see
Eq.(\ref{correlation})) in terms of
functional integrals. For this purpose we  replace the field operators by
classical (Grassmann) integration variables for bosons (fermions) and integrate
the exponential of the action
functional over all configurations of these variables. To obtain well-defined
quantities, a cutoff has to be introduced (the same
for bosons and fermions). Referring to the second quantized Hamiltonian
introduced previously, we define $H^{[\kappa]}$ by replacing $A$ and $V$ by
\begin{eqnarray}
A_{nm}^{[\kappa]} & = & \left\{ \begin{array}{cc} A_{nm} & \mbox{if} \, n,m \le
            \kappa \\   0 & \mbox{, otherwise} \end{array} \right.
  \\ 
V^{[\kappa]}(\Psi^*,\Psi) & = & \left\{\begin{array}{cc} V & \quad\mbox
{all indices} \ \le \kappa
            \\ 0 & \mbox{, otherwise} 
\end{array} \right. 
\end{eqnarray}
respectively.
>From now on, bosons and fermions are treated separately.
For an arbitrary operator $a$ on ${\cal H}$ we define 
\[
   a(\tau) := e^{\tau(H^{[\kappa]}-\mu N)}a e^{-\tau(H^{[\kappa]}-\mu N)}
\]
where $\tau$ denotes {\em imaginary time}. Thus, we obtain $\tau$-dependent
fields $\Psi_n(\tau),\Psi_n^*(\tau)$.
We replace the bosonic field operators by complex variables 
\begin{equation}
\Psi_n(\tau), \Psi_n^*(\tau) \rightarrow\psi_n(\tau), \psi_n^*(\tau)
\end{equation} 
satisfying $\psi_n^*(\tau)=\overline{\psi_n(\tau)}$, with $\tau \in
[0,\beta]$. In terms of the complete orthonormal system $\lbrace h_n
\rbrace_{n=1}^{\infty}$ in 
${\cal H}^{(1)}_\Lambda$, we understand the $\psi_n$'s as modes in the
decomposition  
\begin{equation}
\psi(\tau,\xi) = \sum \psi_n(\tau) h_n(\xi)
\end{equation}
Furthermore, we assume that $A^{[\kappa]} \geq 0$ and
$V^{[\kappa]}[\psi^*,\psi] \ge 0, \; \forall \psi^*,\psi$.

We define the Euclidean, i.e., imaginary-time, bosonic action by 
\begin{equation} \label{Eact}
S^{[\kappa]}[\psi^*,\psi] \equiv \int_0^\beta d\tau \,\left[ \sum_{n\leq\kappa}
(\psi_n^* \frac{\partial}{\partial \tau} \psi_n)(\tau) + 
H^{[\kappa]}(\psi^*(\tau),\psi(\tau))- \mu N(\tau)\right]
\end{equation}
The system is always assumed to be confined
to a compact box $\Lambda$.

As the variable $\tau$ takes values in the interval $[0\, , \beta]$,
we must specify  boundary conditions for the variables
$\psi(\tau), \psi^*(\tau)$ at $\tau=0,\beta$. They are determined by the KMS
condition for correlation functions. 

Let us consider the correlation
function of two bosonic operators $\Psi^*(\tau)$ and $\Psi(0)$.
By analyticity in the complex time variable, the KMS condition
implies that
\be
\rho_{\b}(\Psi^*(\tau) \Psi(0))=\rho_{\b}(\Psi(0) \Psi^*(\tau-\beta))
\ee
The imaginary-time correlation function of operators $a_1,\ldots a_n$ is
defined as
\begin{eqnarray}
  \langle a_1(\tau_1)\cdots a_n(\tau_n)\rangle_{\beta,\mu} & := &
  \frac{1}{\Xi}\mbox{Tr}\,(e^{-\beta (H^{[\kappa]}-\mu N)}\,\mbox{T}\,
  a_1(\tau_1)\cdots a_n(\tau_n))
\end{eqnarray}
The imaginary-time ordering operator ``T'' rearranges the operators in 
decreasing time
order, and the imaginary times $\tau_1,\ldots \tau_n$ are assumed to be 
non-negative, pairwise different and smaller than $\beta$.
Using the KMS condition we obtain the identities
\begin{eqnarray} 
  \lefteqn{\langle\Psi^*(\t) \Psi(0)\rangle_{\b,\mu}=
  \rho_{\b}(\Psi^*(\t)\Psi(0))
  =\rho_{\b}(\Psi(0) \Psi^*(\t-\b))}\nonumber\\
  & & = \langle\Psi(0) \Psi^*(\t-\b)\rangle_{\b,\mu}
  =\langle\Psi^*(\t-\b) \Psi(0)\rangle_{\b,\mu}\label{der}
\end{eqnarray}
This simple consideration implies periodic boundary
conditions for bosonic variables:
\be \label{per}
\psi^*(\t+\b)=\psi^*(\t)\  , \ \psi(\t+\b)=\psi(\t)
\ee

Introducing integration measures
\be
{\cal D}\psi=\prod_{m\le \kappa} d\psi_m  \ \ \ ,\ \ \ 
{\cal D}\psi^*=\prod_{m\le \kappa} d\psi^*_m  
\ee
we have the following result.
\vspace{\baselineskip}\\
\begin{em}{\bf Lemma : } 
The cutoff correlation functions can be written in terms of
functional integrals as follows:
\begin{eqnarray}
  \lefteqn{\langle
  \prod_i \Psi_{n_i}(\tau_i) \prod_j \Psi^*_{m_j}(\sigma_j)
  \rangle_{\beta,\mu}\equiv
  \frac{1}{\Xi} \mbox{Tr} \left(e^{-\beta (H^{[\kappa]}-\mu N)}
  \,\mbox{T}\, 
  \prod_i \Psi_{n_i}(\tau_i) \prod_j \Psi^*_{m_j}(\sigma_j) \right)}
  \nonumber \\
  & & =\frac{1}{\Xi}
  \int {\cal D} \psi^* {\cal D} \psi\, e^{-S^{[\kappa]}[\psi^*,\psi]} 
  \prod_i \psi_{n_i}(\tau_i) \prod_j \psi^*_{m_j}(\sigma_j)
  \hspace{43mm}\label{lem} 
\end{eqnarray}
where $n_i, m_j \le \kappa$. The integral on the r.h.s. is supplied with
periodic boundary conditions, Eq. (\ref{per}). For $\mu <0$,
both sides of the equality are well-defined in the limit $\kappa
\rightarrow \infty$, and the equality continues to hold in this limit.
\end{em}
\vspace{\baselineskip}\\
It is easy to verify this lemma at the level of formal power
series, or by appealing to Schwinger-Dyson equations. A rigorous proof is
somewhat more complicated, but the result is standard.

%%%%%%%%%%%%%%%%%%%%%%%%%%%%%%%%%%%%%%%%%%%%%%%%%%%%%%%%%%%%%%%%%%%%%%%%%%%%
\subsection{Functional integrals: Fermions}

Our discussion for fermions follows the procedure outlined for bosons, but 
some room will
be given to understanding the properties of integration over
anticommuting variables.

Just as in the previous section, we replace
the operators $\Psi, \Psi^*$ by symbols $\psi, \psi^*$
\begin{equation}
\Psi_m, \Psi_n^* \rightarrow \psi_m, \psi^*_n
\end{equation}
where $\psi_n$ and $\psi^*_m$ are independent {\em Grassmann variables}:
\begin{equation}
\{\psi_n,\psi_m\} = \{\psi^*_n,\psi^*_m\} = \{\psi_n,\psi^*_m\} = 0
\end{equation}
To integrate over Grassmann variables, we define the {\em Berezin integral} by
requiring that 
\begin{equation}
\{ d\psi_n^\#,\psi_m^\#\} = \{d\psi_n^\#,d\psi_m^\#\} = 0
\end{equation}
and 
\begin{equation}
\int  d\psi_n\,\psi_n = \int  d\psi_n^*\,\psi_n^* = 1
\end{equation}
\begin{equation}
\int d\psi_n\,1 = \int d\psi^*_n\,1 = 0
\end{equation}
We recall some properties of the Berezin integral that follow easily from
the definition and standard combinatorics: Introducing ${\cal D}^{[n]} \psi =
d\psi_n \ldots d\psi_1$ and ${\cal D}^{[n]} \psi^* = d\psi^*_1 \ldots
d\psi^*_n$, we have that
\be\label{grass1}
1 = \prod_{i=1}^n \int d\psi_i\,\psi_i \prod_{j=1}^n \int d\psi^*_j\,\psi^*_j
  = \int  {\cal D}^{[n]} \psi^* {\cal D}^{[n]}\psi\,
    \psi_1 \ldots  \psi_n\psi^*_n \ldots \psi^*_1 
\ee                     
It is easy to see that, for $M^{[n]} [\psi^*,\psi]$ a monomial in
 $\psi_1, \ldots, \psi_n, \psi^*_1,\ldots,\psi^*_n$,
\begin{equation}\label{grass2}
\int {\cal D}^{[n]} \psi^* {\cal D}^{[n]} \psi\, M^{[n]} [\psi^*,\psi]  = 0
\end{equation} 
unless $M^{[n]}[\psi^*,\psi]=\lambda\psi_1\ldots\psi_n\psi^*_n\ldots\psi^*_1$,
$\lambda\in\extraC$. From Eqs.\ (\ref{grass1}) and (\ref{grass2}) and from
the standard definition of determinants we derive that
\be
\int {\cal D}^{[n]} \psi^* {\cal D}^{[n]} \psi\, 
e^{-(\psi^*,K^{[n]}\psi)}  = \det K^{[n]}  
\ee 
where $K^{[n]}$ is a regular $n\times n$ matrix. In fact, this determinant 
formula may be used as a basis for a
definition of integration over Grassmann variables.

It is convenient to
introduce the left derivative $\frac{\delta}{\delta\psi_i}$ by setting
\begin{eqnarray}
\frac{\delta}{\delta\psi_{i_k}} \psi_{i_1} \ldots \psi_{i_k} \ldots \psi_{i_n}
    &=& (-1)^{(k+1)}  \left( \frac{\delta}{\delta\psi_{i_k}}\psi_{i_k}\right)
    \psi_{i_1}\ldots  \psi_{i_{k-1}}\psi_{i_{k+1}} \ldots \psi_{i_n} 
    \nonumber\\ 
&=& (-1)^{(k+1)} \psi_{i_1} \ldots \psi_{i_{k-1}}\psi_{i_{k+1}} \ldots
     \psi_{i_n}   
\end{eqnarray} 
where we assume that $i_1,\ldots i_n$ are pairwise different. The derivative 
$\frac{\delta}{\delta\psi^*_i}$ is defined similarly. All anticommutators
involving derivatives vanish, except for
\begin{equation}
\{\psi_i,\frac{\delta}{\delta\psi_j}\} =
           \{\psi^*_i,\frac{\delta}{\delta\psi^*_j} \} = \delta_{ij}
\end{equation}

These rules yield an integration by parts formula for Berezin
integrals. Let $F^{[n]}[\psi^*,\psi]$ and $G^{[n]}[\psi^*,\psi]$
be two monomials
in $\psi_1, \ldots, \psi_n, \psi^*_1, \ldots,
\psi^*_n$ of degrees $f$ and $g$. Then
\begin{eqnarray} 
  \lefteqn{\int {\cal D}^{[n]} \psi^* {\cal D}^{[n]} \psi\, 
  F[\psi,\psi^*] \frac{\delta G[\psi,\psi^*]}{\delta \psi_m} 
  =}\nonumber \\
  & & (-1)^{f+1} 
  \int {\cal D}^{[n]} \psi^* {\cal D}^{[n]} \psi\,\frac{\delta
  F[\psi,\psi^*]}{\delta \psi_m} G[\psi,\psi^*]\label{intpart1}  
\end{eqnarray}
A similar formula holds when 
$\frac{\delta}{\delta\psi_m}$ is replaced by $\frac{\delta}{\delta\psi_m^*}$.
In particular, choosing $G$ to be the exponential
$\exp \{-(\psi^*, K^{[n]}\psi)\}$, we arrive at the following identity
\begin{eqnarray} 
  \lefteqn{\int {\cal D}^{[n]} \psi^* {\cal D}^{[n]} \psi\,\psi_n e^{-(\psi^*,
  K^{[n]}\psi)}F[\psi,\psi^*]=}\nonumber \\ 
  & & (K^{[n]\:-1})_{nm}   \int {\cal D}^{[n]} \psi^*{\cal D}^{[n]} \psi\,
  e^{-(\psi^*, K^{[n]}\psi)} \frac{\delta F[\psi,\psi^*]}{\delta \psi^*_m}
  \label{intpart2} 
\end{eqnarray}

We define the Euclidean action for fermions by the same formula as in Eq.
(\ref{Eact}). We find the boundary conditions
at $\tau=0,\beta$ for Grassmann variables $\psi^*, \psi$ by using
the KMS condition. The calculation proceeds along the same lines 
as for commuting variables, but we get an extra
minus sign in the last equality
of (\ref{der}), due to fermionic imaginary-time ordering. Thus, we conclude 
that the boundary
conditions for fermionic variables  are {\em anti-periodic}:
\be
\psi^*(\t+\b)=-\psi^*(\t) \ , \psi(\t+\b)=-\psi(\t) 
\ee 
With these boundary conditions, formula (\ref{lem}) holds
true for fermions. The proof is much easier
than the one for bosons, since, after imposing a cutoff,
it reduces to pure multi-linear algebra involving (\ref{intpart1}) and
(\ref{intpart2}).

In the following chapters we shall not explicitly refer to the finite-volume
(infrared) cutoff $\Lambda$ and the ultraviolet cutoff $(\kappa <
\infty)$ anymore, and we shall work with functional integrals in a formal
way. In most instances it is, however, straightforward to justify our
manipulations. For simplicity, we shall study systems at zero temperature; but
our method can be used for positive temperatures as well.

We shall assume that the reader is familiar with basic notions and results in
the quantum theory of systems of non-relativistic, non-interacting fermions at
positive density, such as the Fermi sphere and -surface, the Fermi momentum
$k_{F}$ and the Fermi velocity $v_{F}=\frac{k_{F}}{m}$. We shall use units
such that $\hbar = 1$ (and thus shall not distinguish between ``wave vector''
and ``momentum'').  

%%%%%%%%%%%%%%%%%%%%%%%%%%%%%%%%%%%%%%%%%%%%%%%%%%%%%%%%%%%%%%%%%%%%%%%%%%%%%
\section{Weakly coupled electron gases}
%%%%%%%%%%%%%%%%%%%%%%%%%%%%%%%%%%%%%%%%%%%%%%%%%%%%%%%%%%%%%%%%%%%%%%%%%%%%%

On a {\em microscopic scale} ($\approx 10^{-10}$m), many systems of 
condensed matter physics 
can be described approximately in terms of non-relativistic electrons ---
which are fermions ---
with two-body interactions, moving in a static background. We are interested
in
studying such systems in thermal equilibrium 
at some temperature $T$ and chemical
potential $\mu$. The Heisenberg equations of motion and the  equations for
equilibrium states (KMS condition) of the microscopic 
system are, i.g., not exactly solvable. Our main interest lies, 
however, in predicting transport quantities like conductivity, 
which only depend on 
physical properties of the system at 
{\em mesoscopic length scales} ($\approx 10^{-6}$m) characterized by
a dimensionless parameter $\lambda \gg 1$ (to be thought of as the 
ratio of meso- to microscopic scale). Such quantities 
are therefore calculable from processes involving 
momenta of order $k_F/\lambda$ around the
Fermi surface, i.e., from properties of the scaling limit (large $\lambda$, low
frequencies) of the system. One can try to extract information on the 
scaling limit of the system 
without explicitly solving the microscopic equations. 
Techniques that sometimes allow one to do this involve a principle of 
{\em dimensional reduction} --- 
i.e.,  the observation that, in the scaling limit, systems of non-relativistic 
(free) electrons in $d$ spatial dimensions behave like systems of multi-flavour
Dirac fermions in 1+1 dimensions --- as well as  
{\em renormalization-group-improved} perturbation theory around the 
non-interacting electron gas. These techniques will be explained in the
following chapters.

%%%%%%%%%%%%%%%%%%%%%%%%%%%%%%%%%%%%%%%%%%%%%%%%%%%%%%%%%%%%%%%%%%%%%%%%%%%%%
\subsection{Free electrons and dimensional reduction}
In this section, we show how a system of non-relativistic 
free electrons in $d$-dimensional space can be approximated 
by multi-flavour relativistic fermions in 1+1 dimensional space-time. 
We start from the action
\begin{equation}\label{eq20}
  S_0(\psi^*,\psi) = \sum_{\sigma}\int\!d^{d+1}x\,\psi^*_{\sigma}(x)
  \bigl(\partial_0-\frac{1}{2m}\Delta -\mu\bigr)\psi_{\sigma}(x)
\end{equation}
with a prescribed chemical potential $\mu$ (related to 
the Fermi momentum $k_F$ by $\mu=\frac{k_F^2}{2m}$, and to the 
particle density by $n=2\tau_d k_F^d/(2\pi)^d$, where $\tau_d$ is the
volume of the $d$-dimensional unit ball, and the factor 2 accounts 
for the spin orientations). For simplicity, we work at 
zero temperature.  

The Euclidean free fermion Green's function is given by 
($x=(t,\vec{x})$, $y=(s,\vec{y})$, $t>s$, with $t$ and $s$ now denoting
{\em imaginary time})
\begin{eqnarray} 
  G^0_{\sigma\sigma'}(x-y) & = & \langle e^{tH}\Psi_{\sigma}(\vec{x})
        e^{-(t-s)H} \Psi^*_{\sigma'}(\vec{y})e^{-sH} \rangle_{\mu} \nonumber\\
  & = & -\delta_{\sigma\sigma'} \int\dbar\,^{d+1}k\, 
        \frac{ e^{-ik_0(t-s)+i\vec{k}(\vec{x}-\vec{y})}}
        {ik_0 -(\frac{k^2}{2m}-\mu)} \label{eq21}\\
  & = & \delta_{\sigma\sigma'} \int\dbar\,^d k\,
        \Theta\left(\frac{k^2}{2m}-\mu\right) e^{-(t-s)(\frac{k^2}{2m}-\mu)} 
        e^{i\vec{k}(\vec{x}-\vec{y})} \label{eq21b}  
\end{eqnarray}
where
\begin{equation}
  \dbar k:=\frac{dk}{2\pi}
\end{equation}
The equality (\ref{eq21b}) follows with the help of the residue theorem. 
In the following, we are interested in physics at 
a mesoscopic length scale, or, more precisely, in the 
scaling limit (very low momenta and frequencies).
Then it suffices to determine 
the leading contribution to $G^0_{\sigma\sigma'}(x-y)$ at arguments $x$ and
$y$ with $v_F|t-s|+|\vec{x}-\vec{y}|\approx \lambda/k_F$ (where
$v_F=\frac{k_F}{m}$ is the Fermi velocity).
This contribution  comes from modes with momenta in 
a shell $S_F^{(\lambda)}$ of thickness $k_F/\lambda$ around the Fermi 
sphere $S_F$:
\begin{eqnarray*}
  S_F              & = & \{\vec{k}\in\extraR^d |\, \vec{k}^2 = k_F^2\} \\ 
  S_F^{(\lambda)}  & = & \left\{\vec{k}\in\extraR^d |\, \mbox{dist}(\vec{k}, 
                         S_F)\leq \frac{k_F}{2\lambda} \right\} 
\end{eqnarray*}
States of low energy describe electron-hole pairs with momenta 
in the vicinity of the Fermi surface of the system. 

Let us rewrite (\ref{eq21}) in terms of new variables $\vec{\omega}$, 
$p_{\parallel}$ and $p_0$, with $k_F\vec{\omega}\in S_F$, $p_0=k_0$, and 
$\vec{k}=(k_F+p_{\parallel})\vec{\omega}$. If $\vec{k}\in S_F^{(\lambda )}$ 
then $|p_{\|}|\ll k_F$, 
and we can approximate the integrand of (\ref{eq21}) by dropping 
the $p_{\|}^2$ term in the denominator. For large $\lambda$ 
and small frequencies,
\begin{equation}\label{propagator}
  G^0_{\sigma\sigma'}(x-y) \approx \delta_{\sigma\sigma'}
  \int\frac{d\sigma(\vec{\omega})}{(2\pi)^{d-1}}\, k_F^{d-1}\,
  e^{ik_F\vec{\omega}(\vec{x}-\vec{y})}  G_c(t-s,\vec{\omega}
  (\vec{x}-\vec{y}))
\end{equation}
where $d\sigma (\vec{\omega})$ is the uniform measure on the unit sphere, and
$G_{c}$ is the two-dimensional, chiral propagator
\begin{equation} 
  G_c(\tau,\xi) = -\int\dbar p_0\dbar p_{\|}\,  
  \frac{e^{-ip_0\tau+ip_{\parallel}\xi}}{ip_0 - v_F p_{\parallel}}     
\end{equation}
We will set $v_F=1$ 
in the remainder of this section. Introducing the 
complex variable $z = i\tau+\xi$ and
its complex conjugate $\bar{z}$, it is easy to verify that 
$G_c(z)$ satisfies 
\[
-2i \bar{\partial} G_c(z) = \delta^{(2)}(z) 
\]
In other words, $G_c$ is the Green's function of a {\em chiral Dirac 
fermion} in 1+1 dimensions. Explicitly,  
\begin{equation} 
  G_c(z) = \frac{i}{\pi z}
\end{equation}
One can further approximate
the $\vec{\omega}$-integration in Eq.\ (\ref{propagator}) by a summation 
over discrete directions 
$\vec{\omega}_j$ by dividing $S_F^{(\lambda)}$ into $N$ small boxes 
$B_{\vec{\omega}_j}$, $j=1,\ldots N$,
of roughly cubical shape: $B_{\vec{\omega}_j}$ is centered at 
$\vec{\omega}_j\in S_F$ and has approximate side length $k_F/\lambda$; 
note that $N \approx\mbox{const}\,\lambda^{d-1}$. Thus 
\begin{equation}\label{eq22}
  G^0_{\sigma\sigma'}(x-y) \approx -\delta_{\sigma\sigma'}
  \sum_{\vec{\omega}_j} e^{i\vec{\omega}_j(\vec{x}-\vec{y})} 
  \int\dbar p_0\dbar p_{\|}\dbar p_{\perp}\,
  \frac{e^{-ip_0(t-s)+i\vec{p}(\vec{x}-\vec{y})}}{ip_0-p_{\parallel}}     
\end{equation}
where $\vec{p}=p_{\parallel}\vec{\omega} + \vec{p}_{\perp}$ is a vector 
in $B_{\vec{\omega}_j}-k_F\vec{\omega}_j$, and $p_{0}\in\extraR$.  

We have shown that, in the scaling limit, the behaviour 
of a $d$-dimensional, non-relativistic, non-interacting electron gas 
is described by $N$ flavours of free, chiral Dirac fermions 
in a $1+1$ dimensional 
space-time. The statement is to be understood in an appropriate sense, 
since the propagator $G_c(t-s,\vec{\omega}(\vec{x}-\vec{y}))$ actually
depends on the 
``flavour index'' $\vec{\omega}$. But the  energy of an electron or a hole
with momentum $\vec{k}$ only depends on $p_{\|}$, where $p_{\|}=\vec{k}
\vec{\omega}-k_F$, and $\vec{\omega}=\frac{\vec{k}}{|\vec{k}|}$; it is
proportional to $|p_{\|}|$, just as for relativistic fermions in $1+1$
dimensions.   

%%%%%%%%%%%%%%%%%%%%%%%%%%%%%%%%%%%%%%%%%%%%%%%%%%%%%%%%%%%%%%%%%%%%%%%%%%%
\subsection{Weakly coupled electrons and the renormalization group (RG)}
\nopagebreak

We have presented a somewhat unusual description of the free fermion 
system, because expressions such as (\ref{eq22}) provide a 
convenient starting point for treating interacting fermions 
by a perturbation expansion.
In particular, one may hope that interesting 
physical quantities have an expansion in powers of $1/\lambda$
analogous to the $1/N$ expansion in the Gross-Neveu model. 
Later we will see that this is the case for weakly coupled systems, 
and, as in the Gross-Neveu model, the large number of flavours $N$ 
encountered in the mesoscopic regime will play an important r\^{o}le in the 
actual calculations. We will, however, 
also see that perturbation expansions usually cannot 
be applied ``naively'', but have to be improved 
by applying renormalization group (RG) methods. 

We consider systems with a Euclidean action of the form
\[
S(\psi^*,\psi)=S_0(\psi^*,\psi)+S_I(\psi^*,\psi)
\]
where $S_0(\psi^*,\psi)$ is the quadratic term (\ref{eq20}), and 
$S_I(\psi^*,\psi)=g_0P(\psi^*,\psi)$ is some higher order polynomial 
interaction to be specified below; we assume that the dimensionless
coupling constant $g_0$ is small. 
Since correlation functions of the fermion fields $\Psi^{\#}(x)$
do not have good scaling behaviour, we first split off the 
oscillatory factors associated with the ``direction index'' $\vec{\omega}$ 
(see, e.g., Eq. (\ref{eq22})) by expanding the fermions
\begin{equation}
  \Psi(x)=\sum_{\vec{\omega}}e^{i\vec{\omega}\vec{x}}\Psi_{\vec{\omega}}(x)
\end{equation}
into quasi-particle operators $\Psi_{\vec{\omega}}(x)$. We are interested in
calculating {\em connected correlators} (temperature-ordered Green functions)
of the form 
\begin{equation}
(\lambda^\alpha)^{2n} \langle \Psi_{\vec{\omega}_1}(\lambda x_1)\cdots
\Psi_{\vec{\omega}_n}(\lambda x_n) \Psi^*_{\vec{\omega}_1'}(\lambda x_1')
\cdots\Psi^*_{\vec{\omega}_n'}(\lambda x_n')\rangle^c_\mu 
\end{equation}
at large distance scales $\lambda$; the factor in front of the 
bracket, involving the scaling dimension $\alpha$ of the 
quasi-particle fields, accounts for the  
scaling behaviour of the quasi-particle operators. 
Strictly speaking, instead of considering the expectation values 
above, one 
ought to ``smear out'' the fields $\Psi^\#_{\vec{\omega}}$ with some test 
function (an approximate $\delta$- function) $h(\vec{\xi})$:
\[
\Psi_{\vec{\omega},\lambda}^{\#}(x)=
(\lambda)^{-(d+1)}
\int d^{d+1}y\,h((\vec{x}-\vec{y})/\lambda)\Psi^{\#}_{\vec{\omega}}(y)
\]
Because we are interested in the large-scale behaviour, we can work  
with test functions $h$ of compact support in momentum space, e.g. with
supp$\,\hat{h}=\{k\in\extraR^{d+1}|\,
\vec{k}^2\leq k_F^2\}$. After Fourier transformation, the 
connected correlation functions of smeared quasi-particle fields 
are given by 
\[
(\lambda^{\alpha-d-1})^{2n}\hat{h}(k_1)
 \cdots \hat h(k_n')\, \widehat{G^c_{2n}}(k_1/\lambda,\ldots
 k_n'/\lambda)
\]
with 
\[
\widehat{G^c_{2n}}=  \langle\hat{\Psi}_{\vec{\omega}_1}(k_1)\cdots
\hat{\Psi}_{\vec{\omega}_n}(k_n)\hat{\Psi}^*_{\vec{\omega}_1'}(k_1')\cdots 
\hat{\Psi}^*_{\vec{\omega}_n'}(k_n')\rangle^c_\mu
\]

According to Chapter 1, we can express this
correlator in terms of a functional integral 
\[ 
\widehat{G^c_{2n}} = 
  \frac{1}{\Xi}\int {\cal D} \hat{\psi}^* {\cal D} \hat{\psi}\,
  e^{-S(\hat{\psi}^*,\hat{\psi})}\hat{\psi}_{\vec{\omega}_1}(k_1)\cdots
  \hat{\psi}^*_{\vec{\omega}_n'}(k_n')
 \]
where $\vec{k}_j\in B_{\vec{\omega}_j}-k_F\vec{\omega}_j$, etc.  (because of
the restriction on the support of the test functions).
We now decompose the fermion fields into ``slow'' and ``fast'' modes 
by writing $\hat{\psi} = \hat{\psi}_<+\hat{\psi}_>$, with
\begin{eqnarray*} 
  \mbox{supp}\,\hat{\psi}_>\subset\extraR\times(\extraR^d \setminus
S_F^{(\lambda)})
    &  \   &  (\mbox{region}\ >)\\
  \mbox{supp}\,\hat{\psi}_<\subset\extraR\times S_F^{(\lambda)}
    &  \   &  (\mbox{region}\ <)
\end{eqnarray*}
\begin{figure}
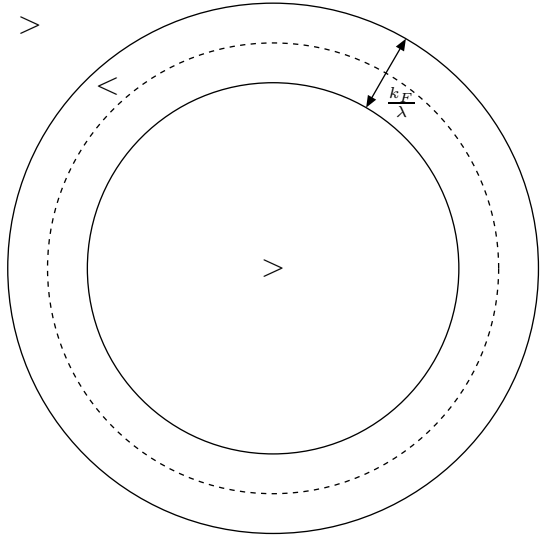

  \center
  \figureBA
  \caption{Integrating out degrees of freedom}
\end{figure}
The large-scale behaviour of the system is described by the effective 
action $S_{\rm eff}(\hat{\psi}^*_<,\hat{\psi}_<)$ given by
\begin{equation}
  e^{-S_{\rm eff}(\hat{\psi}^*_<,\hat{\psi}_<)} = 
  \frac{1}{\Xi_>}\int {\cal D}\hat{\psi}^*_>{\cal D}\hat{\psi}_>\,
   e^{-S(\hat{\psi}^*_>,\hat{\psi}^*_<,\hat{\psi}_>,\hat{\psi}_<)}
\end{equation}
where we perform the functional integral over fast modes only; the
normalization factor $\Xi_>$ is chosen so as to ensure that 
$S_{\rm eff}(0,0)=0$. Obviously, $S_{\rm eff}$ depends on our choice of 
the scale $\lambda$. 

The quadratic part, $S_0$, of the action  
splits into two pieces:  
\[
  S_0(\hat{\psi}^*,\hat{\psi})=S_{0,>}(\hat{\psi}^*_>,\hat{\psi}_>)+
  S_{0,<}(\hat{\psi}^*_<,\hat{\psi}_<)
\]
The first term and the normalization factor can be absorbed into the 
definition of the integration measure
\[
dP(\hat{\psi}^*_>,\hat{\psi}_>)    = 
\frac{1}{\Xi_>}{\cal D}\hat{\psi}^*_>{\cal D}\hat{\psi}_>
e^{-S_{0,>}(\hat{\psi}^*_>,\hat{\psi}_>)}
\]
and we have to calculate
\begin{eqnarray}
  \lefteqn{e^{-S_{\rm eff}(\hat{\psi}^*_<,\hat{\psi}_<)} 
        \equiv e^{-S_{0,<}(\hat{\psi}^*_<,\hat{\psi}_<)} 
        \int dP(\hat{\psi}^*_>,\hat{\psi}_>)
        e^{-S_I(\hat{\psi}^*_>,\hat{\psi}^*_<,\hat{\psi}_>,\hat{\psi}_<)}}
        \nonumber\\ 
  &  &  =\exp\{-S_{0,<}-\langle S_I\rangle_{G^0_>}
        +\frac{1}{2}\langle S_I;S_I\rangle_{G^0_>}
        -\frac{1}{3!}\langle S_I;S_I;S_I\rangle_{G^0_>}-\ldots\}\label{eq23}
\end{eqnarray}
The abbreviations 
\begin{eqnarray*} 
  \langle A;B\rangle   & := & \langle AB\rangle 
                              -\langle A\rangle\langle B\rangle \\
  \langle A;B;C\rangle & := & 
    \langle ABC\rangle-\langle A;B\rangle\langle C\rangle
    -\langle C;A\rangle\langle B\rangle-\langle B;C\rangle\langle A\rangle
    -\langle A\rangle \langle B\rangle\langle C\rangle
\end{eqnarray*}
etc., denote connected correlators;  
the subscript ``$G^0_>$'' indicates that the expectations $\langle (\, .\,) 
\rangle_{G^0_>}$ are 
calculated with the help of the infrared-cutoff free fermion propagator
$G^0_>$, 
in accordance with the functional measure $dP(\hat{\psi}^*_>,\hat{\psi}_>)$. 
The second  
equality in (\ref{eq23}) states that the effective action 
is given by sums over connected diagrams. This is the so-called
{\em linked cluster theorem}. Eq.\
(\ref{eq23}) also shows that $S_{\rm eff}$ contains i.g.\ far more 
interaction terms than $S_I$; for weakly coupled systems, however, 
the original couplings will remain the dominant ones. The proofs of
equation (\ref{eq23}) and of the linked cluster theorem are standard. We
therefore omit them. {\em Feynman rules} 
for the computation of $S_{\rm eff}$
are easily derived (we do not present the  details).

In order to analyze the perturbation series in (\ref{eq23}) further, 
we must specify the interactions in the system: We assume that the 
leading term in $S_I$ is a two-body interaction 
\begin{equation}\label{eq24}
  S_I(\psi^*,\psi)= \frac{g_0}{2}k_F^{1-d}
\sum_{\sigma,\sigma'}\int d^{d+1}x\,
  d^{d+1}y\,
  :\psi^*_{\sigma}(x)\psi_{\sigma}(x) 
  v(\vec{x}-\vec{y})\delta(x^0-y^0)
  \psi^*_{\sigma'}(y)\psi_{\sigma'}(y):
\end{equation}
where $v(\vec{x}-\vec{y})$ is a smooth short-range potential (we choose units
such that $\hbar = 1$; the factor $k_F^{1-d}$ then ensures that $g_0$ is
dimensionless). 
It is useful to estimate how close to $S_F$ one can get with ``naive'' 
perturbation theory: The free propagator, $-(ik_0-\vec{\omega}\vec{k})^{-1}$, 
is regular in region $>$, it is in fact of order $\lambda/k_F$. The 
size of the two-body interaction 
associated with the graph 
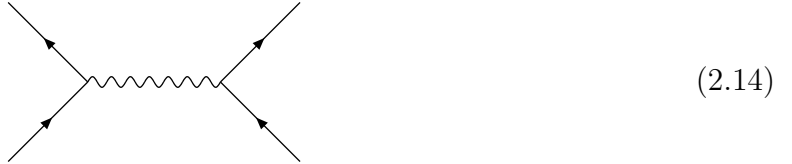
\begin{equation}\figureBC\end{equation}
is therefore roughly equal to $|g_0\lambda^2|$ (which is dimensionless).
The wavy line represents the interaction potential which is of order $g_0$,
the oriented solid lines represent electron ``half-propagators'' each of which
is of order $\lambda^{\frac{1}{2}}$.
As long as $\lambda\ll 1/\sqrt{g_0}$, there are 
no convergence problems with the perturbation series --- 
except that the number of diagrams seemingly grows too fast; 
but their relative signs ensure appropriate cancellations (Pauli principle).

This crude estimate 
shows that we cannot pass to the limit $\lambda\rightarrow\infty$ without
meeting infrared divergence problems in the perturbation series. 
The method to control the scaling limit is RG improved perturbation 
theory. It allows us to analyze the large-scale behaviour of the 
system by carefully keeping track of the (relative) 
growth of the various terms in the effective action (quadratic 
part as well as couplings) when the integration over 
fermion modes approaches those near the Fermi surface. 
In fact, for the interaction (\ref{eq24}), the actual RG computations are not 
terribly involved, since  $S_{\rm eff}$ has  an 
expansion in powers of $1/\lambda$, and focusing on the leading order in
$\frac{1}{\lambda}$
drastically reduces the number of diagrams that have to be 
calculated. 

We shall apply an iterative RG scheme patterned on 
Wilson's approach. In the ``zeroth''
step, we choose some large scale $\lambda_0$ such that  
$\lambda_0\ll 1/\sqrt{g_0}$ and calculate the corresponding 
effective action
$S^{(0)}_{\rm eff}$ perturbatively, as in (\ref{eq23}) --- to 
leading order in $1/\lambda_0$. 
The action $S^{(o)}_{\rm eff}$ depends on a collection of modes corresponding
to wave vectors that are located in a shell of width
$\frac{k_F}{\lambda_0}$ around the Fermi surface $S_F$.
Although not 
essential for our method, we divide this 
shell into $N\approx \mbox{const}\,\lambda_0^{d-1}$ sectors (boxes),
as in the previous section. 
Furthermore, we rescale all momenta so that, instead of belonging to
$B_{\vec{\omega}_j}$, they are contained in boxes 
$\tilde{B}_{\vec{\omega}_j}$ of side length $\approx k_F$.   
Our RG procedure consists in iterating the following 
two steps: 
\begin{description}
\item[Dec] Choose some (fixed) integer $M>1$ and integrate over the 
  fermion modes corresponding to wave vectors in 
  $S_F^{(\lambda)}\setminus S_F^{(M\lambda)}$ (where $\lambda=M^j\lambda_0$,
  for some $j$).
\item[Resc] Divide each sector in $S^{(M\lambda)}$ into $M^{d-1}$
  new sectors to restore the cubical shape of the sectors. Then 
  rescale all momenta by $k\longmapsto \tilde{k}= Mk$.
\end{description}
To determine the RG flow of the couplings, we have to find 
out how the various terms in the action transform under rescaling
and which diagrams contribute, during the integration process, 
to leading order in $1/\lambda$.
\begin{figure}
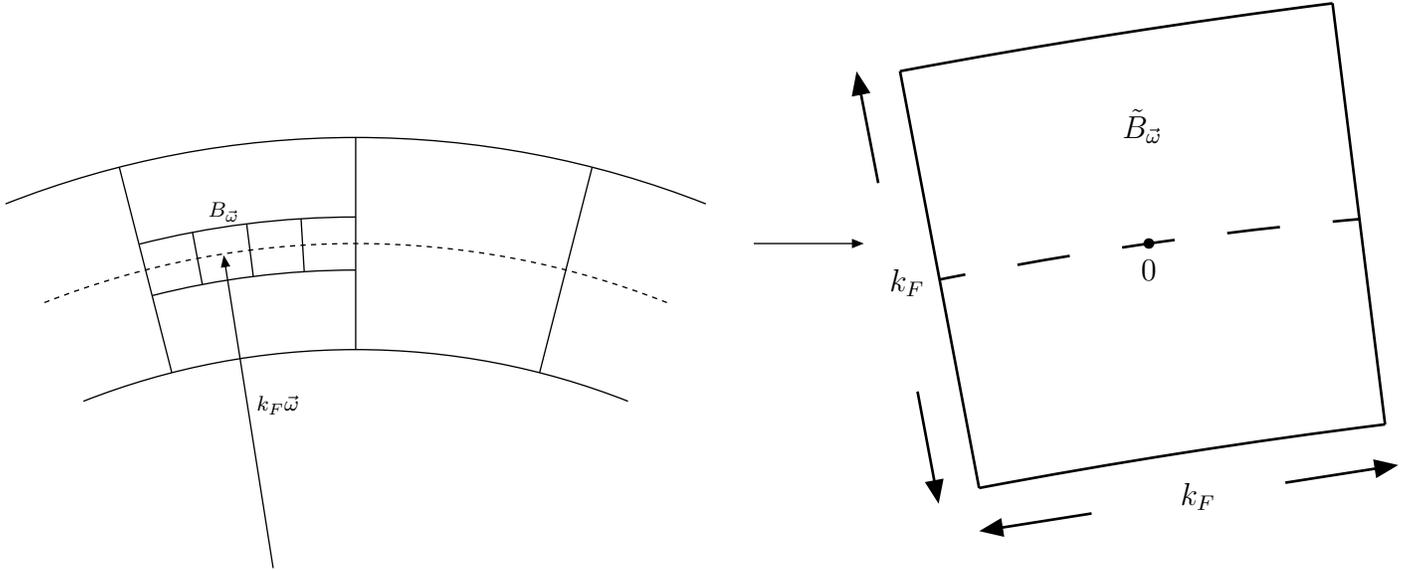
\center
  \figureBB
  \caption{The renormalization group procedure}
\end{figure}

%%%%%%%%%%%%%%%%%%%%%%%%%%%%%%%%%%%%%%%%%%%%%%%%%%%%%%%%%%%%%%%%%%%%%%%%%%%
\section{The renormalization group flow}
%%%%%%%%%%%%%%%%%%%%%%%%%%%%%%%%%%%%%%%%%%%%%%%%%%%%%%%%%%%%%%%%%%%%%%%%%%%

In this chapter we attempt to explicitly implement the renormalization group
procedure described in the last chapter. Our aim is to understand  the
renormalization group flow to lowest order in $\frac{1}{\lambda_0}$ and for
small values of $g_0$.

%%%%%%%%%%%%%%%%%%%%%%%%%%%%%%%%%%%%%%%%%%%%%%%%%%%%%%%%%%%%%%%%%%%%%%%%%%%%
\subsection{Scaling of action and fields}

We shall begin with the discussion of the second step and of the scaling
dimensions of the different terms in the
effective action. Since the decimation of degrees of freedom is most explicitly
described in momentum-space, our renormalization group procedure is
implemented in the momentum-space representation. Thus, we first calculate
the action in terms of the sector fields discussed
in the previous chapter.

Let $\lambda\gg 1$, and assume that the degrees of freedom not lying in the 
shell
$S_F^{(\lambda)}$ have already been integrated out. We divide $S_F^{(\lambda)}$ into boxes
of roughly cubical shape and approximate side length
$\frac{k_F}{\lambda}$. The number of such boxes is
$N\approx\omega_{d-1}\lambda^{d-1}$, where $\omega_{d-1}$ is the surface 
volume of
the unit sphere in $d$ spatial dimensions. Let $B_{\vec{\omega}}$,
$|\vec{\omega}|=1$, denote the box which is centered at $k_F\vec{\omega}$. The
support of those modes
\begin{equation}
  \hat{\psi}_{\sigma}(k)=\int d^{d+1}x\, 
  e^{i(k_0 t-\vec{k}\vec{x})}\psi_{\sigma}(x)
\end{equation}
that have not been integrated out, yet, is contained in 
$\extraR\times S_F^{(\lambda)}$. {\em Sector fields} are 
defined as 
\begin{equation}
  \psi_{\vec{\omega},\sigma}(x):=\int_{\extraR\times B_{\vec{\omega}}}
  \dbar\,^{d+1}k\, e^{-i(k_0 t-(\vec{k}-k_F\vec{\omega})\vec{x})}
  \hat{\psi}_{\sigma}(k)
\end{equation}
The Fourier transform of $\psi_{\vec{\omega},\sigma}(x)$ has support in
$\extraR\times (B_{\vec{\omega}}-k_F\vec{\omega})$, and 
\[
  \psi_{\sigma}(x)=\sum_{\vec{\omega}} e^{ik_F\vec{\omega}\vec{x}}
  \psi_{\vec{\omega},\sigma}(x)
\]

Let us first look at the quadratic part of the effective action of the modes
localized in the shell
$S^{(\lambda)}_F$. We temporarily assume
that this part has the form (\ref{eq20}) (there will be a finite
wave function renormalization which we are neglecting at this point).
Inserting the Fourier transformed fields yields 
\begin{eqnarray}\label{eq30}
  S_0(\psi^*,\psi) & = & -\sum_{\sigma}\int\dbar\,^{d+1}k \dbar\,^{d+1}q\,
    \hat{\psi}_{\sigma}^*(k)\left(iq_0-\frac{\hbar^2|\vec{q}|^2}{2m}+\mu\right)
    \hat{\psi}_{\sigma}(q)(2\pi)^{d+1}\delta (k-q) \nonumber\\
  & = & -\sum_{\vec{\omega},\sigma}\int_{\extraR\times 
    (B_{\vec{\omega}}-k_F\vec{\omega})}\dbar\,^{d+1}p\,
    \hat{\psi}_{\vec{\omega},\sigma}^*(p)\left(ip_0-v_F\vec{\omega}\vec{p}+
    O\left(\frac{1}{\lambda^2}\right)\right)\hat{\psi}_{\vec{\omega},\sigma}(p)
\end{eqnarray}
with $v_F=\frac{k_F}{m}$. In the last line, we have put the term quadratic in
$p$  into an error term $O(\frac{1}{\lambda^2})$. The leading part of the
inverse propagator in $S_0$ is
of order $O(\frac{1}{\lambda})$. Since we always restrict our
analysis to the leading order in $\frac{1}{\lambda}$, we can omit this
error term in what follows.

Next, we consider the quartic term (\ref{eq24}). We insert
the expansion of the electron fields in terms of sector fields and get
\begin{eqnarray}
  \lefteqn{S_I(\psi^*,\psi)= g_0 k_F^{1-d}
    \frac{1}{2}\sum_{\stackrel{\vec{\omega}_1,
    \ldots\vec{\omega}_4}{\sigma,\sigma'}}\int d^{d+1}x d^{d+1}y\,
    e^{-ik_F(\vec{\omega}_1-\vec{\omega}_4)\vec{x}}
    e^{-ik_F(\vec{\omega}_2-\vec{\omega}_3)\vec{y}}} \nonumber\\
  & & \psi^*_{\vec{\omega}_1,\sigma}(x)\psi^*_{\vec{\omega}_2,\sigma'}(y)
    v(\vec{x}-\vec{y})\delta(x_0-y_0)\psi_{\vec{\omega}_3,\sigma'}(y)
    \psi_{\vec{\omega}_4,\sigma}(x) \label{eq31}
\end{eqnarray}
In terms of the Fourier transform of the potential
\[
  v(\vec{x})=\int\dbar\,^d k\,e^{i\vec{k}\vec{x}}\hat{v}(\vec{k})
\]
we can write
\[
  v(\vec{x})\delta (x^0)=\int\dbar\,^{d+1}k\,e^{-i(k_0 x^0-\vec{k}\vec{x})}
  \hat{v}(\vec{k})
\]
Since we assume that $v(\vec{x})$ is a {\em short-range potential},
$\hat{v}(\vec{k})$  is a smooth function of $\vec{k}$. Applying Fourier
transformation to the sector fields in (\ref{eq31}), we obtain
\begin{eqnarray*}
  \lefteqn{S_I= g_0 k_F^{1-d}
  \frac{1}{2}\sum_{\stackrel{\vec{\omega}_1,\ldots\vec{\omega}_4}
  {\sigma,\sigma'}}\int \dbar\,^{d+1}p_1\cdots\dbar\,^{d+1}p_4\int
  \dbar^{d+1}k}\\
  & & 2\pi\delta(p_{1,0}-p_{4,0}-k_0)2\pi\delta(p_{2,0}-p_{3,0}+k_0)\\
  & & (2\pi)^d\delta(\vec{p}_1-\vec{p}_4-\vec{k}+k_F(\vec{\omega}_1-
      \vec{\omega}_2))
      (2\pi)^d\delta(\vec{p}_2-\vec{p}_3-\vec{k}+k_F(\vec{\omega}_2-
      \vec{\omega}_3))\\
  & & \hat{\psi}^*_{\vec{\omega}_1,\sigma}(p_1)
      \hat{\psi}^*_{\vec{\omega}_2,\sigma'}(p_2)
      \hat{v}(\vec{k})
      \hat{\psi}_{\vec{\omega}_3,\sigma'}(p_3)
      \hat{\psi}_{\vec{\omega}_4,\sigma}(p_4)
\end{eqnarray*}
where the $p_i$-integrations range over the boxes $B_{\vec{\omega}_i}-
k_F\vec{\omega}_i$. Performing the $k$-integration, we obtain the expression
\begin{eqnarray*}
  \lefteqn{g_0 k_F^{1-d}
   \frac{1}{2}\sum_{\stackrel{\vec{\omega}_1,\ldots\vec{\omega}_4}
  {\sigma,\sigma'}}\int \dbar\,^{d+1}p_1\cdots\dbar\,^{d+1}p_4\,
  2\pi\delta(p_{1,0}+p_{2,0}-p_{3,0}-p_{4,0})}\\
  & & (2\pi)^d\delta(\vec{p}_1+\vec{p}_2-\vec{p}_3-\vec{p}_4+
      k_F(\vec{\omega}_1+\vec{\omega}_2-\vec{\omega}_3-\vec{\omega}_4))\\
  & & \hat{v}(\vec{p}_1-\vec{p}_4+k_F(\vec{\omega}_1-\vec{\omega_4}))
      \hat{\psi}^*_{\vec{\omega}_1,\sigma}(p_1)
      \hat{\psi}^*_{\vec{\omega}_2,\sigma'}(p_2)
      \hat{\psi}_{\vec{\omega}_3,\sigma'}(p_3)
      \hat{\psi}_{\vec{\omega}_4,\sigma}(p_4)
\end{eqnarray*}
Considering the argument of the second $\delta$-distribution appearing in the
integrand, the
part containing the $\vec{\omega}_i$'s is of order $k_F$, while the part
involving the $p_i$'s is only of order $\frac{k_F}{\lambda}$. Therefore, for
the
argument of this $\delta$-distribution to vanish, the term containing the
$\vec{\omega}_i$'s must be zero (to lowest order in
$\frac{1}{\lambda}$). Furthermore, as the Fourier transform $\hat{v}$ of the
interaction is a smooth function, its value under the integral sign is well
approximated by $\hat{v}(k_F(\vec{\omega}_1-\vec{\omega}_4))$, dropping the
box momenta. Consequently we get 
\begin{eqnarray}\label{eq35}
  \lefteqn{S_I= g_0 k_F^{1-d}
  \frac{1}{2}\sum_{\stackrel{\vec{\omega}_1,\ldots\vec{\omega}_4}
  {\sigma,\sigma'}}
  \hat{v}(k_F(\vec{\omega}_1-\vec{\omega}_4))
  \int \dbar\,^{d+1}p_1\cdots\dbar\,^{d+1}p_4}\nonumber\\
  & & (2\pi)^{d+1}\delta(p_1+p_2-p_3-p_4)
      \hat{\psi}^*_{\vec{\omega}_1,\sigma}(p_1)
      \hat{\psi}^*_{\vec{\omega}_2,\sigma'}(p_2)
      \hat{\psi}_{\vec{\omega}_3,\sigma'}(p_3)
      \hat{\psi}_{\vec{\omega}_4,\sigma}(p_4)\label{eq32}\\
  & & +\mbox{\ terms of higher order in\ }\frac{1}{\lambda}\nonumber
\end{eqnarray}
Instead of studying the quartic term of the original (microscopic) action of
the electron gas, we should actually study the quartic (and higher-degree)
terms of the {\em effective action} at scale $\frac{k_F}{\lambda}$, with
$\lambda=\lambda_0\gg 1$ and $|g_0 \lambda_0^2| \ll 1$. Using cluster
expansions to integrate out the degrees of freedom labeled by momenta
$\vec{k}\not\in S_F^{(\lambda_0)}$, one can show that, for weakly coupled
systems, the quartic term of the effective action at scale
$\frac{k_F}{\lambda_0}$ still has the form given in eq. (\ref{eq35}), except
that $\hat{v}(k_F (\vec{\om}_1 -\vec{\om}_4))$ is replaced by a coupling
function $g(\vec{\om}_1 ,..,\vec{\om}_4) \approx \hat{v}(k_F (\vec{\om}_1
-\vec{\om}_4))$, with $\vec{\om}_1+\vec{\om}_2
=\vec{\om}_3+\vec{\om}_4$. (Moreover, terms of degree larger than four in
$\psi^*$, $\psi$ are very small.)

Next, we determine the scaling dimensions of action and fields. We rescale the
fields
in such a way, that the supports of the Fourier transformed, rescaled sector 
fields are boxes of
roughly cubical shape and with side length approximately equal to $k_F$, and 
such that
the quadratic part of the action remains unchanged to leading order in
$\frac{1}{\lambda}$. The first condition implies, that we should scale momenta
according to $p=\tilde{p}/\lambda$. In configuration space this corresponds to
the scaling $x=\lambda\xi$. Thus the scaled sector fields are
\[
  \tilde{\psi}_{\vec{\omega},\sigma}(\xi)=\lambda^{\alpha}
  \psi_{\vec{\omega},\sigma}(\lambda\xi)
\]
where the {\em scaling dimension} $\alpha$ of the sector fields still has to 
be determined. The Fourier transformed sector fields are
\[
  \hat{\tilde{\psi}}_{\vec{\omega},\sigma}(\tilde{p})=
  \lambda^{\alpha-d-1}\hat{\psi}_{\vec{\omega},\sigma}
  (\frac{\tilde{p}}{\lambda})
\]
The support of $\hat{\tilde{\psi}}_{\vec{\omega},\sigma}$ lies in 
$\tilde{B}_{\vec{\omega}}=\lambda (B_{\vec{\omega}}-k_F\vec{\omega})$. Indeed,
$\tilde{B}_{\vec{\omega}}$ is a roughly cubical box with approximate 
side length $k_F$. Inserting the scaled, Fourier-transformed fields into the
quadratic part (\ref{eq30}) of the action yields
\begin{eqnarray*}
  S_0 & = & -\lambda^{2(d+1-\alpha)}\lambda^{-(d+1)}
    \sum_{\vec{\omega},\sigma}\int_{\extraR\times\tilde{B}_{\vec{\omega}}}
    \dbar\,^{d+1}\tilde{p}\,
    \hat{\tilde{\psi}}^*_{\vec{\omega},\sigma}(\tilde{p})
    \left(\frac{1}{\lambda}(i\tilde{p}_0-v_F\vec{\omega}\vec{\tilde{p}})
    +O\left(\frac{1}{\lambda^2}\right)\right)
    \hat{\tilde{\psi}}_{\vec{\omega},\sigma}(\tilde{p})\\
  & = & -\lambda^{d-2\alpha}
    \sum_{\vec{\omega},\sigma}\int_{\extraR\times\tilde{B}_{\vec{\omega}}}
    \dbar\,^{d+1}\tilde{p}\,
    \hat{\tilde{\psi}}^*_{\vec{\omega},\sigma}(\tilde{p})
    \left(i\tilde{p}_0-v_F\vec{\omega}\vec{\tilde{p}}
    +O\left(\frac{1}{\lambda}\right)\right)
    \hat{\tilde{\psi}}_{\vec{\omega},\sigma}(\tilde{p})
\end{eqnarray*}
For the free part of the action to have the same form as for the unscaled
fields , we have to choose $\alpha=\frac{d}{2}$. Thus the scaled fields are
\begin{eqnarray}
  \tilde{\psi}_{\vec{\omega},\sigma}(\xi) & = & \lambda^{\frac{d}{2}}
    \psi_{\vec{\omega},\sigma}(\lambda\xi)\\
  \hat{\tilde{\psi}}_{\vec{\omega},\sigma}(\tilde{p}) & = & 
    \lambda^{-(\frac{d}{2}+1)}
    \hat{\psi}_{\vec{\omega},\sigma}(\tilde{p}/\lambda)
\end{eqnarray}

Let us determine the scaling behaviour of a term in the action of the form
\begin{eqnarray}
  \lefteqn{S_n=\frac{1}{n!}\sum_{\stackrel{\vec{\omega}_1+\cdots
  +\vec{\omega}_n=\vec{\omega}_{n+1}+\cdots +\vec{\omega}_{2n}}
  {\sigma_1,\ldots\sigma_{2n}}}
  \int\dbar\,^{d+1}p_1\cdots\dbar\,^{d+1}p_{2n}}\nonumber\\
  & & w(p_1,\ldots p_{2n})
      \hat{\psi}^*_{\vec{\omega}_1,\sigma_1}(p_1)\cdots
      \hat{\psi}_{\vec{\omega}_{2n},\sigma_{2n}}(p_{2n})
      (2\pi)^{d+1}\delta (p_1+\cdots -p_{2n})
\end{eqnarray}
The function $w$ is assumed to be homogeneous of degree $\kappa\in\extraR$,
i.e.
\[
  w(sp_1,\ldots sp_{2n})=s^{\kappa}w(p_1,\ldots p_{2n})
\]
Expressing the sector fields in terms of the scaled fields thus yields
\begin{eqnarray}
  \lefteqn{S_n=\frac{1}{n!}\lambda^{-((n-1)d-1+\kappa)}
  \sum_{\stackrel{\vec{\omega}_1+\cdots
  +\vec{\omega}_n=\vec{\omega}_{n+1}+\cdots +\vec{\omega}_{2n}}
  {\sigma_1,\ldots\sigma_{2n}}}
  \int\dbar\,^{d+1}\tilde{p}_1\cdots\dbar\,^{d+1}\tilde{p}_{2n}}\nonumber\\
  & & w(\tilde{p}_1,\ldots \tilde{p}_{2n})
      \hat{\tilde{\psi}}^*_{\vec{\omega}_1,\sigma_1}(\tilde{p}_1)\cdots
      \hat{\tilde{\psi}}_{\vec{\omega}_2n,\sigma_{2n}}(\tilde{p}_{2n})
      (s\pi)^{d+1}\delta (\tilde{p}_1+\cdots -\tilde{p}_{2n})
\end{eqnarray}
The exponent of $\lambda$ is called the {\em scaling dimension} of $S_n$, 
i.e., the
scaling dimension of $S_n$ is $-((n-1)d-1+\kappa)$. With $n=2$ and $\kappa =0$,
the quartic term (\ref{eq32}) turns out to have scaling dimension
$1-d$. Furthermore, we see that quartic terms of higher degree in the momenta
have a
smaller scaling dimension. The same is true for contributions to the
effective action that are of higher degree in the fields. 

The effective action in terms of the rescaled fields thus reads
\begin{eqnarray}\label{action}
  \lefteqn{S_{\rm eff}=Z^{-1}\sum_{\vec{\omega},\sigma}
  \int\dbar\,^{d+1}\tilde{p}\,
  \hat{\tilde{\psi}}^*_{\vec{\omega},\sigma}(\tilde{p})
  (i\tilde{p}_0-v_F\vec{\omega}\vec{\tilde{p}})
  \hat{\tilde{\psi}}_{\vec{\omega},\sigma}(\tilde{p})}\nonumber\\
  & & +\frac{1}{2}\frac{1}{\lambda^{d-1}}Z^{-2}
      \sum_{\stackrel{\vec{\omega}_1+\vec{\omega}_2=\vec{\omega}_3+
      \vec{\omega}_4}{\sigma,\sigma'}}
      g(\vec{\omega}_1,\ldots,\vec{\omega}_4)
      \int \dbar\,^{d+1}\tilde{p}_1\cdots\dbar\,^{d+1}\tilde{p}_4\\
  & & \hat{\tilde{\psi}}^*_{\vec{\omega}_1,\sigma}(\tilde{p}_1)
      \hat{\tilde{\psi}}^*_{\vec{\omega}_2,\sigma'}(\tilde{p}_2)
      \hat{\tilde{\psi}}_{\vec{\omega}_3,\sigma'}(\tilde{p}_3)
      \hat{\tilde{\psi}}_{\vec{\omega}_4,\sigma}(\tilde{p}_4)
      (2\pi)^{d+1}\delta(\tilde{p}_1+\tilde{p}_2-\tilde{p}_3-\tilde{p}_4)
      \nonumber\\
  & & +\mbox{\ terms of higher order in\ }\frac{1}{\lambda}\nonumber
\end{eqnarray}
In  expression (\ref{action}) for $S_{\rm eff}$ we have introduced a wave
function
renormalization constant $Z$, in order to indicate that the quadratic part of
the effective action may flow under renormalization (decimation of degrees of
freedom and rescaling). In the next section we shall derive renormalization
group flow equations for $Z,\ v_F$ and the coupling constants
$g(\vec{\omega}_1,\ldots,\vec{\omega}_4)$. These equations will determine the
dependence of $Z,\ v_F$ and $g(\vec{\omega}_1,\ldots,\vec{\omega}_4)$ on the
scale parameter $\lambda$.

The expression for $S_{\rm eff}$ on the right hand side of (\ref{action}) shows
that the inverse propagator of the sector fields 
$\hat{\tilde{\psi}}^\#_{\vec{\omega},\sigma}(\tilde{p})$ is diagonal in
the sector index $\vec{\om}$ and that it only depends on $p_0$ and $p_\| =
\vec{\om}\vec{p}$ (but {\em not} on $\vec{p}_\bot =
\vec{p}-(\vec{\om}\vec{p})\vec{\om}$). These features are an aspect of the
principle of {\em dimensional reduction} from $d+1$ to $1+1$ dimensions. 
Indeed,
we observe a rather striking formal similarity between expression
(\ref{action}) and the action of the Gross-Neveu model of interacting,
relativistic Dirac fields in $1+1$ space-time dimensions: The sector index
$\vec{\om}$ plays the r\^{o}le of the {\em flavour index} of the Dirac fields 
in the
Gross-Neveu model; the number of distinct sector indices, 
$\approx\mbox{const}\ 
\lambda^{d-1}$, corresponds to the number, $N$, of flavours of fermions in the
Gross-Neveu model. The coupling constants, 
$\lambda^{1-d}g(\vec{\omega}_1,\ldots,\vec{\omega}_4)$,
correspond to the coupling constant, $\frac{g_{GN}}{N}$, of the quartic term
in the Gross-Neveu model. (The correspondence between $\frac{g_{GN}}{N}$ and 
$\lambda^{1-d}g(\vec{\omega},-\vec{\omega},\vec{\omega}',-\vec{\omega}')
\equiv \lambda^{1-d}g_{BCS}(\vec{\omega}\cdot \vec{\omega}')$ is particularly
precise, as discussed in Chapter 4.)

A powerful method to analyze the Gross-Neveu model is the $\frac{1}{N}$-
expansion. This suggests to analyze non-relativistic, interacting electron
gases with the help of a $\frac{1}{\lambda}$- expansion (with
$\frac{1}{\lambda} \sim \frac{1}{N}$, for $d=2$), and this is precisely what
we shall do in the remaining sections, following beautiful ideas of Feldman,
Magnen, Rivasseau and Trubowitz. 
In the Gross-Neveu model, $Z$ and the velocity of light (corresponding to
$v_F$) do {\em not} flow under renormalization, to leading order in
$\frac{1}{N}$. This suggests that, for the electron gas, $Z$ and $v_F$ do not
flow under renormalization to leading order in $\frac{1}{\lambda}$; a
prediction that will turn out to be correct!

In the following sections, we shall always work in momentum space and with
rescaled sector fields. We shall thus omit the ``hat'' and the ``tilde'' from
the rescaled fields on momentum space. Our analysis will be based on the
assumption that $\lambda_0\gg 1$ and that all coupling constants 
$g^{(0)}(\vec{\omega}_1,\ldots,\vec{\omega}_4) = 
g(\vec{\omega}_1,\ldots,\vec{\omega}_4)\mid_{\lambda=\lambda_{0}}\ \ll 1$.
We shall determine the renormalization flow to leading order in
$\frac{1}{\lambda_{0}}$ (sometimes omitting terms that are of leading order
in $\frac{1}{\lambda_{0}}$, but of high order in 
$g^{(0)}(\vec{\omega}_1,\ldots,\vec{\omega}_4)$).
%
%%%%%%%%%%%%%%%%%%%%%%%%%%%%%%%%%%%%%%%%%%%%%%%%%%%%%%%%%%%%%%%%%%%%%%%%%%%%%%%
\subsection{Integrating out modes}
In the last section, we have understood how the different parts of
the effective action of an interacting electron gas behave under rescaling.
Here we turn to the second step of the renormalization group method --- the
decimation of degrees of freedom. Initially, we assume the 
degrees of freedom in $\extraR^d\setminus
S_F^{(\lambda_0)}$ to be integrated out, as discussed in Sect. 2.2. In the
j'th step, the degrees of
freedom localized in $S_{F}^{(M^{j-1} 
\lambda_0)}\backslash S_{F}^{(M^{j}\lambda_0)}$ have to be eliminated, where
$M$ is a positive integer $\geq 2$. Thus the 
scaling factor at scale $j$ is $\lambda_j=\lambda_0 M^j$.
At scale 0 we are given $Z^{(0)}$, $v^{(0)}_F$ and functions
$g^{(0)}(\vec{\omega}_1,\ldots,\vec{\omega}_4)$ of the unit vectors 
$\vec{\omega}_1,\ldots ,\vec{\omega}_4$, with 
$\vec{\omega}_1+\vec{\omega}_2=\vec{\omega}_3+\vec{\omega}_4$. As announced,
we assume that $\lambda_0 \gg 1$ and $\mbox{max}_{\vec{\omega}_1,
\vec{\omega}_2,\vec{\omega}_3}\lbrace
g^{(0)}(\vec{\omega}_1,\ldots,\vec{\omega}_4)\rbrace\ \ll 1$.

>From this point on, we omit the spin indices when no confusion
arises. We propose to first discuss the possible
intersector scattering geometries, as we would like to better understand 
the structure of the coupling constants
$g^{(0)}(\vec{\omega}_1,\ldots,\vec{\omega}_4)$.
How many independent $g^{(0)}(\vec{\omega}_1,\ldots,\vec{\omega}_4)$'s exist ? For $d = 3$ we
suppose that $\vec{\omega}_3 \neq -\vec{\omega}_4$. On the unit sphere,
there are $N^{(0)} \approx \mbox{const}\,\lambda_0^{d-1}$ different 
$\vec{\omega}$'s. All
choices $\vec{\omega}_1$, $\vec{\omega}_2$ with $\vec{\omega}_1
 + \vec{\omega}_2 = \vec{\omega}_3 + \vec{\omega}_4$ lie on a cone containing
$\vec{\omega}_3$ and $\vec{\omega}_4$
with symmetry axis $\vec{\omega}_3 + \vec{\omega}_4$. Therefore there are
$O(\lambda_0^{d-2})$ choices. Only when $\vec{\omega}_3 =
-\vec{\omega}_4$, there are $N^{(0)}\approx\mbox{const}\,
\lambda_0^{d-1}$ choices.
Similarly, in $d=2$, there are exactly two choices for $\vec{\omega}_1$,
$\vec{\omega}_2$ , if $\vec{\omega}_3 \neq -\vec{\omega}_4$, and
$N^{(0)} \approx \mbox{const}\,\lambda_0$ choices if $\vec{\omega}_3 =
-\vec{\omega}_4$.
\begin{figure}[ht]
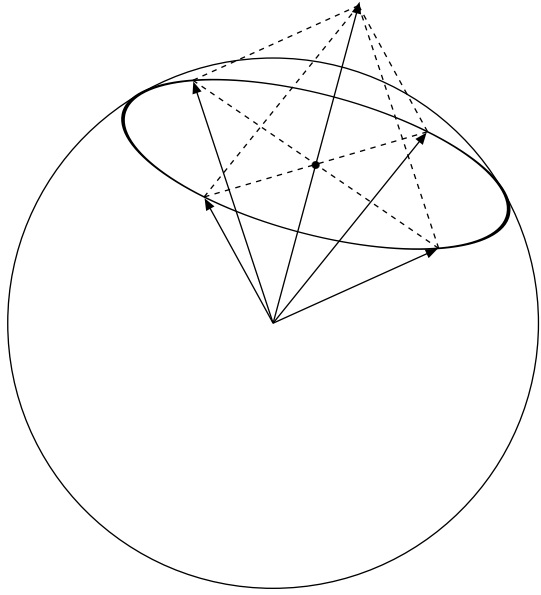

\center\figureCJ
\caption{Sector momentum conservation in $d=3$}
\end{figure}

Couplings involving incoming states with $\vec{\omega}_3 \neq
-\vec{\omega}_4$ shall be denoted by
%
%\begin{equation}
$g^{(0)}(\vec{\omega}_1,\ldots,\vec{\omega}_4)$
%\end{equation}.
Couplings that involve sector indices with
$\vec{\omega}_3 = -\vec{\omega}_4$ or, equivalently,
$\vec{\omega}_1=-\vec{\omega}_2$ shall be denoted by
%
%\begin{equation}
$g_{\,{\rm BCS}}^{(0)}(\vec{\omega}_1,\vec{\omega}_4)$
%\end{equation}
%
(for {\em BCS-scattering}). The latter will prove to be crucial for the 
understanding
of the superconducting state. We observe that, because of
rotation invariance, the coupling $g_{\,{\rm BCS}}$ is only a function of the 
angle between $\vec{\omega}_1$ and $\vec{\omega}_4$.

Technically, we will carry out the decimation of degrees of freedom using
perturbation theory. If, under the renormalization group transformation, 
$g^{(j)}(\vec{\omega}_1,\ldots,\vec{\omega}_4)=
g(\vec{\omega}_1,\ldots,\vec{\omega}_4)|_{\lambda=\lambda_j}$ 
increases with growing
j (and decreasing distance to the Fermi surface), the validity of perturbative
results eventually breaks down.

We start explicit calculations with the renormalization of the
electron propagator when passing from scale $\lambda_0$ to $\lambda_1$. 
We know that every interaction squiggle provides a factor
$\lambda_0^{1-d}g^{(0)}(\vec{\omega}_1,\ldots\vec{\omega}_4)$.
What are the dominant radiative corrections of the electron propagator? The
two possible one loop diagrams are
\begin{equation}\figureCA\label{eq3}\end{equation}
We call them {\em tadpole} and {\em turtle graph}. We shall
estimate their amplitudes. There is one interaction squiggle of order
$\frac{1}{\lambda_0^{d-1}}$ and $N^{(0)}\approx\mbox{const}\,\lambda_0^{d-1}$
choices for the
inner particle sectors, denoted by $\vec{\omega}^{'}$. Therefore, both these
graphs correspond to contributions of  order zero in $\frac{1}{\lambda_0}$. It
will prove to be 
useful to
introduce a simplified graphical notation. Namely, we replace the
squiggle by a point, i.e., each
vertex is represented by a cross with two incoming and two outgoing
lines. The new vertex stands for the sum of the two old vertices that yield
the same new vertex. For example, the diagrams in (\ref{eq3})
are now represented by one graph, namely
\begin{equation}\figureCB\end{equation}

What about combinations of these first order corrections?
\begin{equation}\figureCC\end{equation}
For every vertex corresponding to a factor $\frac{1}{\lambda_0^{d-1}}$, there
are
$N^{(0)}\approx\mbox{const}\,\lambda_0^{d-1}$ loop sectors to choose from. 
Therefore all these graphs correspond to amplitudes of order one. But, as a
matter of fact, one observes
with amazement that all their amplitudes vanish. The reason is
that, in
all these graphs, there is an oriented loop involving two particle lines with
the same
sector label $\vec{\omega}^{'}$ and the same box momenta and energies. For
each such loop, the
integration over the loop momentum yields a factor
  \[\int \dbar k\int\dbar k_0\,\left(
  \frac{1}{ik_0-v_F\vec{\omega}\vec{k}}\right) ^2\]
in the total amplitude. The two poles of the integrand coincide, and the
integrand decays quadratically at infinity. The residue
theorem then tells us that the integral vanishes.

Furthermore, we know that graphs which are not {\em one particle irreducible 
(1PI)}
yield a vanishing amplitude. Indeed, a graph of this type has the following
form
\begin{equation}\figureCD\end{equation}
and one has to sum over $\vec{\omega}^{'}$ and to integrate over $k$ with
$\frac{k_F}{M}\leq|\vec{\omega}'\vec{k}|\leq k_F$. The empty circles stand for 
arbitrary
subdiagrams with the indicated external legs. Because of momentum
conservation, there is a factor $\delta_{\vec{\omega},\vec{\omega}'}\,
\delta(p-k)$. Hence only $\vec{\omega}'=\vec{\omega}$
contributes. But, in this case, the argument of the $\delta$-distribution never
vanishes since $|\vec{\omega}\vec{p}|<\frac{k_F}{M}$. Consequently the 
amplitude is zero.

The graphs not taken into account so far are of order
$O(1/\lambda_0)$ or higher. As an example, let us consider the graph
\begin{equation}\figureCE\end{equation}
Clearly, $\vec{\omega}_3$ is determined by $\vec{\omega}$, $\vec{\omega}_1$ 
and $\vec{\omega}_2$. Thus there remain two summations, over $\vec{\omega}_1$
and $\vec{\omega}_2$. If $\vec{\omega}_1$ is different 
from $-\vec{\omega}$, then the number of $\vec{\omega}_2$'s satisfying the
relations $|\vec{\omega}+\vec{\omega}_2-\vec{\omega}_1|=1$ is of order
O($\lambda_0^{d-2}$). Hence the number of summands is at most of order
O($\lambda_0^{2d-3}$). Since each of the squiggles gives a factor
$\frac{1}{\lambda_0^{d-1}}$, the amplitude corresponding to this graph is of
order
O$(1/\lambda_0$), as claimed. 

Conclusion: The corrections of lowest order in $\frac{1}{\lambda_0}$ to 
the electron propagator arise
from the tadpole and the turtle graph. In order to identify the
renormalized quantities more easily, we work with a renormalized action in its
most general form (dropping terms of higher order in $\frac{1}{\lambda}$, 
$\lambda = \lambda_0,\lambda_1,\lambda_2,...$)
\begin{eqnarray}\label{eq316}
  \lefteqn{S_{\rm eff}=-Z^{-1}\sum_{\vec{\omega},\sigma}\int \dbar\, ^{d+1}k\,
  \psi^*_{\vec{\omega}\sigma}(k)(ik_0 -v_F\vec{\omega}\vec{k})
  \psi_{\vec{\omega},\sigma}(k)}\nonumber\\
  & & -\lambda Z^{-1}\sum_{\vec{\omega},\sigma}\int \dbar\, ^{d+1}k\,
  \psi^*_{\vec{\omega},\sigma}(k)\delta\mu\psi_{\vec{\omega},\sigma}(k)
  +\  O(\frac{1}{\lambda})
  \nonumber\\
  & & +\frac{1}{2}\frac{1}{\lambda^{d-1}}Z^{-2}
  \sum_{\stackrel{\vec{\omega}_1+\vec{\omega}_2=\vec{\omega}_3+
  \vec{\omega}_4}{\sigma,\sigma^{'}}}g(\vec{\omega}_1,\ldots\vec{\omega}_4)
  \int \dbar\, ^{d+1}k_1\cdots\,\dbar\, ^{d+1}k_4\label{eq2} \\ 
  & & \psi^*_{\vec{\omega}_1,\sigma}(k_1)\psi^*_{\vec{\omega}_2,\sigma^{'}}
  (k_2)\psi_{\vec{\omega}_3,\sigma^{'}}(k_3)\psi_{\vec{\omega}_4,\sigma}(k_4)
  (2\pi)^{d+1}\delta (k_1+k_2-k_3-k_4)\nonumber\\
  & & + \ O(\frac{1}{\lambda^d})
\end{eqnarray}
On the right side of (\ref{eq316}), the wave vectors $k = (k_0,\vec{k})$ are
constrained to belong to $\extraR \times \tilde{B}_{\vec{\om}}$, for a given
sector momentum $\vec{\om}$.
The factor $\lambda$ in front of the second term expresses the
fact that this term has scaling dimension $1$. Denoting the inverse propagator
by $\Gamma$, the sum of all quadratic terms in the effective action can be
written as
\begin{equation}
  \sum_{\sigma,\vec{\omega}}\int \dbar\, ^{d+1}k\,\psi^*_{\sigma,
  \vec{\omega}}(k)\Gamma_{\vec{\omega}}(k)\psi_{\sigma,\vec{\omega}}(k)
\end{equation}
where
\begin{eqnarray}
  \Gamma_{\vec{\omega}}(0) & = & -\lambda Z^{-1}\,\delta\mu \\
  \frac{\partial}{\partial p_0}\Gamma_{\vec{\omega}}(p)\mid_{p=0} & 
  = & -i\,Z^{-1} \\
  \vec{\omega}(\vec{\nabla}_{\vec{p}}\Gamma_{\vec{\omega}}
  (p)\mid_{p=0}) & = & v_F Z^{-1}
\end{eqnarray}
Because the tadpole and the turtle graph correspond to $p$-independent
amplitudes, there are no corrections 
to $Z$ and $v_F$ in leading order in $\frac{1}{\lambda_0}$:
\begin{eqnarray}
  Z^{(1)} & = & Z^{(0)}+O((g^{(0)})^2/\lambda_0) \\
  v_F^{(1)} & = & v^{(0)}_{F}+O((g^{(0)})^2/\lambda_0)
\end{eqnarray}

Let us compute the amplitude of the tadpole graph renormalizing the
chemical potential:
\begin{equation}
  \frac{1}{\lambda_0^{d-1}}\sum_{\vec{\omega}^{'},\sigma^{'}}
  g^{(0)}(\vec{\omega},\vec{\omega}^{'},\vec{\omega}^{'},
  \vec{\omega}) \int\dbar
  k_{\perp}\int_{\frac{k_F}{M}\leq |k_{\parallel}|\leq k_F}\dbar k_{\parallel}
  \ \lim_{\tau\downarrow 0}\int\dbar k_0\, 
  \frac{\mbox{e}^{-i\tau k_0}}{ik_0-v_F k_{\parallel}}
\end{equation}
Applying the residue theorem, the $k_0$-integration yields a constant equal to
$-1$ for $k_\|>0$ and zero for $k_\|<0$. 
Thus the 
$k$-integration gives a result of order 1. The factor $\lambda_0^{1-d}$  
is cancelled by the summation over sectors. Therefore the 
tadpole amplitude is $O(g^{(0)})$. Direct
computation shows that the turtle diagram is of the same order.
Comparing with (\ref{eq2}) we obtain
\begin{equation}
  \delta\mu^{(1)}=O(g^{(0)}/\lambda_0)
\end{equation}
The renormalization of the chemical potential deforms the singular surface in
the propagator --- the Fermi surface --- and modifies the original form
(\ref{action}) of the effective action. This will be cured by a
change of variables discussed at the end of this section.

Let us now turn to the renormalization of the 
couplings $g$. To determine $g^{(1)}$, we have to sum over all connected
graphs with four external legs. In fact, the graphs that are not 1PI give
a vanishing contribution. These graphs are of the form
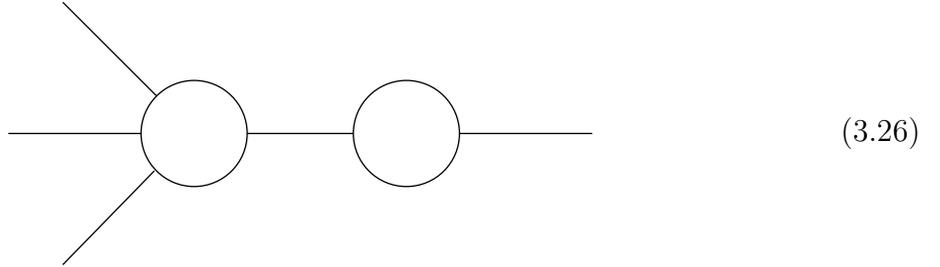
\begin{equation}\figureCF\end{equation}
As in the renormalization of the electron propagator, the amplitude of such a 
graph vanishes, as a consequence of momentum conservation.

The tree level contribution  to $g^{(1)}(\vec{\omega}_1,\ldots,\vec{\omega}_4)$
is just $g^{(0)}(\vec{\omega}_1,\ldots,\vec{\omega}_4)$. The one-loop
correction is described by the diagrams
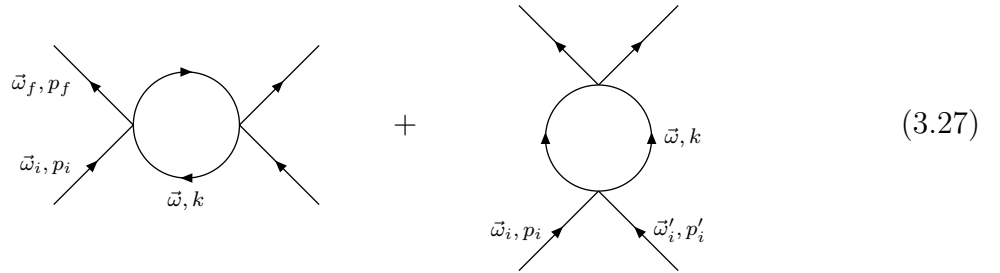
\begin{equation}\label{eq326}\figureCG\end{equation}
The first diagram corresponds to the sum of the following
three graphs 
\begin{equation}\figureCH\end{equation}
The amplitude is of order at most O($(g^{(0)})^2/\lambda_0^d$). In fact, for
$\vec{\omega}_i\neq\vec{\omega}_f $, the number of $\vec{\omega}$'s
contributing to the $\vec{\omega}$-summation is
$O(\lambda_0^{d-2})$, due to sector momentum conservation. Consequently the
amplitude 
is of order $O((g^{(0)})^2/\lambda_0^d)$. But when $\vec{\omega}_i=
\vec{\omega}_f$ the number of terms in the sum is actually 
$O(\lambda_0^{d-1})$.
We thus must take a closer look at the integration involved in the calculation
of
the amplitudes. The $\vec{k}$-integration extends over $\vec{k}$'s satisfying
$\frac{k_F}{M}<|\vec{\omega}\vec{k}|<k_F$ and $\frac{k_F}{M}<|\vec{\omega}
(\vec{k}+\vec{p}_i-\vec{p}_f)|<k_F$. 
Explicitly the loop integral is given by
  \[\int\dbar\,^d k\int\dbar k_0\,\frac{1}{ik_0-v_F\vec{\omega}\vec{k}}\ \  
  \frac{1}{i(k_0+p_{i,0}-p_{f,0})-v_F\vec{\omega}(\vec{k}+\vec{p_i}-
  \vec{p_f})}\]
In order for the integral to be different from zero, the two poles must not  
lie
in the same (upper or lower) half plane, i.e., $\vec{\omega}
\vec{k}$ and $\vec{\omega}(\vec{k}+\vec{p}_i-\vec{p}_f)$ must have opposite
signs. The set of $\vec{k}$'s satisfying all three conditions is of measure
zero 
since $|\vec{\omega}(\vec{p}_i-\vec{p}_f)|<\frac{2k_F}{M}$. Hence the integral
vanishes. Thus
the amplitude of the graph is of order $O((g_{(0)})^2/\lambda_0^d)$, as
claimed, except for exceptional configurations of external sector momenta 
($\mid \vec{\omega}_i - \vec{\omega}_f \mid \sim \frac{1}{\lambda_0}$).

Next, we turn to the second diagram in (\ref{eq326}) corresponding to the graph
\begin{equation}\figureCI\end{equation}
When $\vec{\omega}_i\neq -\vec{\omega}_i'$, the number of 
nonvanishing terms in the $\vec{\omega}$-summation is $O(\lambda_0^{d-2})$, and
the amplitude is at most of order $O((g^{(0)})^2/\lambda_0^d)$. But, for the
BCS configuration, 
$\vec{\omega}_i=-\vec{\omega}'_i$, the situation is completely different.
There are now 
$O(\lambda_0^{d-1})$ possible choices for the internal particle momenta, and 
no miracle makes the amplitude vanish. In fact, the loop integration for 
$\vec{p}_i=\vec{p}_f=0$ is
\begin{equation}
  \int_{\tilde{B}_{\vec{\om}}}
  \dbar\,^d k\int\dbar k_0\,\frac{1}{k_0^2+(v_F\vec{\omega}\vec{k})^2}
\end{equation}
which is strictly positive. Thus the amplitude is of order
$O((g^{(0)})^2/\lambda_0^{d-1})$.

Inserting the correct scale factors, we get the following flow equations for
the
quartic couplings:
\begin{eqnarray}
  g^{(1)} & = & g^{(0)}+O((g^{(0)})^2/\lambda_0)\ ,\ \mbox{for}
  \ \vec{\omega}_i\neq -\vec{\omega}_i' 
\\
  g_{\,{\rm BCS}}^{(1)} & = & g_{\,{\rm BCS}}^{(0)}
                              +O((g_{\,{\rm BCS}}^{(0)})^2)\ \ ,\ \mbox{for}
 \ \vec{\omega}_i= -\vec{\omega}_i'
\end{eqnarray}
Thus, for the BCS couplings, the tree level does not yield the complete
contribution to lowest order in $\frac{1}{\lambda_0}$! In order to understand
the flow of the BCS couplings, we have to
investigate the loop corrections in more detail. This will be 
done in the next section.

The iteration step from scale $j$ to scale $j+1$ is analyzed in a similar 
manner as from scale $0$ to scale $1$. Neglecting vertices of degree $> 4$ in
$\psi^*$  and $\psi$, we obtain
\begin{eqnarray}
  Z^{(j+1)} & = & Z^{(j)}+O((g^{(j)})^2/\lambda_j)\\
  v_F^{(j+1)} & = & v_F^{(j)}+O((g^{(j)})^2/\lambda_j)\\
  \delta\mu^{(j+1)} & = & O(g^{(j)}/\lambda_j) \\
  g^{(j+1)} & = & g^{(j)}+O((g^{(j)})^2/\lambda_j)
  \ ,\ \mbox{for}
  \ \vec{\omega}_i\neq -\vec{\omega}_i' \label{eq335}
\\
  g_{{\rm BCS}}^{(j+1)} & = & g_{{\rm BCS}}^{(j)}+O((g^{(j)}_{{\rm BCS}})^2)
  \ ,\ \mbox{for}
  \ \vec{\omega}_i= -\vec{\omega}_i' 
\end{eqnarray}
We observe that the couplings $g^{(j)}(\vec{\om}_1,...,\vec{\om}_4)$, for
$\vec{\omega}_i\neq -\vec{\omega}_i'$, do essentially not
flow. But the BCS couplings may change considerably
under the renormalization group flow. 
If the BCS channel is turned off, perturbation theory is valid, and the
system approaches a Landau-Fermi liquid. If $g_{{\rm BCS}}^{(j)}$ grows in 
$j$ then
perturbation theory breaks down. In this situation the system becomes a
superconductor, as studied in more detail in the next section and in Chapter 4.
The unlimited growth of $g_{{\rm BCS}}^{(j)}$ in $j$ then reflects the
fact that we are performing a perturbative analysis about a state that is
not a ground state. As a matter of fact, superconductors do not possess a 
Fermi surface.

At this point, a comment on contributions of degree $> 4$ in $\psi^*$, $\psi$
to the effective action is appropriate. The decimation of degrees of freedom
(i.e., the integration over degrees of freedom) in an iteration step $j$,
$j=0,1,2,...$, of the renormalization group procedure produces terms like
\be
\figureCS\label{note1}
\ee
of scaling dimension $-(d-1)(k-1)$ in the effective action
$S_{\rm eff}^{(j)}$. The form and renormalization flow of the dominant
contributions to these terms can be studied quite explicitly. In the next
iteration step, from $j$ to $j+1$, these terms yield contributions to the
dimensionless coupling constants $g^{(j+1)}$ of the terms of degree 4 in 
$S_{\rm eff}^{(j+1)}$ that turn out to be of order $1=(\frac{1}{\lambda_j})^0$,
even for incoming sector momenta 
$\vec{\omega}_i\neq -\vec{\omega}_i'$,
\be
\figureCT\label{note2}
\ee
and hence may be important. They are, however, of higher order in 
$g^{(j)},g^{(j-1)},...$ . A careful analysis (presented elsewhere) reveals
that these contributions induce a finite flow of the couplings $g^{(j)}$
towards RPA-(random phase approximation) type fixed points. For short-range
two-body interactions, and if $g^{(0)} \ll 1$, this represents an {\em 
unimportant}
modification of (\ref{eq335}). However, for long-range (e.g. Coulomb) two-body
interactions, or if $g^{(0)}$ is not small, the modification in the
renormalization flow of the $g^{(j)}$'s due to the terms of degree $> 4$ in
$\psi^*$, $\psi$ in the effective action is {\em essential} and is intimately
connected with the phenomenon of {\em screening}. 
A detailed discussion of these
matters would go beyond the scope of these notes.
 
Next, we analyze the renormalization of the chemical
potential. At scale $j$, the effective action is given by (we use unscaled
fields) 
\begin{eqnarray*}
  S^{(j)}_{\rm eff} & = & -Z^{(j)\,-1}\sum_{\sigma,\vec{\omega}}
  \int\dbar\,^{d+1}k\,
  \psi_{\vec{\omega},\sigma}^*(k)(ik_0-v_F^{(j)}\vec{\omega}\vec{k}+
  \delta\mu^{(j)})
  \psi_{\vec{\omega},\sigma}(k)\\
  & & +\mbox{\ higher degree terms}
\end{eqnarray*}
We may perform a shift, $k_{\parallel}\rightarrow k_{\parallel}-\frac{\delta
\mu^{(j)}}{v_F^{(j)}}$, of the $k_{\parallel}$-variable. Note that 
$\frac{\delta\mu^{(j)}}{v_F^{(j)}}=k_F O(\frac{g^{(j)}}{\lambda_j})$. Thus, if
the
coupling $g^{(j)}$ remains approximately constant under the RG flow and
sufficiently small, the shift of the $k_{\|}$-variable will always be  smaller
than $\frac{k_F}{\lambda_j}$. 
The integration
measure is invariant under this coordinate transformation. The transformed
action reads
\[S_{\rm eff}^{(j)}=-Z^{(j)\,-1}\sum_{\sigma,\vec{\omega}}\int\dbar\,^{d+1}k\,
  \psi_{\vec{\omega},\sigma}^*(k)(ik_0-v_F^{(j)}k_{\parallel})
  \psi_{\vec{\omega},\sigma}(k)
  +\mbox{\ higher order terms}\]
Hence $S_{\rm eff}^{(j)}$ is again of the form (\ref{action}) (except that
the domain over which $\vec{k}$ is integrated has changed slightly).
\begin{figure}
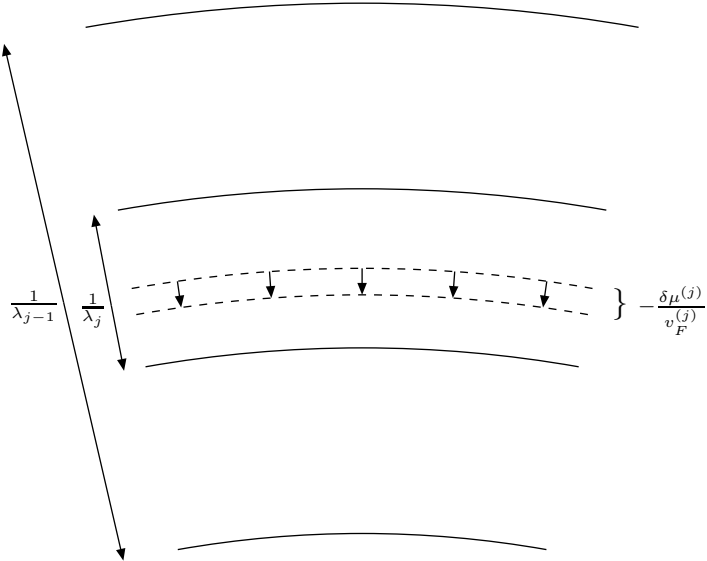

  \center\figureCR
  \caption{Renormalization of the Fermi surface}
\end{figure} 

Our results motivate the following interpretation of the renormalization group
transformation: Assuming that the interacting system has a Fermi surface
(i.e., that the BCS channel is turned off),
the RG procedure enables us to
approach the renormalized Fermi surface stepwise, starting from the Fermi
surface of the noninteracting
system. In the
limit $j\rightarrow\infty$, we reach the physical Fermi surface of the
interacting
system. Imposing renormalization conditions on $Z^{(\infty)}$, 
$v_F^{(\infty)}$, $g^{(\infty)}$  
amounts to solving a ``final
condition problem'' in the space of running coupling constants. 
The RG recursion
generates the discrete dynamics of coupling constants, as the scale parameter
$j$ is varied, but instead of initial conditions, experiment fixes
final conditions. What we actually do is that we integrate the equations for
the running couplings backwards from the Landau-Fermi liquid fixed point
($j=\infty$) down
to $j=0$. The values
$\mu^{(0)}$, $Z^{(0)}$, $v_F^{(0)}$, and $g^{(0)}$  
parametrize a {\em microscopic system}, whereas  
$\mu^{(\infty)}$, $Z^{(\infty)}$, 
$v_F^{(\infty)}$ and $g^{(\infty)}$  parametrize a {\em universality 
class of scaling limits describing macroscopic states}.

%%%%%%%%%%%%%%%%%%%%%%%%%%%%%%%%%%%%%%%%%%%%%%%%%%%%%%%%%%%%%%%%%%%%%%%%%%%%%
\subsection{The BCS channel}

In the previous section we observed that the one-loop correction in the 
renormalization group flow of $g_{\,{\rm BCS}}$ is of order zero in $\frac{1}{
\lambda}$; $g_{\,{\rm BCS}}$ might be a {\em relevant coupling}. 
To reach a better understanding of $g_{\,{\rm BCS}}$,
we determine the flow of $g_{\,{\rm BCS}}$, taking into account all 
corrections of order zero in $\frac{1}{\lambda}$ (but ignoring terms of degree
$>4$ in $\psi^*$ and $\psi$ in the effective actions; see (\ref{note1}), 
(\ref{note2})). We omit the scale index when no confusion may arise.

First, we have to determine all four-legged diagrams corresponding to
corrections to 
$g_{\,{\rm BCS}}$ of order zero in $\frac{1}{\lambda}$. These are precisely 
those
diagrams whose amplitude is of order $O(\frac{1}{\lambda^{d-1}})$. Every
graph with four or more external legs can be constructed from a uniquely
determined diagram containing no two-legged subgraphs. This is done by 
replacing each inner line by an appropriate graph with two external legs. 
We denote by
\begin{equation}\label{eq4}\figureCK\end{equation}
the sum  of all graphs with two external legs; i.e., in a diagram
we have to replace each  such symbol by the sum of all two-legged diagrams.
Let us consider an arbitrary 
four- or more legged diagram with no two-legged subdiagrams, with all inner 
lines carrying this modified electron propagator line. We shall
compute the amplitude of such a graph by just replacing the electron
propagator by the amplitude of the modified electron propagator line. (To see
why this is correct, we expand each modified propagator into
the corresponding sum of terms. Every summand is the amplitude of a specific
diagram 
to which a combinatorial factor is assigned that is, in general, smaller than the combinatorial
factor  belonging to that diagram. But one easily proves that
summing up all terms corresponding to a given diagram yields the correct 
factor.)

What are the connected diagrams with four external legs whose amplitude is of 
order $O(\frac{1}{\lambda^{d-1}})$? Consider a four-legged graph with n
interaction
squiggles. The squiggles provide a factor of order 
$\lambda^{-n(d-1)}$. There are $2n-2$ inner lines. But the $\delta$-functions
of sector momentum conservation  restrict the sum over the $\vec{\omega}$'s
to $n-1$ independent sector momenta. Thus, in order for the amplitude
to be of order $O(\frac{1}{\lambda^{d-1}})$, {\em all} the $n-1$ sector 
momentum
summations must  extend over the entire $(d-1)$-dimensional Fermi sphere! In
this case, the diagrams
\begin{equation}\label{eq3b}\figureCL\end{equation}
with $n\geq 0$, are of order $O(\frac{1}{\lambda^{d-1}})$, unless a miracle 
happens that makes some of them vanish. Using the above argument, one can
prove that the graphs (\ref{eq3b}) are the {\em only} four-legged 
diagrams with no two-legged subdiagrams which are of order $O(\frac{1}{
\lambda^{d-1}})$.

We conclude that if we do not include terms of degree $>4$ in $\psi^*$, $\psi$
in the effective action then all the diagrams that yield a correction to
$g_{BCS}$ of 
order zero in $\frac{1}{\lambda}$ are contained in the set of graphs built
from (\ref{eq3b}) by replacing all the inner lines by (\ref{eq4}). In fact,
in (\ref{eq4}) only the graphs of order zero in $\frac{1}{\lambda}
$ have to be taken into account. Let us consequently define
\begin{equation}\label{eq5}\figureCM\end{equation}
as the sum of all two-legged graphs corresponding to corrections of order zero
in $\frac{1}{\lambda}$ to the quadratic part of the effective action. The
diagrams that have to be computed are
\begin{equation}\label{eq6}\figureCN\end{equation}
with $n\in\extraN$. In our computation we just have to replace the electron
propagators in (\ref{eq3b}) by the renormalized ones in (\ref{eq5}). It is
well known that the renormalized electron propagator is given by
\begin{equation}
  \mbox{amplitude of\ }\figureCOa = -(ik_0-v_F\vec{\omega}\vec{k}+
  \mbox{amplitude of\ }\figureCOb )^{-1}
\end{equation}
where only amputated 1PI graphs with amplitude of order zero in 
$\frac{1}{\lambda}$
are included on the right hand side. The results discussed in the last section
then show that 
\begin{equation}
  \mbox{amplitude of\ }\figureCP = -(ik_0-v_F\vec{\omega}\vec{k}+\lambda
  \delta\mu_1)^{-1}
\end{equation}
where $\delta\mu_1$ is the renormalization of the chemical potential, and
only contributions of first
order in $\frac{1}{\lambda}$ are taken into account; $\delta\mu_1$ depends on
$Z$, $v_F$ and $g$.

Now we are able to compute the renormalized value of $g_{\,{\rm BCS}}$. The 
amplitude
of (\ref{eq6})  with zero incoming and outgoing box momenta is given by
\[
  \left( \frac{1}{\lambda^{d-1}}\right)^{n+1}
  \sum_{\vec{\omega}_1,\ldots\vec{\omega}_n}
  (-1)^n\beta^n
  g_{\,{\rm BCS}}(\vec{\omega},\vec{\omega}_n)
  g_{\,{\rm BCS}}(\vec{\omega}_n,\vec{\omega}_{n-1})\cdots
  g_{\,{\rm BCS}}(\vec{\omega}_1,\vec{\omega}')
\]
and $\beta$ is a strictly positive number given by
\begin{eqnarray}
  \beta  &  =  &  \int\dbar k_{\perp}\,\dbar k_{\|}\,\dbar k_0\,
                  \frac{1}{ik_0-v_F k_{\|}+\lambda\delta\mu_1}\ \ 
                  \frac{1}{-ik_0-v_F k_{\|}+\lambda\delta\mu_1}
                  \nonumber\\
         &  =  &  \int\dbar k_{\perp}\,\dbar k_{\|}\,\dbar k_0\,
                  \frac{1}{k_0^2+(v_Fk_{\|}-\lambda\delta\mu_1)^2} > 0
\end{eqnarray}
We find that the renormalized value of $g_{\,{\rm BCS}}$ is
\begin{eqnarray}
  \lefteqn{g^{(j+1)}_{\,{\rm BCS}}(\vec{\omega},\vec{\omega}')=}\nonumber\\  
  & &  g^{(j)}_{\,{\rm BCS}}(\vec{\omega},\vec{\omega}')+
       \sum_{n=1}^{\infty}\left(\frac{1}{\lambda_j^{d-1}}\right)^n
       \sum_{\vec{\omega}_1,\ldots\vec{\omega}_n}
       (-1)^n\beta_j^n g^{(j)}_{\,{\rm BCS}}(\vec{\omega},\vec{\omega}_n)\cdots
       g^{(j)}_{\,{\rm BCS}}(\vec{\omega}_1,\vec{\omega}')
       \label{eq7}\\
  & &  +O\left(\frac{g^{(j)}}{\lambda_j}\right) \nonumber
\end{eqnarray}
In order to get a more explicit expression for the flow equation (\ref{eq7}),
it is
useful  to expand the BCS couplings, 
$g_{\,{\rm BCS}}(\vec{\omega},\vec{\omega}') \equiv
g_{\,{\rm BCS}}(\angle (\vec{\omega},\vec{\omega}'))$, into spherical harmonics
\begin{equation}
  g_{\,{\rm BCS}}(\vec{\omega},\vec{\omega}')=\sum_{l=0}^{\infty}g_l h_l(
  \vec{\omega},\vec{\omega}')
\end{equation}
with
\begin{equation}
  h_l (\vec{\omega},\vec{\omega}'):=\left\{
  \begin{array}{l@{\ \ \ \ }c}
    \frac{1}{\pi (1+\delta_{l,0})}\cos (l\angle (\vec{\omega},\vec{\omega}'))
    &  d=2 \\
    \frac{2l+1}{4\pi}P_l\left(\frac{\vec{\omega}\cdot\vec{\omega}'}
    {|\vec{\omega}||\vec{\omega}'|}\right)  &  d=3
  \end{array}\right.
\end{equation}
$P_l$, $l\in\extraN$, are the Legendre polynomials. For $d=2$, we assume
O(2)-invariance of the potentials (instead of only SO(2)-invariance). The
normalizations  of the functions $h_l$ have been chosen in such a way that
\begin{equation}
  \int_{|\vec{\omega}|=1}d\sigma(\vec{\omega}_1)\,
  h_l(\vec{\omega},\vec{\omega}_1)
  h_{l'}(\vec{\omega}_1,\vec{\omega}')=\delta_{l,l'}h_l(\vec{\omega},
  \vec{\omega}')
\end{equation}
Because
\begin{equation}
  \frac{1}{\lambda^{d-1}}\sum_{\vec{\omega}}(\,\cdot\,)=\int_{|\vec{\omega}|=1}
  d\sigma(\vec{\omega})\,(\,\cdot\,)\ +O\left(\frac{1}{\lambda}\right)
\end{equation}
the r.h.s. of the flow equation (\ref{eq7}) becomes (we omit the sub- and
superscripts $j$)
\begin{eqnarray*}
  \lefteqn{g_{\,{\rm BCS}}(\vec{\omega},\vec{\omega}')+
  \sum_{n=1}^{\infty}(-1)^n\beta^n
  \int_{|\vec{\omega}_i|=1}d\sigma(\vec{\omega}_1)\,\cdots 
  d\sigma(\vec{\omega}_n)\,
  g_{\,{\rm BCS}}(\vec{\omega},\vec{\omega}_n)\cdots 
  g_{\,{\rm BCS}}(\vec{\omega}_1,\vec{\omega})}\\
  & = & \sum_{l=0}^{\infty}g_l h_l(\vec{\omega},\vec{\omega}')
        +\sum_{n=1}^{\infty}(-1)^n\beta^n\sum_{l_i \geq 0}
        g_{l_0}\cdots g_{l_n}\delta_{l_0,l_1}\cdots\delta_{l_{n-1},l_n}
        h_{l_0}(\vec{\omega},\vec{\omega}')\\
  & = & \sum_{\l=0}^{\infty}(\sum_{n=0}^{\infty}(-1)^n\beta^n g_l^{n+1})
        h_l(\vec{\omega},\vec{\omega}') \\
  & = & \sum_{l=0}^{\infty}\frac{g_l}{1+\beta g_l}h_l(\vec{\omega},
        \vec{\omega}')
\end{eqnarray*}
up to terms of order $\frac{1}{\lambda}$. The flow equation for the BCS
couplings hence takes the form
\begin{equation}\label{eq8}
  g^{(j+1)}_l = \frac{g_l^{(j)}}{1+\beta_j g_l^{(j)}}
  +O\left(\frac{1}{\lambda_j}\right)\ \ \ ,\ l\in\extraN
\end{equation}
where $\beta_j$ is positive and approximately independent of $j$; it depends
on $Z^{(j)}$, $v_F^{(j)}$ and $g^{(j)}$.
The flow equations for different angular momenta, $l$, decouple to lowest
order in
$\frac{1}{\lambda}$. But at order
$\frac{1}{\lambda}$, the flow equations for different $l$'s are coupled
({\em Kohn-Luttinger effect}).

Before we analyze the flow of the couplings, we  study the r\^{o}le
played by electron spin. The potential $v(\vec{x})$ was supposed to be
independent
of spin. Thus the couplings $g^{(j)}(\vec{\omega}_1,\ldots,\vec{\omega}_4)$
are spin-independent, too. We thus divide the quartic term in the effective
action into a spin-singlet and a spin-triplet part. We define
\begin{equation}\label{pairing}
  \phi^{\stackrel{\raisebox{-2pt}{\scriptsize $s$}}{t}}_{\vec{\omega}_1,
  \vec{\omega}_2,\sigma,\sigma'}(k_1,k_2):=\frac{1}{2}
  (\psi_{\vec{\omega}_1,\sigma}(k_1)\psi_{\vec{\omega}_2,\sigma'}(k_2)\mp
  \psi_{\vec{\omega}_1,\sigma'}(k_1)\psi_{\vec{\omega}_2,\sigma}(k_2))
\end{equation}
Clearly $\phi^{\stackrel{\raisebox{-2pt}{\scriptsize $s$}}{t}}_{
\vec{\omega}_1,\vec{\omega}_2,\sigma,\sigma'}=\mp\phi^{\stackrel{
\raisebox{-2pt}{\scriptsize $s$}}{t}}_{\vec{\omega}_1,\vec{\omega}_2,\sigma',
\sigma}$, and hence $\phi^s_{\vec{\omega}_1,\vec{\omega}_2,\sigma,
\sigma}=0$ ($\phi^s$ will correspond to spin-singlet pairing, and $\phi^t$ to
spin-triplet pairing). The quartic term in the action becomes
\begin{eqnarray*}
  \lefteqn{\frac{1}{2}\frac{1}{\lambda^{d-1}}Z^{-2}
  \sum_{\stackrel{\vec{\omega}_1 +\vec{\omega}_2 = 
  \vec{\omega}_3+\vec{\omega}_4}{\sigma,\sigma'}}
  \int\dbar\,^{d+1}k_1\,\cdots\dbar\,^{d+1}k_4\,
  (2\pi)^{d+1}\delta(k_1+k_2-k_3-k_4)} \\
  & & \{\frac{1}{2}
      (g(\vec{\omega}_1,\vec{\omega}_2,\vec{\omega}_3,\vec{\omega}_4)+
      g(\vec{\omega}_1,\vec{\omega}_2,\vec{\omega}_4,\vec{\omega}_3))
      \psi^*_{\vec{\omega}_1,\sigma}(k_1)\psi^*_{\vec{\omega}_2,\sigma'}(k_2)
      \phi^s_{\vec{\omega}_3,\vec{\omega}_4,\sigma',\sigma}(k_3,k_4) \\
  & & +\frac{1}{2}
      (g(\vec{\omega}_1,\vec{\omega}_2,\vec{\omega}_3,\vec{\omega}_4)-
      g(\vec{\omega}_1,\vec{\omega}_2,\vec{\omega}_4,\vec{\omega}_3))
      \psi^*_{\vec{\omega}_1,\sigma}(k_1)\psi^*_{\vec{\omega}_2,\sigma'}(k_2)
      \phi^t_{\vec{\omega}_3,\vec{\omega}_4,\sigma',\sigma}(k_3,k_4)\}
\end{eqnarray*}
Thus, we define the singlet and the triplet couplings as
\begin{equation}
  g^{\stackrel{\raisebox{-2pt}{\scriptsize $s$}}{t}}
  (\vec{\omega}_1,\vec{\omega}_2,\vec{\omega}_3,\vec{\omega}_4):=
  \frac{1}{2}(g(\vec{\omega}_1,\vec{\omega}_2,\vec{\omega}_3,\vec{\omega}_4)\pm
  g(\vec{\omega}_1,\vec{\omega}_2,\vec{\omega}_4,\vec{\omega}_3))
\end{equation}
Because of the property that
  $g(\vec{\omega}_1,\vec{\omega}_2,\vec{\omega}_3,\vec{\omega}_4)=
   g(\vec{\omega}_2,\vec{\omega}_1,\vec{\omega}_4,\vec{\omega}_3)$
we see that
\begin{equation}
  g^{\stackrel{\raisebox{-2pt}{\scriptsize $s$}}{t}}
  (\vec{\omega}_1,\vec{\omega}_2,\vec{\omega}_3,\vec{\omega}_4)=\pm
  g^{\stackrel{\raisebox{-2pt}{\scriptsize $s$}}{t}}
  (\vec{\omega}_2,\vec{\omega}_1,\vec{\omega}_3,\vec{\omega}_4)=\pm
  g^{\stackrel{\raisebox{-2pt}{\scriptsize $s$}}{t}}
  (\vec{\omega}_1,\vec{\omega}_2,\vec{\omega}_4,\vec{\omega}_3)
\end{equation}
Using these symmetry properties, the quartic term in the effective action is
found to be
\begin{eqnarray}
  \lefteqn{\frac{1}{2}\frac{1}{\lambda^{d-1}}Z^{-2}
  \sum_{\stackrel{\vec{\omega}_1 +\vec{\omega}_2 = 
  \vec{\omega}_3+\vec{\omega}_4}{\sigma,\sigma'}}
  \int\dbar\,^{d+1}k_1\,\cdots\dbar\,^{d+1}k_4\,
  (2\pi)^{d+1}\delta(k_1+k_2-k_3-k_4)} \nonumber\\
  & & \{g^s (\vec{\omega}_1,\vec{\omega}_2,\vec{\omega}_3,\vec{\omega}_4)
      \phi^{s\ *}_{\vec{\omega}_2,\vec{\omega}_1,\sigma',\sigma}(k_1,k_2)
      \phi^{s}_{\vec{\omega}_3,\vec{\omega}_4,\sigma',\sigma}(k_3,k_4) \\
  & & +\{g^t (\vec{\omega}_1,\vec{\omega}_2,\vec{\omega}_3,\vec{\omega}_4)
      \phi^{t\ *}_{\vec{\omega}_2,\vec{\omega}_1,\sigma',\sigma}(k_1,k_2)
      \phi^{t}_{\vec{\omega}_3,\vec{\omega}_4,\sigma',\sigma}(k_3,k_4)\}
      \hspace{3cm}\nonumber
\end{eqnarray}
i.e., the quartic term is the sum  of a singlet and a triplet part. The
coupling of the singlet
part is $g^s$, and the coupling of the triplet part $g^t$. In terms of the
original BCS couplings, $g^s$ and $g^t$ are given by
\begin{equation}
  g^{\stackrel{\raisebox{-2pt}{\scriptsize $s$}}{t}}_{\,{\rm BCS}}
  (\vec{\omega},\vec{\omega}')=\frac{1}{2}
  (g_{\,{\rm BCS}}(\vec{\omega},\vec{\omega}')\pm g_{BCS}
  (\vec{\omega},-\vec{\omega}'))
\end{equation}
Using that $\angle (\vec{\omega},-\vec{\omega}')=\angle (\vec{\omega},
\vec{\omega}')+\pi$, the functions $h_l$ can be seen to satisfy
$h_l (\vec{\omega},
-\vec{\omega}')=(-1)^l h_l(\vec{\omega},\vec{\omega}')$. Consequently the
singlet and triplet BCS couplings are
\begin{eqnarray}
  g^s_{\,{\rm BCS}}(\vec{\omega},\vec{\omega}')  &  =  &  
  \sum_{q=0}^{\infty}g_{2q}h_{2q}(\vec{\omega},\vec{\omega}')\\
  g^t_{\,{\rm BCS}}(\vec{\omega},\vec{\omega}')  &  =  &  
  \sum_{q=0}^{\infty}g_{2q+1}h_{2q+1}(\vec{\omega},\vec{\omega}')
\end{eqnarray}
In the expansion of $g^s_{\,{\rm BCS}}$, only {\em even} angular momenta 
appear,
and 
the expansion of $g^t_{\,{\rm BCS}}$ only involves {\em odd} angular momenta, 
as required by the Pauli principle.

We now return to the analysis of the renormalization group flow. The flow
equations (\ref{eq8}) can be written in the form
\begin{equation}\label{eq9}
  g^{(j+1)}_l -g^{(j)}_l = -\frac{\beta_j g^{(j)\ 2}_l}{1+\beta_j g^{(j)}_l}
  +O\left( \frac{g^{(j)\:2}}{\lambda_j}\right)
\end{equation}
Instead of studying this difference equation we propose to investigate the
corresponding differential equation. The differential flow equation is
obtained in the limit where the size, $M-1$, of the scale change in an
iteration step tends to zero. Let us write $g^{(j)}_l\equiv g^{(\lambda_j)}_l$,
and similarly
for all the other running coupling constants. We define
\begin{equation}
  g_l(t):=g_l^{(e^t\lambda_0)}\ \ ,\ \ g(t):=g^{(e^t\lambda_0)}
  \ \ ,\ \ Z(t):=Z^{(e^t\lambda_0)}\ \ ,\ \ v_F(t):=v_F^{(e^t\lambda_0)}
\end{equation}
Consider a scale $\lambda = e^t\lambda_0$ and set $M=e^{t'-t}$,
$t'>t$. We divide both sides of (\ref{eq9}) by $t'-t$ and take the limit $t'
\searrow t$. The l.h.s. yields $\frac{d}{dt}g_l(t)$. The coefficient
$\beta = \beta(t',t)$
vanishes in the limit $t'\searrow t$, and only terms linear in
$t'-t$  have to be kept on the r.h.s. of (\ref{eq9}). Thus we linearize
$\beta(t',t)$ :
\begin{eqnarray}
  & & \beta (t',t)=(t'-t)\gamma(t)+O((t'-t)^2)\\
  & & \gamma (t)=\frac{\partial}{\partial t'}\beta(t',t)\mid_{t'=t}\ >0
\end{eqnarray}
The positivity of $\gamma$ follows from the monotone growth of $\beta(t',t)$
in $t'$. Contributions to $\beta$ corresponding to diagrams with  two or more
loops are of
order $O((t'-t)^2)$. Thus, for the calculation of $\gamma$, only one-loop
diagrams 
have to be taken into account; in particular, we don't have to calculate
corrections to the electron propagator as we did
in the calculation of $\beta$.  Taking the limit $t'\searrow t$, the r.h.s.
of (\ref{eq9}), divided by $t'-t$ becomes
\[
  -\gamma g_l(t)^2+O(e^{-t}g(t)^2)
\]
For the error term the same argument as above applies, and we conclude that
only one-loop diagrams contribute. We thus obtain the flow 
equations
\begin{equation}\label{eq10}
  \frac{d}{dt}g_l(t)=-\gamma g_l(t)^2+O(e^{-t}g(t)^2)
\end{equation}
where $\gamma = \gamma(t,Z(t),v_F(t),g(t))$ is positive, independent of $l$,
and approximately independent of $t$.
Of course, we could have found these flow equations by just looking at all 
possible one-loop graphs, namely
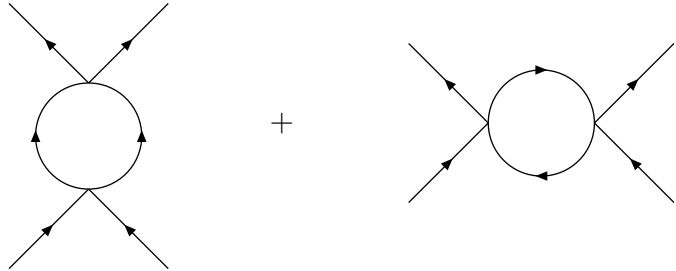
\[\figureCQ\]
The first one yields the first term on the right side in the flow equation and
the second graph
the error term.

We propose to determine the flow described by (\ref{eq10}), neglecting the
error term, and, in accordance with the results of the previous section,
assuming that $v_F$, $Z$ and $g$
essentially do not flow. Then $\gamma$ may be approximated by a positive
constant $\gamma_0$. Thus we have to solve the differential equation
\begin{equation}\label{eq11}
  \frac{d}{dt}g_l = -\gamma_0 g_l^2
\end{equation}
The solution of this equation is easily found to be
\begin{equation}
  g_l(t)=\frac{g_l(0)}{1+\gamma_0 g_l(0)t}
\end{equation}
Note that if $g_l(0) < 0$  the solution blows up at a scale
\begin{equation}\label{blowup}
  t=-\frac{1}{\gamma_0 g_l(0)}\ \ \ , \mbox{i.e. , for}\ \ \ 
  \lambda =e^{-\frac{1}{\gamma_0 g_l(0)}}\lambda_0
\end{equation}
Thus if there is an angular momentum channel, $l$, with attractive
interactions ($g_l(0)<0$) the flow seems to diverge at a finite value of t.
But this just means that perturbation theory breaks down when $t\approx
-(\gamma_0 g_l(0))^{-1}$, and we shall have to employ nonperturbative methods.
The failure of perturbative methods is due to the circumstance that we are
expanding around the wrong state !

Seemingly, everything is fine if $g_l(0)\geq 0 \ \forall l$. However, we 
have
to remember that in (\ref{eq11}) we have omitted the error term which couples
the flows of the running couplings $g_l$ for different values of $l$. Without
fine-tuning of the microscopic two-body potential, it will typically happen 
that $g_l(t)<0$, for some $l$, at some scale $t$. At that point, $g_l(t)$ will
start to grow
untill perturbation theory breaks down, and the ground state of the system
will be superconducting. Thus, generically, a rotationally invariant system
of non-relativistic interacting electrons will be a superconductor for small
enough temperatures. This is the  celebrated Kohn-Luttinger effect first
studied in this fashion by Feldman, Magnen, Rivasseau and Trubowitz.
 
%%%%%%%%%%%%%%%%%%%%%%%%%%%%%%%%%%%%%%%%%%%%%%%%%%%%%%%%%%%%%%%%%%%%%%%%%%%%%
\section{Spontaneous breaking of gauge invariance, and superconductivity}
%%%%%%%%%%%%%%%%%%%%%%%%%%%%%%%%%%%%%%%%%%%%%%%%%%%%%%%%%%%%%%%%%%%%%%%%%%%%%

In the last chapter, we have argued that if some BCS coupling $g_l^{(j_0)}$
becomes {\em negative} at some scale $\lambda_{j_0}$ then $g_l^{(j)}$ grows in
$j$, for $j > j_0$, untill it becomes so large that our perturbative treatment
breaks down. This phenomenon is the signal for an instability of the RG fixed
point around which we are doing perturbation theory. Indeed, the Landau-Fermi
liquid state is {\em not} the true ground state of the system anymore, and we
expect that the RG flow drives the system towards a new RG fixed point
describing a {\em superconductor}. The global U(1)-symmetry of non-relativistic
many-body theory (gauge invariance of the first kind) turns out to be
spontaneously broken in the new (stable) ground state of the system.

The purpose of this chapter is to analyze superconducting ground states and
the associated breaking of gauge invariance. This is not an entirely simple
story, and we therefore focus our attention on the simplest example, that of
an {\em s-wave (BCS) superconductor}.

Thus, we consider a system with the property that, at some scale $\lambda_0
\gg 1$ (with $|g_0 \lambda_0^2| \ll 1$), $g_{l=0}^{(0)} < 0$ and
$|g_{l}^{(0)}| \ll -g_0^{(0)}$, for $l=1,2,3,...$ . According to the results
of Chapter 3, there is then some $j_{sc}\approx - \frac{1}
{\gamma_0 g_0^{(0)}\ln M}$, see (\ref{blowup}), such that at scale
$\lambda_{j_{sc}}>\lambda_0$ $g_{0}^{(j_{sc})} \ll 0$, $|g_{l}^{(j_{sc})}| \ll
-g_0^{(j_{sc})}$, for $l=1,2,3,...$ , and (for $\vec{\om}_1 \neq -
\vec{\om}_2)$ $|g^{(j_{sc})}(\vec{\om}_1,\ldots,\vec{\om}_4)|
\ll -g_0^{(j_{sc})}$.
This suggests that we neglect all terms of degree $\geq 4$ (in $\psi^*,\
\psi$) in the effective action of the system at scale $\lambda_{j_{sc}}$,
{\em except the s-wave BCS term} with coupling constant
$\frac{g_0^{(j_{sc})}}{\lambda_{j_{sc}}^{d-1}}$. The resulting effective field
theory is the one first considered by Nambu and Gorkov.

There is some useful notation to be introduced:
Let $[\vec{\omega}]$ be the ray through the origin containing
$\vec{\omega}$ and  $-\vec{\omega}$. We might think of $\vec{\om}$ and
$-\vec{\om}$ as the two chiralities of a $1+1$ dimensional system of
relativistic fermions. We define field variables
\begin{eqnarray}
  \psi_{[\vec{\omega}]\uparrow} :=
  \left(\begin{array}{c}
    \psi_{\vec{\omega}\uparrow}\\
    \psi_{-\vec{\omega}\uparrow}
  \end{array} \right) & , & 
  \psi_{[\vec{\omega}]\downarrow} :=
  \left(\begin{array}{c}
    \psi_{\vec{\omega}\downarrow}^*\\
    \psi_{-\vec{\omega}\downarrow}^*
  \end{array} \right)\\
  \bar{\psi}_{[\vec{\omega}]\uparrow} :=
  (\psi_{-\vec{\omega}\uparrow}^*,
  \psi_{\vec{\omega}\uparrow}^*) & , & 
  \bar{\psi}_{[\vec{\omega}]\downarrow} :=
  (\psi_{-\vec{\omega}\downarrow},
  \psi_{\vec{\omega}\downarrow})
\end{eqnarray}
We thus group together field variables belonging to  sectors  on the same
ray. Ordered according to
spin indices,  we consider them as entries of a four-component 
quasi-particle field
\begin{eqnarray}\label{eq43}
  \psi_{[\vec{\omega}]} := 
  \left(\begin{array}{c}
    \psi_{[\vec{\omega}]\uparrow}\\
    \psi_{[\vec{\omega}]\downarrow}
  \end{array} \right) & , & 
  \bar{\psi}_{[\vec{\omega}]} := 
  (\bar{\psi}_{[\vec{\omega}]\uparrow},
  \bar{\psi}_{[\vec{\omega}]\downarrow})
\end{eqnarray}
named after Nambu and Gorkov. Let ${\cal V}_{[\vec{\om}]}$ denote the
two-dimensional, complex vector space whose elements are of the form 
$\left(\begin{array}{c}
    \psi_{\vec{\omega}}\\
    \psi_{-\vec{\omega}}
  \end{array} \right)$, and let ${\cal V}_{\rm spin}$ denote the two-dimensional,
complex vector space whose elements are SU(2) spinors 
$\left(\begin{array}{c}
    \psi_{\uparrow}\\
    \psi_{\downarrow}
  \end{array} \right)$. We can think of the four-component object 
$\psi_{[\vec{\omega}]}$ defined in (\ref{eq43}) as being an element of 
${\cal V}_{\rm spin}\otimes{\cal V}_{[\vec{\om}]}$. Stressing analogies    to 1+1
dimensional, relativistic models (``dimensional reduction''), we also define
gamma matrices
\begin{eqnarray}
\gamma^{0} :=
1_{2}\otimes \sigma_{1} & , &
\gamma^{1} :=
1_{2}\otimes \sigma_{2} 
\end{eqnarray}
($\sigma_{1} = \left(\begin{array}{cc}
0 & 1 \\
1 & 0
\end{array} \right)$
and $\sigma_{2} = \left(\begin{array}{cc}
0 & -i \\
i & 0
\end{array} \right)$ are Pauli matrices). Then we see that
\begin{eqnarray}
  \bar{\psi}_{[\vec{\omega}]\uparrow}=\psi_{[\vec{\omega}]\uparrow}^*
  \sigma_1 & , & 
  \bar{\psi}_{[\vec{\omega}]\downarrow}=
  \psi_{[\vec{\omega}]\downarrow}^*\sigma_1
\end{eqnarray}
\be
  \bar{\psi}_{[\vec{\omega}]}=\psi_{[\vec{\omega}]}^*\gamma^0
\ee

The effective action at scale $j_{sc}$, simplified by omitting all subleading
terms of degree 2 (curvature of Fermi surface) and $\geq 4$ ($g^{(j_{sc})},\ 
g^{(j_{sc})}_{l>0}\approx 0 \ll -g_{l=0}^{(j_{sc})}$\ ,\ldots) is given by
\begin{eqnarray}
  \lefteqn{S_{\rm eff}(\bar{\psi},\psi) =
  \sum_{[\vec{\omega}]}
  \int[\bar{\psi}_{[\vec{\omega}]}
  (\gamma^{0}\partial_{t}-v_{F}\gamma^{1}\vec{\omega}\vec{\nabla})
  \psi_{[\vec{\omega}]}]d^{d+1}x}\nonumber\\
  && +\ \frac{g^{(j_{sc})}_0}{4\lambda_{j_{sc}}^{d-1}}
  \sum_{[\vec{\omega}],[\vec{\omega}']}
  \int [ \bar{\psi}_{[\vec{\omega}]}(\sigma_{1}\otimes 1_{2})
  \psi_{[\vec{\omega}]}  
  \bar{\psi}_{[\vec{\omega}']}(\sigma_{1}\otimes 1_{2})  
  \psi_{[\vec{\omega}']}\nonumber\\
  && +\ \bar{\psi}_{[\vec{\omega}]}(\sigma_{2}\otimes 1_{2})
  \psi_{[\vec{\omega}]}
  \bar{\psi}_{[\vec{\omega}']}(\sigma_{2}\otimes 1_{2})
  \psi_{[\vec{\omega}']} ]\label{action1}\,
  d^{d+1}x 
\end{eqnarray}
It only includes the s-wave BCS interactions and 
has the following basic features:
\renewcommand{\theenumi}{\arab{enumi})}
\begin{enumerate}
\item[\rm 1)] The action strongly resembles the one of a 1 + 1 dimensional,
  relativistic quantum field theory with $N = \frac{N^{(j_{sc})}}{2}$
  fermion flavours and  quartic self-interaction, such as the chiral
  Gross-Neveu model. Its infrared
  properties are closely related to those of 
  the chiral Gross-Neveu model. The perturbative infrared renormalization
  around the Fermi surface and the $\frac{1}{N}$-expansion are virtually
  identical.
\item[\rm 2)] The action, written as in (\ref{action1}), exhibits a manifest
  global
  U(1)-symmetry, given by
  \begin{eqnarray*}
  \psi_{[\vec{\omega}]} \rightarrow 
  e^{i \alpha (\sigma_{3} \otimes 1_{2})}\psi_{[\vec{\omega}]}\\
  \bar{\psi}_{[\vec{\omega}]} \rightarrow 
  \bar{\psi}_{[\vec{\omega}]}e^{-i \alpha (\sigma_{3} \otimes 1_{2})}
  \end{eqnarray*}
  with $\alpha\in [0,2\pi)$.
\end{enumerate}
The generator of this symmetry is the particle number operator $N$; it is the
usual gauge symmetry of the first kind. In the superconducting phase of the
system, this symmetry is spontaneously broken. We recall that the
{\em Mermin-Wagner theorem} states that, for a field theory model in $d+1=2$
dimensions with a continuous symmetry such as the chiral Gross-Neveu model,
the continuous symmetry {\em cannot} be broken spontaneously. In the chiral
Gross-Neveu model, continuous symmetry breaking only occurs in the
$N\rightarrow\infty$ limit. In spite of the formal similarities between
nonrelativistic many-body theory in $d\geq 2$ space dimensions, in the
Nambu-Gorkov approximation, and the chiral Gross-Neveu model in one space
dimension (``dimensional reduction''), the Mermin-Wagner theorem does
actually {\em not} apply to the former, and spontaneous breaking of the U(1)
gauge symmetry {\em is} possible in many-body theory (for $d=2$ at temperature
$T=0$, and for $d\geq 3$ at sufficiently small temperatures) !

By direct calculation we see that
\begin{eqnarray}
  \bar{\psi}_{[\vec{\omega}]}(\sigma_{1}\otimes 1_{2})
  \psi_{[\vec{\omega}]} & = &
  \lbrace  
  \psi_{-\vec{\omega}\uparrow}^{*}
  \psi_{\vec{\omega}\downarrow}^{*} +
  \psi_{\vec{\omega}\uparrow}^{*}
  \psi_{-\vec{\omega}\downarrow}^{*} +
  \psi_{-\vec{\omega}\downarrow}
  \psi_{\vec{\omega}\uparrow} +
  \psi_{\vec{\omega}\downarrow}
  \psi_{-\vec{\omega}\uparrow}
  \rbrace\nonumber\\
  & = & 2\{\phi^s_{\vec{\omega},-\vec{\omega},\downarrow,\uparrow}(x,x)
        +\phi^{s\ *}_{\vec{\omega},-\vec{\omega},\downarrow,\uparrow}(x,x)\}
  \label{BCSparm1}
\end{eqnarray}
and
\begin{eqnarray}
  \bar{\psi}_{[\vec{\omega}]}(\sigma_{2}\otimes 1_{2}) 
  \psi_{[\vec{\omega}]} & = & 
  - i \lbrace
  \psi_{-\vec{\omega}\uparrow}^{*} 
  \psi_{\vec{\omega}\downarrow}^{*} +
  \psi_{\vec{\omega}\uparrow}^{*}
  \psi_{-\vec{\omega}\downarrow}^{*} -
  \psi_{-\vec{\omega}\downarrow}
  \psi_{\vec{\omega}\uparrow} -
  \psi_{\vec{\omega}\downarrow}
  \psi_{-\vec{\omega}\uparrow}
  \rbrace\nonumber\\
  & = & 2i\{\phi^s_{\vec{\omega},-\vec{\omega},\downarrow,\uparrow}(x,x)
        -\phi^{s\ *}_{\vec{\omega},-\vec{\omega},\downarrow,\uparrow}(x,x)\}
  \label{BCSparm2}
\end{eqnarray}
where $\phi^s$ has been defined in (\ref{pairing}). Thus, these quadratic
expressions correspond to real and imaginary part of the BCS order parameter,
respectively.

The action $S_{\rm eff}(\bar{\psi},\psi)$ defined in (\ref{action1}) can be
replaced by a more convenient, {\em equivalent} action, $\tilde{S}$, that is
quadratic in $\bar{\psi}$ and $\psi$ and depends on a complex Lagrange
multiplier field  $\phi = \phi_{1} + i\phi_{2}$. The new action is given by
\begin{eqnarray}\label{action2}
  \lefteqn{\tilde{S}(\bar{\psi},\psi,\bar{\phi},\phi) =
  \sum_{[\vec{\omega}]} 
  \int[\bar{\psi}_{[\vec{\omega}]}
  (\gamma^{0}\partial_{t}-v_{F}\gamma^{1}\vec{\omega}\vec{\nabla})
  \psi_{[\vec{\omega}]}]d^{d+1}x}\nonumber\\
  && + g
  \sum_{[\vec{\omega}]}
  \int[\bar{\psi}_{[\vec{\omega}]}(\sigma_{1}\otimes 1_{2})
  \psi_{[\vec{\omega}]}
  \phi_{1} -
  \bar{\psi}_{[\vec{\omega}]}(\sigma_{2}\otimes 1_{2})
  \psi_{[\vec{\omega}]} \phi_{2}]
  d^{d+1}x  
  \\
  && +\frac{1}{2}\int(\phi_{1}^{2} + \phi_{2}^{2})d^{d+1}x \nonumber
\end{eqnarray}
where $2 g^{2} = - \frac{g^{(j_{sc})}_0}{\lambda_{j_{sc}}^{d-1}}>0$. 
We note that, under the U(1) symmetry discussed in remark 2) above, the field
$\phi$ transforms as $\phi\rightarrow e^{2i\alpha}\phi$.
We emphasize that $S_{\rm eff}(\bar{\psi},\psi)$ and
$\tilde{S}(\bar{\psi},\psi,\bar{\phi},\phi)$ are equivalent in terms of their
physical content: It is easily checked that, after functionally integrating
out the $\phi$-field,
\begin{eqnarray*}
  \int {\cal D}\phi_{1} {\cal D}\phi_{2} 
  e^{-\tilde{S}(\bar{\psi},\psi,\bar{\phi},\phi)} 
  = \mbox{const}\  e^{-S_{\rm eff}(\bar{\psi},\psi)}
\end{eqnarray*}
the original action $S_{\rm eff}(\bar{\psi},\psi)$ is restored in the
exponent. 
But $\tilde{S}(\bar{\psi},\psi,\bar{\phi},\phi)$ is
much easier to work with, because it is quadratic in the fields 
$\bar{\psi}$ and $\psi$. The Bose field $\phi$ will turn out to describe the
{\em Cooper pairs} of electrons.

The interaction between fermion- and Lagrange
multiplier fields is described by
\begin{equation}
g
\sum_{[\vec{\omega}]}
\int[\phi(x) \bar{\psi}_{[\vec{\omega}],\uparrow}
\psi_{[\vec{\omega}],\downarrow}
 +
\bar{\phi}(x) \bar{\psi}_{[\vec{\omega}],\downarrow}
\psi_{[\vec{\omega}],\uparrow}]\,
d^{d+1}x 
\end{equation}
This expression shows how interaction vertices between electrons (holes) and 
bosons $\phi$, $\bar{\phi}$ are
organized in the Nambu-Gorkov theory. One Nambu-Gorkov vertex is equivalent to
the following four vertices (each arrow points to a
$\psi_{\vec{\omega},\sigma}$):
\begin{equation}
\parbox[c]{300pt}{
\begin{picture}(300,80)(45,0)
\multiput(0,0)(80,0){4}{\Vertex(60,20){3}}
\multiput(0,0)(160,0){2}{\ArrowLine(30,20)(60,20)}
\multiput(0,0)(160,0){2}{\ArrowLine(90,20)(60,20)}
\multiput(0,0)(160,0){2}{\ArrowLine(140,20)(170,20)}
\multiput(0,0)(160,0){2}{\ArrowLine(140,20)(110,20)}
\multiput(0,0)(80,0){4}{\DashLine(60,55)(60,20){5}}
\multiput(60,10)(80,0){4}{g}
\multiput(65,50)(160,0){2}{$\bar{\phi}$}
\multiput(145,50)(160,0){2}{$\phi$}
\put(30,25){$\vec{\omega},\uparrow$}
\put(65,25){$-\vec{\omega},\downarrow$}
\put(110,25){$-\vec{\omega},\downarrow$}
\put(145,25){$\vec{\omega},\uparrow$}
\put(190,25){$\vec{\omega},\downarrow$}
\put(225,25){$-\vec{\omega},\uparrow$}
\put(270,25){$-\vec{\omega},\uparrow$}
\put(305,25){$\vec{\omega},\downarrow$}
\end{picture}}
\end{equation}
One can join only the first two vertices with each other and the second two
vertices with each other. There are two possibilities of joining the first two
vertices :
\begin{equation}
\parbox[c]{300pt}{
\begin{picture}(300,80)(45,0)
\multiput(0,0)(80,0){4}{\Vertex(60,20){3}}
\multiput(0,0)(160,0){1}{\ArrowLine(30,20)(60,20)}
\multiput(0,0)(160,0){1}{\ArrowLine(140,20)(170,20)}
\multiput(0,0)(160,0){1}{\ArrowLine(140,20)(60,20)}
\multiput(160,0)(160,0){1}{\ArrowLine(30,20)(60,20)}
\multiput(160,0)(160,0){1}{\ArrowLine(140,20)(170,20)}
\multiput(160,0)(160,0){1}{\ArrowLine(140,20)(60,20)}
\multiput(0,0)(80,0){4}{\DashLine(60,55)(60,20){5}}
\multiput(60,10)(80,0){4}{g}
\multiput(65,50)(160,0){2}{$\bar{\phi}$}
\multiput(145,50)(160,0){2}{$\phi$}
\put(30,25){$\vec{\omega},\uparrow$}
\put(85,25){$-\vec{\omega},\downarrow$}
\put(145,25){$\vec{\omega},\uparrow$}
\put(190,25){$-\vec{\omega},\downarrow$}
\put(245,25){$\vec{\omega},\uparrow$}
\put(305,25){$-\vec{\omega},\downarrow$}
\end{picture}}
\end{equation}
and two possibilities of joining the second two vertices, obtained by
exchanging $\uparrow$ and $\downarrow\,$.
Along a string of electron (hole)
propagator lines, the different sector labels $\vec{\omega}$, $-\vec{\omega}$
and boson fields $\phi$, $\bar{\phi}$ have to occur in an alternating pattern.
Because the first two and the second two vertices never intercombine, 
the spins can be omitted if a factor of 2 is attached to each loop. 

In the following, we focus our attention on the boson fields, and we will
attempt to eliminate the Nambu-Gorkov fermions. Because
$\tilde{S}(\bar{\psi},\psi,\bar{\phi},\phi)$ is quadratic in $\bar{\psi}$ and
$\psi$, one can perform the fermionic functional integration explicitly:
\begin{eqnarray}
  \lefteqn{\int {\cal D}\bar{\psi} {\cal D}\psi\,
  e^{-\tilde{S}(\bar{\psi},\psi,\bar{\phi},\phi)} =
  \exp\left(-\frac{1}{2}\int d^{d+1}x\, |\phi(x)|^{2}\right) }\cdot
  \nonumber\\
  && \cdot\ \det[(\gamma^{0}\partial_{t}-v_{F}\gamma^{1}\vec{\omega}
  \vec{\nabla}) +
  g
  \left(\begin{array}{cc}
  0 & \phi(x) \\
  \bar{\phi}(x) & 0
  \end{array} \right)\otimes 1_{2}]
\end{eqnarray}
After normalizing  the right hand side by division
through $\det[\gamma^{0}\partial_{t}-v_{F}\gamma^{1}\vec{\omega}\vec{\nabla}]$,
the determinant
\begin{equation}
  \det[1 +
  \frac{1}{\gamma^{0}\partial_{t}-v_{F}\gamma^{1}\vec{\omega}\vec{\nabla}}
  \lbrace g
  \left(\begin{array}{cc}
  0 & \phi(x) \\
  \bar{\phi}(x) & 0 
  \end{array} \right)\otimes 1_{2}\rbrace]
\end{equation}
can be evaluated using the identity
\begin{equation}\label{opid}
  \det(1 + A) = \exp[\sum_{n=1}^{\infty}
  \frac{(-1)^{n+1}}{n}{\rm Tr}(A^{n})] \label{eq:det}
\end{equation}
The term $\mbox{Tr}\,(A^n)$ is the amplitude of the n-th
order one-loop diagram
\begin{equation}
\parbox[c]{300pt}{
\begin{picture}(300,120)(55,-20)
\multiput(150,0)(0,0){1}{
\begin{picture}(300,120)(0,0)
\multiput(0,0)(230,0){1}{\CArc(50,40)(35,0,360)}
\multiput(0,0)(230,0){1}{\Vertex(15,40){2}}
\multiput(0,0)(230,0){1}{\Vertex(85,40){2}}
\multiput(0,0)(230,0){1}{\Vertex(67.5,70.31){2}}
\multiput(0,0)(230,0){1}{\Vertex(67.5,9.69){2}}
\multiput(0,0)(230,0){1}{\Vertex(32.5,9.69){2}}
\multiput(0,0)(230,0){1}{\Vertex(32.5,70.31){2}}
\DashLine(67.5,70.31)(72.5,78.97){2}
\DashLine(15,40)(5,40){2}
\DashLine(85,40)(95,40){2}
\DashLine(32.5,70.31)(27.5,78.97){2}
\DashLine(32.5,9.69)(27.5,1.03){2}
\DashLine(67.5,9.69)(72.5,1.03){2}
\end{picture}}
\end{picture}}
\end{equation}
with a factor of 
$\lbrace g
\left(\begin{array}{cc}
0 & \phi(x) \\
\bar{\phi}(x) & 0
\end{array} \right)\otimes 1_{2}\rbrace$
on each external line and a Nambu-Gorkov propagator on each segment of the
loop. From our discussion of the possible pairings of Nambu-Gorkov
vertices we conclude that loops with an odd number of vertices vanish.
The expression in the exponent on the right side of (\ref{opid}) reduces to
$- \frac{1}{2}\sum_{n=1}^{\infty}
\frac{1}{n}Tr(A^{2n})$. Calculating $\mbox{Tr}\,(A^{2n})$ in terms of the 
original electron (hole) propagators, the spin summation can be absorbed in
an overall factor of 2 cancelling the $\frac{1}{2}$ in the
exponent. For a fixed  $[\vec{\omega}]$, a 2n-loop looks like
\begin{equation}\label{loopg1}
\parbox[c]{300pt}{
\begin{picture}(300,120)(55,-20)
\multiput(150,0)(0,0){1}{
\begin{picture}(300,120)(0,0)
\multiput(0,0)(230,0){1}{\ArrowArcn(50,40)(17,90,-270)}
\multiput(0,0)(230,0){1}{\Text(48,40)[l]{k}}
\multiput(0,0)(230,0){1}{\CArc(50,40)(35,0,360)}
\multiput(0,0)(230,0){1}{\Vertex(15,40){3}}
\multiput(0,0)(230,0){1}{\Vertex(85,40){3}}
\multiput(0,0)(230,0){1}{\Vertex(67.5,70.31){3}}
\multiput(0,0)(230,0){1}{\Vertex(67.5,9.69){3}}
\multiput(0,0)(230,0){1}{\Vertex(32.5,9.69){3}}
\multiput(0,0)(230,0){1}{\Vertex(32.5,70.31){3}}
\DashLine(67.5,70.31)(72.5,78.97){2}
\DashLine(15,40)(5,40){2}
\DashLine(85,40)(95,40){2}
\DashLine(32.5,70.31)(27.5,78.97){2}
\DashLine(32.5,9.69)(27.5,1.03){2}
\DashLine(67.5,9.69)(72.5,1.03){2}
\Text(78,82)[l]{$\bar{\phi}(p_{2n})$}
\Text(101,38)[l]{$\phi(p_{1})$}
\Text(78,1)[l]{$\bar{\phi}(p_{2})$}
\multiput(0,0)(230,0){1}{\Text(87,60)[l]{$\vec{\omega}$}}
\multiput(0,0)(230,0){1}{\Text(82,22)[l]{$-\vec{\omega}$}}
\multiput(0,0)(230,0){1}{\Text(44,83)[l]{$-\vec{\omega}$}}
\multiput(0,0)(230,0){1}{\Text(58,62)[l]{g}}
\multiput(0,0)(230,0){1}{\Text(58,17)[l]{g}}
\multiput(0,0)(230,0){1}{\Text(75,38)[l]{g}}
\end{picture}}
\end{picture}}
\end{equation}
inserting alternatingly  electron (hole) sector labels $\vec{\omega}$, $-
\vec{\omega}$ and boson fields $\phi$ and $\bar{\phi}$ along the loop line.
In order to compute the full amplitude of
such a graph in momentum space, we have to integrate over the loop momentum
$(k_{0},\vec{k})$ for arbitrary external momenta $(p_{j,0},\vec{p}_{j})$,
$j=1,...,2n$, which is a difficult task. 

It will turn out to be useful to expand the amplitude corresponding to
(\ref{loopg1}) into a sum of terms that look somewhat more
manageable: Anticipating spontaneous symmetry breaking, we assume that, in a
(superconducting, extremal) ground state of the system, the Bose field $\phi$
has a {\em non-zero} expectation value $\phi_c$. The modulus of $|\phi_c|$
is determined by the values of physical parameters (the density of the
system, the  strength of $g_0^{(j_{sc})}$, etc.), while the {\em phase} 
of $\phi_c$,
an angle in $[0,2\pi)$, is only fixed after suitable symmetry breaking
{\em boundary conditions} have been imposed --- as usual in the study of 
systems
with spontaneously broken symmetries. We thus decompose the Bose field $\phi$
into a constant part, $\phi_c$, and a fluctuation field, $\chi(x)$:
\begin{eqnarray}
  \phi(x) & = & \phi_{c} + \chi(x) \nonumber\\
  \bar{\phi}(x) & = & \bar{\phi}_{c} + \bar{\chi}(x)
\end{eqnarray}
The field $\chi(x)$ describes small fluctuations of the 
Cooper-pair condensate
around $\phi_c$. The decomposition  of $\phi(x)$ induces a
decomposition of the amplitude corresponding to (\ref{loopg1}) into a
sum of monomials in $\chi$ and
$\bar{\chi}$. For each fixed n and
$[\vec{\omega}]$, this decomposition, a binomial series
in $\bar{\chi}(p)$ and  $\chi(p)$, can be described pictorially, as follows:
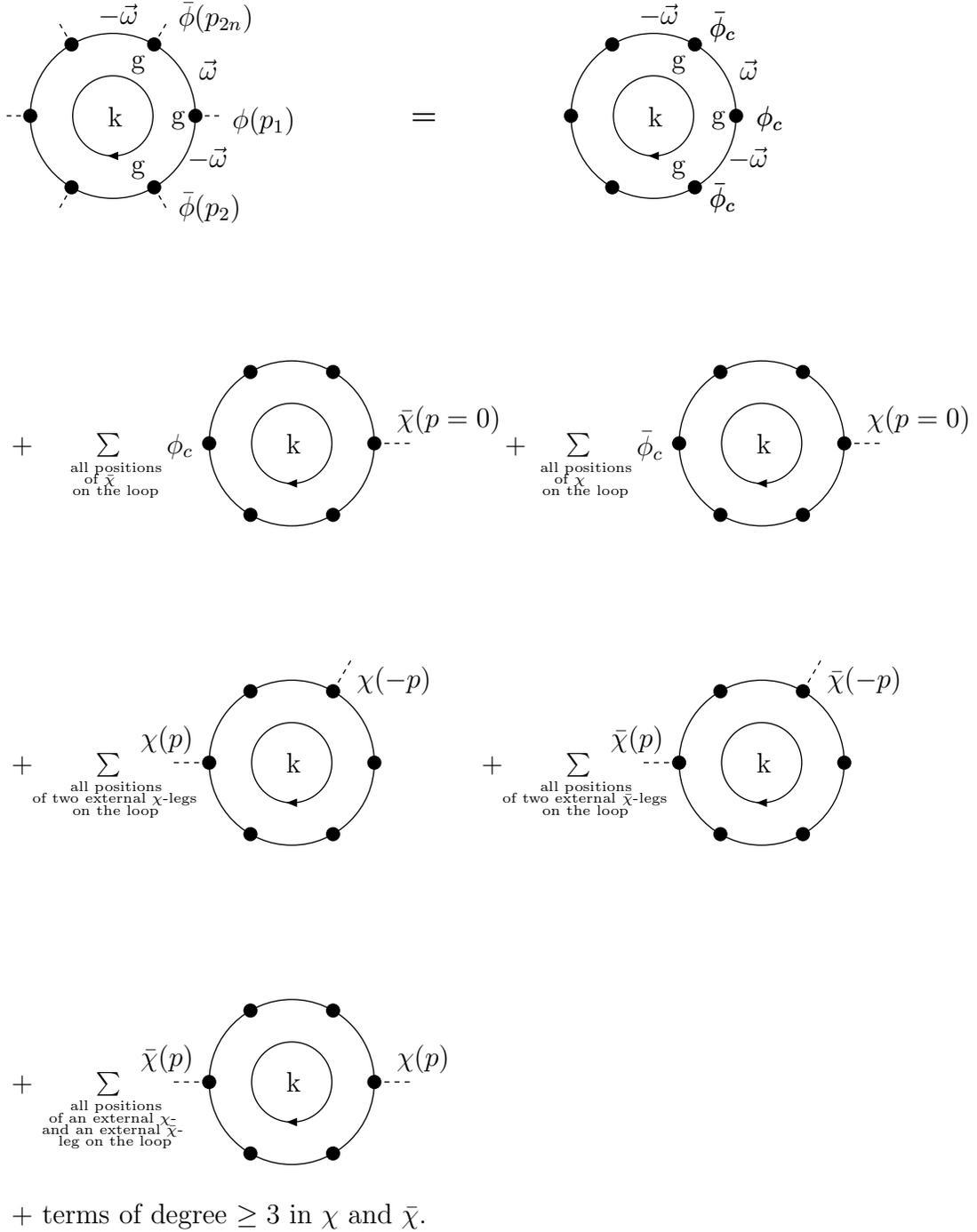
\[\parbox[c]{300pt}{
\begin{picture}(300,120)(80,0)
\multiput(20,0)(0,0){1}{
\begin{picture}(300,120)(0,0)
\multiput(0,0)(230,0){2}{\ArrowArcn(50,40)(17,90,-270)}
\multiput(0,0)(230,0){2}{\Text(48,40)[l]{k}}
\multiput(0,0)(230,0){2}{\CArc(50,40)(35,0,360)}
\multiput(0,0)(230,0){2}{\Vertex(15,40){3}}
\multiput(0,0)(230,0){2}{\Vertex(85,40){3}}
\multiput(0,0)(230,0){2}{\Vertex(67.5,70.31){3}}
\multiput(0,0)(230,0){2}{\Vertex(67.5,9.69){3}}
\multiput(0,0)(230,0){2}{\Vertex(32.5,9.69){3}}
\multiput(0,0)(230,0){2}{\Vertex(32.5,70.31){3}}
\DashLine(67.5,70.31)(72.5,78.97){2}
\DashLine(15,40)(5,40){2}
\DashLine(85,40)(95,40){2}
\DashLine(32.5,70.31)(27.5,78.97){2}
\DashLine(32.5,9.69)(27.5,1.03){2}
\DashLine(67.5,9.69)(72.5,1.03){2}
\Text(78,82)[l]{$\bar{\phi}(p_{2n})$}
\Text(101,38)[l]{$\phi(p_1)$}
\Text(78,1)[l]{$\bar{\phi}(p_2)$}
\multiput(0,0)(230,0){2}{\Text(87,60)[l]{$\vec{\omega}$}}
\multiput(0,0)(230,0){2}{\Text(82,22)[l]{$-\vec{\omega}$}}
\multiput(0,0)(230,0){2}{\Text(44,83)[l]{$-\vec{\omega}$}}
\multiput(0,0)(230,0){2}{\Text(58,62)[l]{g}}
\multiput(0,0)(230,0){2}{\Text(58,17)[l]{g}}
\multiput(0,0)(230,0){2}{\Text(75,38)[l]{g}}
\multiput(230,0)(0,0){2}{
\Text(74,78)[l]{$\bar{\phi}_{c}$}
\Text(94,39)[l]{$\phi_{c}$}
\Text(74,5)[l]{$\bar{\phi}_{c}$}}
\end{picture} }
\multiput(200,16)(0,0){2}{\put(1,20){\large $=$}}
\end{picture}}
\]
\[\parbox[c]{300pt}{
\begin{picture}(300,130)(80,0)
\multiput(100,0)(200,0){2}{\ArrowArcn(50,40)(17,90,-270)}
\multiput(100,0)(200,0){2}{\Text(48,40)[l]{k}}
\multiput(100,0)(200,0){2}{\CArc(50,40)(35,0,360)}
\multiput(100,0)(200,0){2}{\Vertex(15,40){3}}
\multiput(100,0)(200,0){2}{\Vertex(85,40){3}}
\multiput(100,0)(200,0){2}{\Vertex(67.5,70.31){3}}
\multiput(100,0)(200,0){2}{\Vertex(67.5,9.69){3}}
\multiput(100,0)(200,0){2}{\Vertex(32.5,9.69){3}}
\multiput(100,0)(200,0){2}{\Vertex(32.5,70.31){3}}
\multiput(100,0)(200,0){2}{\DashLine(85,40)(100,40){2}}
\Text(297,39)[l]{$\bar{\phi}_{c}$}
\Text(97,39)[l]{$\phi_{c}$}
\Text(195,50)[l]{$\bar{\chi}(p=0)$}
\Text(395,50)[l]{$\chi(p=0)$}
\multiput(45,16)(200,0){2}{
\begin{picture}(50,30)(0,0)
\put(18,19){{\Large $\sum$}}
\put(7,12){{\tiny all positions}}
\put(8,2){{\tiny on the loop}}
\end{picture} }
\multiput(45,16)(200,0){1}{
\put(17,7){{\tiny of $\bar{\chi}$}}
\put(217,7){{\tiny of $\chi$}} }
\multiput(30,16)(210,0){2}{\put(1,20){$+$}}
\end{picture}}
\]
\[\parbox[c]{300pt}{
\begin{picture}(300,130)(80,0)
\multiput(100,0)(200,0){2}{\ArrowArcn(50,40)(17,90,-270)}
\multiput(100,0)(200,0){2}{\Text(48,40)[l]{k}}
\multiput(100,0)(200,0){2}{\CArc(50,40)(35,0,360)}
\multiput(100,0)(200,0){2}{\Vertex(15,40){3}}
\multiput(100,0)(200,0){2}{\Vertex(85,40){3}}
\multiput(100,0)(200,0){2}{\Vertex(67.5,70.31){3}}
\multiput(100,0)(200,0){2}{\Vertex(67.5,9.69){3}}
\multiput(100,0)(200,0){2}{\Vertex(32.5,9.69){3}}
\multiput(100,0)(200,0){2}{\Vertex(32.5,70.31){3}}
\multiput(100,0)(200,0){2}{\DashLine(67.5,70.31)(75,83.3){2}}
\multiput(100,0)(200,0){2}{\DashLine(0,40)(15,40){2}}
\Text(86,50)[l]{$\chi(p)$}
\Text(178,75)[l]{$\chi(-p)$}
\Text(286,50)[l]{$\bar{\chi}(p)$}
\Text(378,75)[l]{$\bar{\chi}(-p)$}
\multiput(45,16)(0,0){1}{
\begin{picture}(50,30)(0,0)
\put(18,19){{\Large $\sum$}}
\put(7,12){{\tiny all positions}}
\put(-9,7){{\tiny of two external $\chi$-legs}}
\put(8,2){{\tiny on the loop}}
\end{picture} }
\multiput(245,16)(0,0){1}{
\begin{picture}(50,30)(0,0)
\put(18,19){{\Large $\sum$}}
\put(7,12){{\tiny all positions}}
\put(-9,7){{\tiny of two external $\bar{\chi}$-legs}}
\put(8,2){{\tiny on the loop}}
\end{picture} }
\multiput(30,16)(200,0){2}{\put(1,20){$+$}}
\end{picture}}
\]
\[\parbox[c]{300pt}{
\begin{picture}(300,130)(80,0)
\multiput(100,0)(200,0){1}{\ArrowArcn(50,40)(17,90,-270)}
\multiput(100,0)(200,0){1}{\Text(48,40)[l]{k}}
\multiput(100,0)(200,0){1}{\CArc(50,40)(35,0,360)}
\multiput(100,0)(200,0){1}{\Vertex(15,40){3}}
\multiput(100,0)(200,0){1}{\Vertex(85,40){3}}
\multiput(100,0)(200,0){1}{\Vertex(67.5,70.31){3}}
\multiput(100,0)(200,0){1}{\Vertex(67.5,9.69){3}}
\multiput(100,0)(200,0){1}{\Vertex(32.5,9.69){3}}
\multiput(100,0)(200,0){1}{\Vertex(32.5,70.31){3}}
\multiput(100,0)(200,0){1}{\DashLine(85,40)(100,40){2}}
\multiput(100,0)(200,0){1}{\DashLine(0,40)(15,40){2}}
\Text(86,50)[l]{$\bar{\chi}(p)$}
\Text(195,50)[l]{$\chi(p)$}
\multiput(45,16)(200,0){1}{
\begin{picture}(50,30)(0,0)
\put(18,19){{\Large $\sum$}}
\put(7,12){{\tiny all positions}}
\put(-2,7){{\tiny of an external $\chi$-}}
\put(-5,2){{\tiny and an external $\bar{\chi}$-}}
\put(2,-3){{\tiny leg on the loop}}
\end{picture} }
\multiput(30,16)(200,0){1}{\put(1,20){$+$}}
\end{picture}}
\]
\[\parbox[c]{300pt}{
\begin{picture}(300,50)(80,0)
\multiput(30,16)(0,0){1}{\put(1,20){$+$ terms of degree $\geq 3$ in $\chi$
and $\bar{\chi}$.}}
\end{picture}}\]
Each dot without an external leg stands for a factor
$g\phi_{c}(2\pi)^{d+1}\delta(p)$ or for its
complex conjugate. The external momenta flowing into the diagram at all such
dots {\em vanish}, because $\phi_c$ is constant in $x$-space.
The graphs
with one external leg are summed over all $2n$ possible positions of either
$\chi(0)$ or $\bar{\chi}(0)$
on the loop. The corresponding amplitudes (which are linear in $\bar{\chi}$
or $\chi$) vanish if $\phi_c$ is the expectation value of $\phi$ in the
ground state of the system. 
The effective action for the field $\chi(p)$ is thus given by a power series
in $\bar{\chi}$ and $\chi$ :
\be\label{action3}
  S(\bar{\chi},\chi) = S^{(0)}+
  S^{(1)}(\bar{\chi},\chi)+
  S^{(2)}(\bar{\chi},\chi)+ ...
\ee
where $S^{(r)}$, $r=0,1,2,...$, is a sum of monomials of degree r in 
$\bar{\chi}$ and
$\chi$. We remark that the first three terms on the right side of
(\ref{action3}) also contain the contribution 
\begin{eqnarray}
  \lefteqn{\frac{1}{2} \int \dbar\,^{d+1}p\,(\bar{\phi}_{c}\phi_{c}
  ((2\pi)^{d+1}\delta(p))^2}\nonumber\\ 
  && +\phi_{c}\bar{\chi}(p)(2\pi)^{d+1}\delta(p) + 
     \bar{\phi}_{c}\chi(p)(2\pi)^{d+1}\delta(p) + 
     \bar{\chi}(p)\chi(p))
\end{eqnarray}
from the part of the original action $\tilde{S}$ ( defined in (\ref{action2})
) quadratic in $\bar{\phi}$ and $\phi$.

After dividing by the total volume of the system, i.e., by
$(2\pi)^{d+1}\delta(0)$, the amplitude of the $2n$-th order loop without any
external $\bar{\chi}$- and $\chi$-legs, and for a fixed ray $[\vec{\om}]$,
\begin{equation}
\parbox[c]{300pt}{
\begin{picture}(300,100)(65,0)
\multiput(80,0)(0,0){1}{
\begin{picture}(300,120)(10,0)
\put(150,50){\circle{40}}
\put(130,50){\circle*{5}}
\put(170,50){\circle*{5}}
\put(160,67.32){\circle*{5}}
\put(140,32.68){\circle*{5}}
\put(140,67.32){\circle*{5}}
\put(160,32.68){\circle*{5}}
\put(165,67.32){$g \phi_{c}$}
\put(175,50){$g \bar{\phi}_{c}$}
\put(165,30.68){$g \phi_{c}$}
\ArrowArcn(150,50)(12,90,-270)
\put(148,47){k}
\end{picture} }
\end{picture}}
\end{equation}
is given by
\begin{equation}
2\int_{\extraR\times \tilde{B}_{\vec{\om}}} \dbar\,^{d+1}k\,
(\frac{-g^{2}|\phi_{c}|^{2}}{k_{0}^{2} +
v_{F}^{2}(\vec{\omega}\vec{k})^{2}})^{n}
\end{equation}
Summing over all orders $2n$ ( see (\ref{opid}) ) and all rays
$[\vec{\omega}]$ , our result for $S^{(0)}$ is found to be
\begin{eqnarray}
  S^{(0)} & = &
  \frac{1}{2}\int_{\extraR\times \tilde{B}_{\vec{\om}}} \dbar\,^{d+1}k\,
  |\phi_{c}|^{2} ((2\pi)\delta(k))^2\nonumber\\
  && +2\sum_{[\vec{\omega}]}
  \int_{\extraR\times \tilde{B}_{\vec{\om}}} \dbar\,^{d+1}k\,  
  \sum_{n=1}^{\infty}
  \frac{1}{n}
  \left(\frac{-g^{2}| \phi_{c} |^{2}}{k_{0}^{2} +
  v_{F}^{2}(\vec{\omega}\vec{k})^{2}}\right)^{n}(2\pi)^{d+1}\delta(p=0)
  \\
  & = & (2\pi)^{d+1}\delta(p=0)\left(\frac{1}{2}|\phi_{c}|^{2} -
  2\sum_{[\vec{\omega}]}\int_{\extraR\times \tilde{B}_{\vec{\om}}}
  \dbar\,^{d+1}k\,  
  \log\left(1 +
  \frac{g^{2}| \phi_{c} |^{2}}{k_{0}^{2} +
  v_{F}^{2}(\vec{\omega}\vec{k})^{2}}\right) \right)\hspace{10mm}\nonumber
\end{eqnarray}
The effective potential, $U_{\rm eff}(\bar{\phi_c},\phi_c)$, is defined as
the density of
$S^{(0)}$, i.e., as $S^{(0)}$ divided by the total volume of the system. Thus 
\begin{equation}\label{effpot}
  U_{\rm eff}(\phi_{c}) =
  \frac{1}{2} 
  | \phi_{c} |^{2} -
  2\sum_{[\vec{\omega}]}
  \int_{\extraR\times \tilde{B}_{\vec{\om}}} 
  \dbar\,^{d+1}k\,
  \log\left(1 +
  \frac{g^{2}| \phi_{c} |^{2}}{k_{0}^{2} +
  v_{F}^{2}(\vec{\omega}\vec{k})^{2}}\right) 
\end{equation}

In the analysis of spontaneous symmetry breaking, the effective potential plays
an important r\^{o}le. In the approximation of mean field theory, the
expectation value of the field $\phi$ in an arbitrary, extremal ground state of
the system is  given by a minimum of $U_{\rm eff}(\bar{\phi_c},\phi_c)$. Thanks
to the minus sign in front of the integral on the right side of
(\ref{effpot}), the graph of the effective potential has the shape of a
Mexican hat. The minima of  $U_{\rm eff}$ are obtained by setting the
derivative of $U_{\rm eff}$ with respect to $\mid\phi_{c}\mid^2$ to zero. The
result is that (for small values of $g^2 k_F^{d-1}$)
\be\label{gsphi}
\mid\phi_{c}\mid\approx \frac{k_{F} v_F}{g}
\exp(-\frac{\pi v_F}{g^2 (k_F\lambda_{j_{sc}})^{d-1}})
\ee
where $v_F$ is dimensionless (in our units), $k_F$ has a dimension of
inverse length, and $g$ has a dimension of $(\mbox{length})^{\frac{d-1}{2}}$,
so that the dimension of $|\phi_c|$ is that of 
$(\mbox{length})^{-\frac{d+1}{2}}$,
as it should be in view of the last term on the right side of
(\ref{action2}).

We recall that $2g^2 = -\frac{g_0^{(j_{sc})}}{\lambda_{j_{sc}}^{d-1}}$ (see
(\ref{action2})) and hence
\be
\mid\phi_{c}\mid\approx \frac{\sqrt{2}k_{F} v_F
\lambda_{j_{sc}}^{\frac{d-1}{2}}} 
{\sqrt{-g_0^{j_{sc}}}}
\exp(-\frac{2 \pi v_F}{g_0^{(j_{sc})} k_F^{d-1}})
\ee
for small values of $|g_0^{(j_{sc})}|
\left(\frac{k_F}{\lambda_{j_{sc}}}\right)^{d-1}$.

At values of $\phi_c$ minimizing $U_{\rm eff}(\bar{\phi_c},\phi_c)$, the terms
in $S(\bar{\chi},\chi)$ {\em linear} in $\bar{\chi}$ or $\chi$ must 
vanish,
i.e. $S^{(1)}=0$. (The equations $\frac{\partial
U_{\rm eff}}{\partial(\mid\phi_c\mid^2)}=0$ and $S^{(1)}=0$ are, of course,
equivalent; the solution is given by (\ref{gsphi}).) From now on, $\phi_c$
will denote a minimum of  $U_{\rm eff}$.

The term $S^{(2)}(\bar{\chi},\chi)$ consists of three
different contributions, proportional to $\chi^2$, $\bar{\chi}^2$ and
$\bar{\chi}\chi$, respectively. They can be found by calculating the
amplitudes corresponding to the following sums of diagrams: 
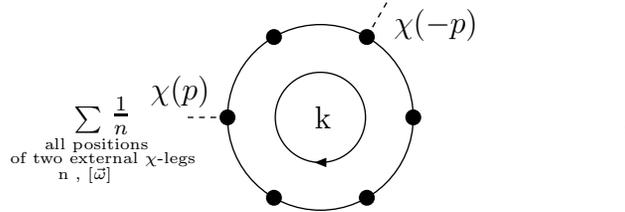
\[
\parbox[c]{300pt}{\begin{picture}(300,130)(55,-20)
\multiput(100,0)(0,0){1}{
\begin{picture}(300,130)(55,-20)
\multiput(100,0)(200,0){1}{\ArrowArcn(50,40)(17,90,-270)}
\multiput(100,0)(200,0){1}{\Text(48,40)[l]{k}}
\multiput(100,0)(200,0){1}{\CArc(50,40)(35,0,360)}
\multiput(100,0)(200,0){1}{\Vertex(15,40){3}}
\multiput(100,0)(200,0){1}{\Vertex(85,40){3}}
\multiput(100,0)(200,0){1}{\Vertex(67.5,70.31){3}}
\multiput(100,0)(200,0){1}{\Vertex(67.5,9.69){3}}
\multiput(100,0)(200,0){1}{\Vertex(32.5,9.69){3}}
\multiput(100,0)(200,0){1}{\Vertex(32.5,70.31){3}}
\multiput(100,0)(200,0){1}{\DashLine(67.5,70.31)(75,83.3){2}}
\multiput(100,0)(200,0){1}{\DashLine(0,40)(15,40){2}}
\Text(86,50)[l]{$\chi(p)$}
\Text(178,75)[l]{$\chi(-p)$}
\multiput(35,16)(200,0){1}{
\begin{picture}(50,30)(0,0)
\put(18,19){{\Large $\sum$}}
\put(7,12){{\tiny all positions}}
\put(-6,7){{\tiny of two external $\chi$-legs}}
\put(12,1){{\tiny n , $[\vec{\omega}]$}}
\put(32,22){$\frac{1}{n}$}
\end{picture} }
\put(265,36){,}
\end{picture}}
\end{picture}}
\]
\[
\parbox[c]{300pt}
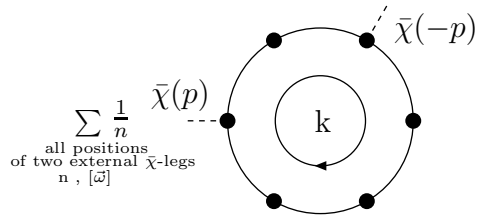
{\begin{picture}(300,130)(55,-20)
\multiput(100,0)(0,0){1}{
\begin{picture}(300,130)(55,-20)
\multiput(100,0)(200,0){1}{\ArrowArcn(50,40)(17,90,-270)}
\multiput(100,0)(200,0){1}{\Text(48,40)[l]{k}}
\multiput(100,0)(200,0){1}{\CArc(50,40)(35,0,360)}
\multiput(100,0)(200,0){1}{\Vertex(15,40){3}}
\multiput(100,0)(200,0){1}{\Vertex(85,40){3}}
\multiput(100,0)(200,0){1}{\Vertex(67.5,70.31){3}}
\multiput(100,0)(200,0){1}{\Vertex(67.5,9.69){3}}
\multiput(100,0)(200,0){1}{\Vertex(32.5,9.69){3}}
\multiput(100,0)(200,0){1}{\Vertex(32.5,70.31){3}}
\multiput(100,0)(200,0){1}{\DashLine(67.5,70.31)(75,83.3){2}}
\multiput(100,0)(200,0){1}{\DashLine(0,40)(15,40){2}}
\Text(86,50)[l]{$\bar{\chi}(p)$}
\Text(178,75)[l]{$\bar{\chi}(-p)$}
\multiput(35,16)(200,0){1}{
\begin{picture}(50,30)(0,0)
\put(18,19){{\Large $\sum$}}
\put(7,12){{\tiny all positions}}
\put(-6,7){{\tiny of two external $\bar{\chi}$-legs}}
\put(12,1){{\tiny n , $[\vec{\omega}]$}}
\put(32,22){$\frac{1}{n}$}
\end{picture} }
\end{picture}}
\end{picture}}
\]
and
\[
\parbox[c]{300pt}
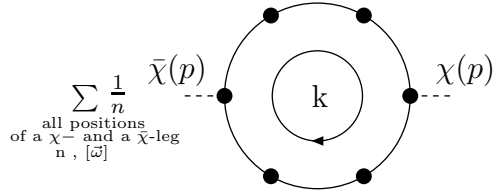
{\begin{picture}(300,130)(55,-20)
\multiput(100,0)(0,0){1}{
\begin{picture}(300,130)(55,-20)
\multiput(100,0)(200,0){1}{\ArrowArcn(50,40)(17,90,-270)}
\multiput(100,0)(200,0){1}{\Text(48,40)[l]{k}}
\multiput(100,0)(200,0){1}{\CArc(50,40)(35,0,360)}
\multiput(100,0)(200,0){1}{\Vertex(15,40){3}}
\multiput(100,0)(200,0){1}{\Vertex(85,40){3}}
\multiput(100,0)(200,0){1}{\Vertex(67.5,70.31){3}}
\multiput(100,0)(200,0){1}{\Vertex(67.5,9.69){3}}
\multiput(100,0)(200,0){1}{\Vertex(32.5,9.69){3}}
\multiput(100,0)(200,0){1}{\Vertex(32.5,70.31){3}}
\multiput(100,0)(200,0){1}{\DashLine(85,40)(100,40){2}}
\multiput(100,0)(200,0){1}{\DashLine(0,40)(15,40){2}}
\Text(86,50)[l]{$\bar{\chi}(p)$}
\Text(195,50)[l]{$\chi(p)$}
\multiput(35,16)(200,0){1}{
\begin{picture}(50,30)(0,0)
\put(18,19){{\Large $\sum$}}
\put(7,12){{\tiny all positions}}
\put(-5,7){{\tiny of a $\chi-$ and a $\bar{\chi}$-leg}}
\put(12,1){{\tiny n , $[\vec{\omega}]$}}
\put(32,22){$\frac{1}{n}$}
\end{picture} }
\end{picture}}
\end{picture}}
\]
We are interested in calculating the Taylor series expansions of the
amplitudes corresponding to these sums of diagrams in $p$ around $p=0$. The
terms constant in $p$ yield the coefficients of the ``mass terms'' of the
fluctuation field $\chi$. These coefficients are found to be given by
\begin{equation}
  2g^{4} \bar{\phi}_{c}^{2}  
  \sum_{[\vec{\omega}]} \int_{\extraR\times\tilde{B}_{\vec{\om}}}
  \dbar\,^{d+1}k\,
  \frac{1}
  {(k_{0}^2 +v_{F}^{2}(\vec{\omega}\cdot\vec{k})^{2}+g^{2}|\phi_{c}|^{2})^{2}}
\end{equation}
(coefficient of term proportional to $\chi^2$), the complex conjugate of
this expression (coefficient of term proportional to $\bar{\chi}^2$), and by 
\be
\frac{1}{2} - 2g^{2}\sum_{[\vec{\omega}]}
\int\dbar\,^{d+1}k\,
\frac{k_{0}^2 +v_{F}^{2}(\vec{\omega}\vec{k})^{2}}
{(k_{0}^2 +v_{F}^{2}(\vec{\omega}\cdot\vec{k})^{2} + g^{2}|\phi_{c}|^{2})^{2}}
\ee
(coefficient of term proportional to $\bar{\chi}\chi$), where the
``$\frac{1}{2}$'' comes from the term 
$\frac{1}{2}\int d^{d+1}x\,\bar{\chi}\chi$
in the original action $\tilde{S}$ (see (\ref{action2})).

It is convenient to introduce polar coordinates in field space by setting
\be
\phi(x) = | \phi_{c} | e^{i\theta} + (\chi_{t}(x)+i\chi_{l}(x))
e^{i\theta}
\ee
where the phase $\theta$ of the ground state expectation value of $\phi(x)$,
$\approx\phi_{c}$, is fixed by the boundary conditions imposed 
on the system. The component $\chi_{t}$ of $\chi$ is parallel to
$\phi_{c}$, i.e., it is {\em transversal} to the manifold of minima of
$U_{\rm eff}$, a circle, at $\phi_{c}$. The component $\chi_{l}$ is
perpendicular to $\phi_{c}$, hence {\em tangential} to the manifold of minima 
of $U_{\rm eff}$ at $\phi_{c}$. 

A straightforward calculation (involving (\ref{gsphi}), i.e., the fact that
$\phi_c$ minimizes $U_{\rm eff}$) now shows that the mass terms for the field
$\chi$ combine to 
\be
\frac{M^2}{2}\int d^{d+1}x\,\chi_t^2
\ee
where 
\be
  M^2 = 16
  g^{4} |\phi_{c}|^{2}  
  \sum_{[\vec{\omega}]} \int_{\extraR\times\tilde{B}_{\vec{\om}}}
  \dbar\,^{d+1}k\,
  \frac{1}
  {(k_{0}^2 +v_{F}^{2}(\vec{\omega}\cdot\vec{k})^{2}+g^{2}|\phi_{c}|^{2})^{2}}
\ee
In accordance with the Goldstone theorem, the effective field theory for
$\chi$ contains a massless field $\chi_l$ describing the Goldstone bosons and
a seemingly massive field $\chi_t$ of mass $M$, where $M$ is equal to the
square root of the curvature of $U_{\rm eff}$ in the radial direction
(parallel to $\phi_c$) at $\phi_c$. Actually, it will turn out that the degree
3 terms, $S^{(3)}$, of the effective action of $\chi$ couple $\chi_t$ to the
second power of $\chi_l$. Thus the field quanta of $\chi_t$ are resonances
that decay into pairs of Goldstone bosons.

We note that {\em all} terms in the effective action for the field $\chi$ 
that are
local and do not contain any derivatives of $\chi$ and $\bar{\chi}$ can be
determined from $U_{\rm eff}(\bar{\phi_c}+\bar{\chi},\phi_c+\chi)$, with
$\chi(x)=\mbox{const}$, by expanding in powers of $\chi$ and $\bar{\chi}$.

We can also compute the kinetic terms of the effective
action of $\chi$. Let us denote their inverse propagators by 
$\Pi_{\chi,\chi}(p)$,
$\Pi_{\bar{\chi}, \bar{\chi}}(p)$, $\Pi_{\bar{\chi},\chi}(p)$ and
$\Pi_{\chi,\bar{\chi}}(p)$,  where p is the external momentum, and the indices
stand for the external legs. For example, the coefficients,
$\alpha_{\bar{\chi},\chi}$ and  $\beta_{\bar{\chi},\chi}$, of the contribution
$\int\dbar^{d+1}p\  \bar{\chi}(p)(\alpha_{\bar{\chi},\chi}p_{0}^2
+ \beta_{\bar{\chi},\chi}| \vec{p} |^2) \chi(p)$ to the effective action are
given by
\begin{eqnarray}
  \alpha_{\bar{\chi},\chi} & = & \frac{1}{2}\frac{\partial^{2}}{\partial
  p_{0}^2}\mid_{p=0} \Pi_{\bar{\chi},
  \chi}(p)\\
  \beta_{\bar{\chi},\chi} & = & \frac{1}{2}\frac{\partial^{2}}{\partial
  p_{i}^2}\mid_{p=0} \Pi_{\bar{\chi},
  \chi}(p) 
\end{eqnarray}
$p_{i}$ ($i = 1,..,d$) is the $i$-th component of $\vec{p}$. Due to
rotational symmetry, $\beta_{\bar{\chi},\chi}$ is independent of the choice
of $i$. All other coefficients can be expressed in terms of derivatives of
$\Pi_{\chi,\chi}$ and $\Pi_{\bar{\chi}, \bar{\chi}}$. In $d=2$ space
dimensions and for small values of $g^2 k_F$, we obtain the
following results:
\begin{eqnarray}
  \alpha_{\bar{\chi},\chi} & \equiv & \alpha_{\chi,\bar{\chi}}\approx
  \frac{k_F\lambda_{j_{sc}}}{12 \pi v_F|\phi_{c}|^{2}}
  \\
  \beta_{\bar{\chi},\chi} & \equiv & \beta_{\chi,\bar{\chi}} \approx
  \frac{v_F k_F\lambda_{j_{sc}}}{48 \pi|\phi_{c}|^{2}}
  \\
  \alpha_{\chi,\chi} & \equiv & \alpha_{\bar{\chi},\bar{\chi}}\approx
  - \frac{k_F\lambda_{j_{sc}}}{24 \pi v_F|\phi_{c}|^{2}}
  \\
  \beta_{\chi,\chi} & \equiv & \beta_{\bar{\chi},\bar{\chi}} \approx
  - \frac{v_F k_F\lambda_{j_{sc}}}{96 \pi|\phi_{c}|^{2}}
\end{eqnarray}

Splitting
$\chi$ into a transversal and a tangential mode, we arrive at the expression
\begin{eqnarray}
  \lefteqn{\int\dbar\,^{3}p\,
  \chi_{t}(p)(
  \frac{k_F\lambda_{j_{sc}}}{12 \pi v_F|\phi_{c}|^{2}}
  p_{0}^2 +
  \frac{v_F k_F\lambda_{j_{sc}}}{48 \pi|\phi_{c}|^{2}}
  | \vec{p} |^2) \chi_{t}(p)} 
  \nonumber\\
  && +\int\dbar\,^{3}p\,
  \chi_{l}(p)(
  \frac{k_F\lambda_{j_{sc}}}{4 \pi v_F|\phi_{c}|^{2}}
  p_{0}^2 +
  \frac{v_F k_F\lambda_{j_{sc}}}{16 \pi |\phi_{c}|^{2}}
  | \vec{p} |^2) \chi_{l}(p)
\end{eqnarray}
for the kinetic part of the effective action. For $g^2 k_F$ small and
$d = 2$, the
contributions to $S^{(3)}(\bar{\chi},\chi)$ of order zero in the momenta 
turn out to be
\begin{eqnarray}
  \lefteqn{\int\dbar\,^{3}p_1\dbar\,^{3}p_2\,
  \frac{1}{3!}\frac{k_F^{3} v_F\lambda_{j_{sc}}}{\pi|\phi_{c}|^3}
  (-\chi_{l}(p_1)\chi_{t}(-p_2)\chi_{l}(p_2-p_1)}
  \nonumber\\
  && + \chi_{t}(p_1)\chi_{t}(-p_2)\chi_{t}(p_2-p_1))\hspace{20mm}
\end{eqnarray}

Rescaling $\chi_{t}$ by a factor
$(\frac{k_F\lambda_{j_{sc}}}{12 \pi v_F|\phi_{c}|^{2}})^{\frac{1}{2}}$
and $\chi_{l}$ by a factor  
$(\frac{k_F\lambda_{j_{sc}}}{4 \pi v_F|\phi_{c}|^{2}})^{\frac{1}{2}}$
, the
lowest order terms in the effective action for the  
$\chi$-field take the following standard form :
\begin{eqnarray}\label{action4}
  \lefteqn{S(\chi,\bar{\chi})=
  \int\dbar\,^{3}p\,
  \chi_{t}(p)(p_{0}^2 +
  \frac{1}{4} v_{F}^{2}
  | \vec{p} |^2) \chi_{t}(-p)} 
  \nonumber\\
  && +\int\dbar\,^{3}p\,
  \chi_{l}(p)(p_{0}^2 +
  \frac{1}{4} v_{F}^{2}
  | \vec{p} |^2) \chi_{l}(-p)
  \nonumber\\
  && +\int\dbar\,^{3}p\, \ \ 
  12 v_F^2 k_F^2\ \ 
  \chi_{t}(p)\chi_{t}(-p)\\
  && +\int\dbar\,^{3}p_1\dbar\,^{3}p_2\,\ \ 
  4 v_F^2 k_F^2\ \ 
  \sqrt{\frac{\pi v_F}{3 k_F \lambda_{j_{sc}}}}
  \nonumber\\
  && \left[
  - \chi_{l}(p^{(1)})\chi_{t}(-p^{(2)})\chi_{l}(p^{(2)}-p^{(1)})
  + 3 \chi_{t}(p^{(1)})\chi_{t}(-p^{(2)})\chi_{t}(p^{(2)}-p^{(1)})
  \right]
  \nonumber
\end{eqnarray}

At tree level, the propagator $\langle\chi_{t}(0)\chi_{t}(x)\rangle$
of the $\chi_t$-field in $x$-space decays exponentially in $|x|$, with
decay rate M. Using expression (\ref{action4}) to calculate radiative
corrections, the behaviour of the propagator of $\chi_t$ changes
drastically. This is due to the vertex in $S(\bar{\chi},\chi)$
proportional to $\chi_l^2 \chi_t$. The dominant one-loop radiative correction
to the propagator of $\chi_t$ is proportional to
\begin{eqnarray}\label{corrprop}
  \int d^{d+1}y d^{d+1}z\, \langle\chi_{t}(0)\chi_{t}(y)\rangle_0 \ 
  \langle\chi_{l}(y)\chi_{l}(z)\rangle_0^2\ 
  \langle\chi_{t}(z)\chi_{t}(x)\rangle_0
\end{eqnarray}
where $\langle(\cdot)\rangle_0$ indicates that the expectation value is
calculated at tree level. Because the propagator 
$\langle\chi_{t}(0)\chi_{t}(x)\rangle_0$ decays exponentially in $|x|$,
while $\langle\chi_{l}(0)\chi_{l}(x)\rangle_0\approx |x|^{1-d}$, expression
(\ref{corrprop}) is proportional to 
\be
\langle\chi_{l}(0)\chi_{l}(x)\rangle_0\sim\mid x
\mid^{2-2d}
\ee
Thus the seemingly massive field quanta of the field $\chi_t$ are resonances
that decay into pairs of massless Goldstone bosons, as announced.

The form (\ref{action4}) of the effective action of the $\chi$-field shows
that in {\em one} space dimension the fluctuations of $\chi_l$ are
{\em logarithmically divergent} (logarithmic infrared divergence of
$\int\dbar^2p\frac{1}{p_0^2 + {\rm const}\ p_1^2}$). The prediction of mean
field theory that the continuous U(1) gauge symmetry is spontaneously broken
is therefore wrong (Mermin-Wagner theorem). The same conclusion is reached
in two space dimensions at {\em positive} temperature, $T$. The propagator of
$\chi_l$ in momentum space is then proportional to
\be
\frac{1}{(k_B T n)^2 + \mbox{const}\,(p_1^2 + p_2^2)}
\ee
where the {\em Matsubara frequencies} $n$ are integers. The term corresponding
to
$n=0$ yields logarithmically divergent fluctuations of $\chi_l$, and, again,
the U(1) symmetry is restored. However, these systems exhibit a {\em
Kosterlitz-Thouless transition}, as $g$ is varied.

But, for $d=2$ and $T=0$, or for $d\geq 3$ and at sufficiently small 
temperatures,
the predictions of mean field theory concerning spontaneous symmetry breaking
are qualitatively correct !

The form (\ref{action2}) of the effective action
$\tilde{S}(\bar{\psi},\psi,\bar{\phi},\phi)$ and Eqs. (\ref{BCSparm1}) and
(\ref{BCSparm2}) show that
the field quanta of the $\chi$-field have electric charge $\pm 2e$, (where e
is the elementary electric charge). If the Coulomb interactions between
charged quasi-particles are incorporated in our theory then the {\em Goldstone
bosons acquire a  positive mass (``Anderson-Higgs mechanism'')}, as
briefly discussed in Chapter 5 of Part I.

The prediction of our theory for the mass, $\approx g|\phi_c|$ with
$|\phi_c|$ given by (\ref{gsphi}), of quasi-particles of charge $\pm e$
agrees with the solution of the standard BCS gap equation.

\end{document}